\documentclass[hyper,a4paper]{JHEP3} 

\usepackage{epsfig}
\usepackage{latexsym}
\usepackage{graphicx}
\usepackage{amsmath}
\usepackage{amsfonts}   
\usepackage{amssymb}    

\def\e3{$\epsilon_3$}

\def\ch2{$\chi^2$}

\def\co#1{{\ifmmode{\cal O}_{#1}\else${\cal O}_{#1}$\fi}}

\newdimen\unit
\def\point#1 #2 #3{\vbox to0pt{\kern-#2\unit
  \hbox{\kern#1\unit#3}\vss}
 \nointerlineskip}

\newcommand{\be}{\begin{equation}}
\newcommand{\ee}{\end{equation}}
\newcommand{\bea}{\begin{eqnarray}}
\newcommand{\eea}{\end{eqnarray}}

\newcommand\ps{\mbox{ ps}} 
\newcommand{\mev}{\mbox{ MeV}}
\newcommand{\gev}{\mbox{ GeV}}
\newcommand{\tev}{\mbox{ TeV}}

\newcommand{\cl}{\text{CL}}

\newcommand{\alphaemmz}{\alpha_{\text{em}}(M_Z)^{\overline{MS}}}
\newcommand{\alphas}{\alpha_s(M_Z)^{\overline{MS}}}

\newcount\hour
\newcount\minute
\newtoks\amorpm
\hour=\time\divide\hour by60 \minute=\time{\multiply\hour by60
\global\advance\minute by- \hour}
\edef\standardtime{{\ifnum\hour<12 \global\amorpm={am}%
    \else\global\amorpm={pm}\advance\hour by-12 \fi
    \ifnum\hour=0 \hour=12 \fi
    \number\hour:\ifnum\minute<100\fi\number\minute\the\amorpm}}
\edef\militarytime{\number\hour:\ifnum\minute<100\fi\number\minute}
\def\bold#1{\setbox0=\hbox{$#1$}%
     \kern-.025em\copy0\kern-\wd0
     \kern.05em\copy0\kern-\wd0
     \kern-.025em\raise.0433em\box0 }

\newcommand{\newc}{\newcommand}
\newc\eg{{\rm {e.g.}}}  \newc\etal{{\rm {et al.}}} \newc\ie{{\rm i.e.}}
\newc\etc{{\rm {etc}}}
\newcommand\lsim{\mathrel{\rlap{\lower4pt\hbox{\hskip1pt$\sim$}}
    \raise1pt\hbox{$<$}}}
\newcommand\gsim{\mathrel{\rlap{\lower4pt\hbox{\hskip1pt$\sim$}}
    \raise1pt\hbox{$>$}}}
\newc{\mhalf}{m_{1/2}}      \newc{\mzero}{m_0}
\newc{\tanb}{\tan\beta}
\newc{\azero}{A_0}
\newc{\at}{A_t} \newc{\ab}{A_b} \newc{\atau}{A_\tau}
\newc{\bmu}{B\mu}           \newc{\sgn}{{\rm sgn}}
\newc{\mone}{M_1}           \newc{\mtwo}{M_2}

\newc{\charone}{\chi_1^\pm} \newc{\mcharone}{m_{\chi_1^\pm}}

\newc{\hl}{h}               \newc{\mhl}{m_{\hl}}   \newc{\gammahl}{\Gamma_{\hl}}
\newc{\hh}{H}               \newc{\mhh}{m_{\hh}}   \newc{\gammahh}{\Gamma_{\hh}}
\newc{\ha}{A}               \newc{\mha}{m_{\ha}}   \newc{\gammaha}{\Gamma_{\ha}}
\newc{\hpm}{H^{\pm}}        \newc{\mhpm}{m_{\hpm}} \newc{\gammahpm}{\Gamma_{\hpm}}
\newc{\hp}{H^{+}} \newc{\mhp}{m_{\hp}} \newc{\hm}{H^{-}}
\newc{\mhm}{m_{\hm}}
\newc{\xt}{X_{t}}           \newc{\xb}{X_{b}}

\newc{\qzero}{Q_0}          \newc{\qstop}{Q_{\widetilde t}}
\newc{\amu}{a_{\mu}}        \newc{\amususy}{a_{\mu}^{\text{SUSY}}}
\newc{\amuexpt}{a_{\mu}^{\text{expt}}}        \newc{\amusm}{a_{\mu}^{\text{SM}}}
\newc{\deltaamususy}{\delta a_{\mu}^{\text{SUSY}}}
\newc\gmtwo{(g-2)_{\mu}} \newc\deltaamu{\Delta a_{\mu}}
\newc{\msbar}{\overline{MS}} \newc{\drbar}{\overline{DR}}
\newc{\yt}{h_t} \newc{\yb}{h_b} \newc{\ytau}{h_{\tau}}

\newc{\mtop}{m_t}               \newc{\mtpole}{M_t}
\newc{\mtaupole}{m_{\tau}^{\text{pole}}}
\newc{\mtmtsmmsbar}{m_t(m_t)^{\msbar}_{{\text{SM}}}}
\newc{\mtmtsmdrbar}{m_t(m_t)^{\drbar}_{{\text{SM}}}}
\newc{\mtmtmssmdrbar}{m_t(m_t)^{\drbar}_{{\text{SUSY}}}}

\newc{\mbmbmsbar}{m_b(m_b)^{\msbar} }

\newc{\mbmbsmmsbar}{m_b(m_b)^{\msbar}_{{\text{SM}}}}
\newc{\mbmzsmmsbar}{m_b(\mz)^{\msbar}_{{\text{SM}}}}
\newc{\mbmzsmdrbar}{m_b(\mz)^{\drbar}_{{\text{SM}}}}
\newc{\mbmzmssmdrbar}{m_b(\mz)^{\drbar}_{{\text{SUSY}}}}

\newc{\mtaumzsmmsbar}{m_{\tau}(\mz)^{\msbar}_{{\text{SM}}}}
\newc{\mtaumzsmdrbar}{m_{\tau}(\mz)^{\drbar}_{{\text{SM}}}}
\newc{\mtaumzmssmdrbar}{m_{\tau}(\mz)^{\drbar}_{{\text{SUSY}}}}

\newc{\mgut}{M_{\rm GUT}}
\newc{\mplanck}{M_{\rm P}}      \newc{\mpl}{M_{\text{Pl}}}
\newc{\msusy}{M_{\rm SUSY}}      \newc{\ms}{M_{\text{S}}}
\newc{\jxf}{J({\xf})}
\newc{\jxfexact}{J_{\rm exact}({\xf})}  \newc{\jxfexp}{J_{\rm exp}({\xf})}
\newc{\VEV}[1]{\langle #1 \rangle}
\newc{\xf}{x_f}
\newc\vrel{v_{\rm rel}}
\newcommand\mchi{m_{\chi}}              
\newc\sell{{\widetilde e}_L}      \newc\msell{m_{\sell}}
\newc\selr{{\widetilde e}_R}      \newc\mselr{m_{\selr}}
\newc\snue{{\widetilde \nu}_e}      \newc\msnue{m_{\snue}}
\newc\snutau{{\widetilde \nu}_\tau}      \newc\msnutau{m_{\snutau}}
\newc\supl{{\widetilde u}_L}      \newc\msupl{m_{\supl}}
\newc\supr{{\widetilde u}_R}      \newc\msupr{m_{\supr}}
\newc\sdl{{\widetilde d}_L}      \newc\msdl{m_{\sdl}}
\newc\sdr{{\widetilde d}_R}      \newc\msdr{m_{\sdr}}

\newcommand\stopone{{\widetilde t}_1}   \newcommand\mstopone{m_{\stopone}}
\newcommand\stoptwo{{\widetilde t}_2}   \newcommand\mstoptwo{m_{\stoptwo}}

\newcommand\gluino{\widetilde g}

\newc\sfermion{\tilde f}  \newc\msfermion{m_{\sfermion}}
\newc\cmeter{{\rm cm}} \newc\meter{{\rm m}} \newc\kmeter{{\rm km}}
\newc\second{{\rm sec}}
\newc{\gstar}{g_\ast}           \newc{\gsstar}{g_{s\ast}}
\newc{\geff}{g_{\rm eff}}
\newcommand\mz{m_{Z}}

\newc{\sthw}{\sin\theta_W}              \newc{\cthw}{\cos\theta_W}
\newc{\bino}{\widetilde B}              \newc{\wino}{\widetilde W_30}
\newc{\higgsinob}{{\widetilde H}^0_b}   \newc{\higgsinot}{{\widetilde H}^0_t}
\newc{\abund}{\Omega h^2}
\newc{\abundchi}{\Omega_\chi h^2}
\newc{\abundcdm}{\Omega_{\text{CDM}} h^2}
\newc{\omegam}{\Omega_{M}}       \newc{\abundm}{\Omega_{M} h^2}
\newc{\omegab}{\Omega_{b}}       \newc{\abundb}{\Omega_{b} h^2}
\newc{\omegacdm}{\Omega_{CDM}}
\newc{\omegatot}{\Omega_{TOT}}
\newc{\rhocrit}{\rho_{crit}}
\newc{\rhochi}{\rho_{\chi}}
\newcommand\pb{\,\mbox{pb}} \newcommand\fb{\,\mbox{fb}}

\newc\BR{BR}
\newc\bsgamma{b\rightarrow s \gamma }
\newc\bxsgamma{\overline{B}\rightarrow X_{s}\gamma}
\newc\brbsgamma{\BR(\overline{B}\rightarrow X_s\gamma)}

\newcommand\brbsmumu{\BR(\overline{B}_s\to\mu^+\mu^-)}

\newcommand\bbbarmix{\overline{B}_s\mbox{-}B_s}      
\newcommand\delmbs{\Delta M_{B_s}}

\newc{\beq}{\begin{equation}}
\newc{\eeq}{\end{equation}}

\newc\stoponetwo{{\widetilde t}_{1,2}}
\newc\sbotonetwo{{\widetilde b}_{1,2}}
\newc\stauonetwo{{\widetilde \tau}_{1,2}}

\newc{\sigsip}{\sigma^{SI}_{p}} \newc{\sigsin}{\sigma^{SI}_{n}}
\newc{\sigsiN}{\sigma^{SI}_{N}}
\newc{\sigsdp}{\sigma^{SD}_{p}} \newc{\sigsdn}{\sigma^{SD}_{n}}
\newc{\sigsiA}{\sigma^{SI}_{A}}

\newc\xilim{\xi_{\rm lim}} 
\newc\tlim{t_{\rm lim}} 
\newc\zetalim{\zeta_{\rm lim}} 

\newc\zetah{\zeta_h}
\newc{\relprobone}[1]{p({#1} \vert d)}
\newc{\relprobtwo}[2]{p({#1},{#2} \vert d)}

\long\def\begincomment#1\endcomment{%
        \begingroup\sf\baselineskip12pt#1\endgroup}

\newcommand{\squishlist}{
   \begin{list}{$\bullet$}
    { \setlength{\itemsep}{0pt}      \setlength{\parsep}{3pt}
      \setlength{\topsep}{3pt}       \setlength{\partopsep}{0pt}
      \setlength{\leftmargin}{1.em} \setlength{\labelwidth}{1em}
      \setlength{\labelsep}{0.5em} } }
\newcommand{\squishend}{
    \end{list}  }
        \newcommand\mW{m_{W}}

\newcommand{\data}{d}
\newcommand{\nuis}{\psi}
\newcommand{\params}{\theta}
\newcommand{\basis}{m}
\newcommand{\derived}{\xi}

\newcommand{\sineff}{\sin^2 \theta_{\text{eff}}}

\newcommand\jcap[3]   { 
		{{\it JCAP\ }{\bf #1} (#2) #3}}

\newcommand\mnras[3]   { 
		{{\it Mon.\ Not.\ R.\ Astron.\ Soc.\ }{\bf #1} (#2) #3}}

\title{Implications for the Constrained MSSM\\
  from a new prediction for {\boldmath $\bsgamma$}}
\author{Leszek Roszkowski\\
        Department of Physics and Astronomy, University of Sheffield,\\
        Sheffield S3 7RH, England, and\\
Theory Division, CERN, CH-1211 Geneva 23, Switzerland\\
        E-mail: \email{L.Roszkowski@sheffield.ac.uk}}
\author{Roberto Ruiz de Austri\\
        Departamento de F\'{\i}sica Te\'{o}rica C-XI
        and Instituto de F\'{\i}sica Te\'{o}rica C-XVI,\\
        Universidad Aut\'{o}noma de Madrid, Cantoblanco,
        28049 Madrid, Spain\\
        E-mail: \email{rruiz@delta.ft.uam.es}}
\author{Roberto Trotta\\
        Astrophysics Department, Oxford University \\
        Denys Wilkinson Building,  Keble Road, Oxford OX1 3RH, United Kingdom\\
         E-mail: \email{rxt@astro.ox.ac.uk}}

\abstract{\small We re-examine the properties of the Constrained MSSM
  in light of updated constraints, paying particular attention to the
  impact of the recent substantial shift in the Standard Model
  prediction for $\brbsgamma$. With the help of a Markov Chain Monte
  Carlo scanning technique, we vary all relevant parameters
  simultaneously and derive Bayesian posterior probability maps.  We
  find that the case of $\mu>0$ remains favored, and that for $\mu<0$
  it is considerably more difficult to find a good global fit to
  current constraints.  In both cases we find a strong preference for
  a focus point region. This leads to improved prospects for detecting
  neutralino dark matter in direct searches, while superpartner
  searches at the LHC become more problematic, especially when
  $\mu<0$. In contrast, prospects for exploring the whole mass range
  of the lightest Higgs boson at the Tevatron and the LHC remain very
  good, which should, along with dark matter searches, allow one to
  gain access to the otherwise experimentally challenging focus point
  region. An alternative measure of the mean quality-of-fit which we
  also employ implies that present data are not yet constraining
  enough to draw more definite conclusions. We also comment on the
  dependence of our results on the choice of priors and on some other
  assumptions.}

\begin{document}

\section{Introduction}\label{sec:intro}

Among various possible sets of boundary conditions that one can impose
on the multi-dimensional parameter space of the effective, low-energy
Minimal Supersymmetric Standard Model (MSSM),~\cite{susy-reviews} the
by far most popular choice is the so-called Constrained MSSM
(CMSSM)~\cite{kkrw94}.\footnote{One well-known implementation of the
CMSSM is the minimal supergravity model~\cite{sugra-reviews}.}  In the
CMSSM, at the GUT scale the soft masses of all the sleptons, squarks
and Higgs bosons have a common scalar mass $\mzero$, all the gauginos
unify at the common gaugino mass $\mhalf$, and so all the tri-linear
terms assume a common tri-linear mass parameter $\azero$. In addition,
at the electroweak scale one selects $\tanb$, the ratio of Higgs
vacuum expectation values and $\text{sign}(\mu)$, where $\mu$ is the
Higgs/higgsino mass parameter whose square is computed from the
conditions of radiative electroweak symmetry breaking (EWSB).

The small number of parameters makes the CMSSM a popular framework for
exploring SUSY phenomenology. Conversely, collider data provides
useful constraints on the parameter space (PS) of the CMSSM. In the
presence of R-parity the lightest neutralino is often the lightest
supersymmetric particle (LSP). Assuming it to be the dominant
component of cold dark matter (DM) in the Universe, allows one to
apply the DM relic density determination by WMAP and other experiments
as a strong constraint on the CMSSM PS~\cite{susy-dm-reviews}.

Another important constraint on the CMSSM comes from the process
$\bsgamma$. An approximate agreement of the Standard Model (SM)
prediction for $\brbsgamma$ with an experimental determination
requires the sum of SUSY contributions, which enter at the same 1-loop
level, to be strongly suppressed. While the experimental world average
has for over a year remained at $(3.55 \pm 0.26) \times
10^{-4}$~\cite{bsgexp}, the recently re-evaluated SM value of
$\brbsgamma$, as obtained by Misiak \etal,
in~\cite{ms06-bsg,mm-prl06}, has moved quite substantially from $(3.60
\pm 0.30) \times 10^{-4}$ down to $(3.15 \pm 0.23) \times
10^{-4}$.\footnote{A further slight decrease to $(2.98 \pm 0.26)
\times 10^{-4}$ after including some additional partial effects due to
a treatment of a photon energy cut $E_\gamma>1.6\gev$ was obtained in
ref.~\cite{bn06}. Note that the above values do depend on the choice
of the cut in $E_\gamma$.}  The main shift was caused by including new
partial NNLO SM contributions, most importantly an approximate
evaluation of the charm mass effects. The new SM value leads to some
discrepancy, at the level of $1.2\sigma$, with the experimental
average.

From the perspective of SUSY corrections, much more important than
this slight discrepancy is the fact that the SM central value has now
moved from above to below the experimental one.  In the case of
minimal flavor violation, which is applicable to the CMSSM, dominant
SUSY contributions come from the charged Higgs/top loop, which always
adds constructively to the SM contribution, and from the chargino/stop
loops, whose sign is opposite to that of $\mu$. With the previous SM
value the $\bsgamma$ branching ratio, this was used as an argument for
assuming $\mu$ to be positive. Indeed, for $\mu<0$ one had to push
superpartner masses into the multi-TeV range in order for the
chargino/stop loop correction to become suppressed, while in the
opposite case much smaller masses were allowed.  However, the recent
shift in the SM predictions for $\brbsgamma$ makes the argument for
selecting $\mu>0$ questionable, and has in fact motivated us to
perform this analysis.

Another argument that is often used in favor of $\mu>0$ is based on a
persistent discrepancy of $\amuexpt-\amusm= (28 \pm 8.1)\times
10^{-10}$ between the experimental value and the SM prediction of the
anomalous magnetic moment of the muon $\gmtwo$~\cite{pdg06}.  Taking
the nearly $3.5\sigma$ difference as being due to SUSY contributions,
$\deltaamususy\equiv \amuexpt-\amusm$, (whose sign is the same as that
of $\mu$), implies $\mu>0$.

However, such conclusions are based on a somewhat oversimplified
treatment of both theoretical and experimental uncertainties, which is
common practice in fixed-grid scans of a SUSY PS. In such studies, a
``step-function'' approach is usually adopted: regions of the PS where
contributions to a given observable are within the $\pm1\sigma$ (or
some other) range around the experimental central value are
treated as fully allowed, while those even slightly outside are
treated as completely ruled out. The same applies to experimental
limits, e.g., on Higgs or superpartner masses. Instead, it seems more
justified to assign varying ``weights'' to different points in a PS,
depending on how well, or how poorly, a prediction for a given
observable matches its experimental determination.  Furthermore, in
the usual approach theoretical errors are typically neglected, and so are
residual uncertainties in relevant SM parameters, simply for the
reason of practicality.

Recently a more refined procedure has been developed which allows
one to overcome these shortcomings. It is based on a statistical
Bayesian analysis linked with a Markov Chain Monte Carlo (MCMC)
scanning techniques~\cite{mcmc,bg04}, and is becoming increasingly
popular in studying SUSY
phenomenology~\cite{al05,allanach06,rtr1,alw06,rrt2,Trotta:2006ew}.
The MCMC technique allows one to make a thorough scan of a model's
full multi-dimensional PS. Additionally maps of probability
distributions can be drawn both for the model's parameters and
also for all the observables (and their combinations) included in
the analysis. In this approach, sharply defined ``allowed
regions'' drawn up in fixed-grid scans are replaced by more
informative probability distributions.

The MCMC Bayesian approach to studying properties of ``new physics''
models, like the CMSSM, is superior in the sense of treating the
impact of different experimental data with their proper weights. It
allows to make global scans of the PS and to derive its global
properties and predictions. When (hopefully) discoveries are
eventually made at current or future experiments, the approach will
provide invaluable in assessing their implications for a given
theoretical model.

In this work we apply the MCMC Bayesian formalism to explore the
impact on the CMSSM's properties from mostly the recent change in the
SM value for $\brbsgamma$. As we will see, regions of the highest
posterior probability, will move rather dramatically to the focus
point (FP)~\cite{focuspoint-fmm} region of the CMSSM PS. This in turn will lead
to a significant shift in prospects for superpartner searches at the
LHC (generally for worse) and in direct searches for DM neutralinos in
the Milky Way (generally for better), while chances of finding $\hl^0$
at the Tevatron will remain good.

In order to assess the robustness of the results obtained in
Bayesian language, following our previous work~\cite{rtr1,rrt2} we
also apply an alternative measure of a mean quality-of-fit, which
is similar to a popular $\chi^2$-measure, which is singles out
(possibly limited) ranges of parameters that give the best fit to
the data.

We consider both signs of $\mu$. In the probabilistic approach, the
case $\mu<0$ cannot be treated anymore as ruled out, but merely as
disfavored, by the $\gmtwo$ result. The relative weight of this
constraint has to be compared with that of other observables in a
proper statistical way. In a recent similar study of Allanach
\etal,~\cite{alw06} (although done with the old values of $\brbsgamma$
and $\deltaamususy$) fits for both signs of $\mu$ were performed. It
was concluded that the the case $\mu<0$ was only marginally
disfavored, with the ratio of probabilities estimated at
$P(\mu<0)/P(\mu>0)=0.07 -0.16$. 
In our study we also find that the case of $\mu<0$ gives a worse fit
to the data than the opposite sign of $\mu$, although the level of
preference for $\mu>0$ is difficult to quantify.
On the other hand, unlike
in~\cite{alw06}, we are not interested in comparing the relative
probabilities of the two cases $\mu<0$ and $\mu>0$, but rather
emphasize different implications of each one for various observables
of interest.

In this paper we include, and update when applicable, all relevant
experimental constraints from collider direct searches and from rare
processes, and also from cosmology on the relic abundance of the
lightest neutralino $\abundchi$. We further take into account residual
error bars in relevant SM parameters. Details of our analysis will be
given below.

We adopt flat priors on the usual CMSSM parameters: $\mhalf$,
$\mzero$, $\azero$ and $\tanb$. We do this primarily for the sake
of comparing our results with the literature (in particular with
the fixed-grid scan approach) where this parametrization is
usually assumed. Our specific results will accordingly depend on
the choice, as we discuss later.

Implications from the current analysis (assuming $\mu>0$) for the
Higgs bosons have already been presented in~\cite{rrt2} where we
showed that, with our choice of priors, the lightest Higgs boson
$\hl^0$ mass is confined to $115.4\gev < \mhl<120.4\gev$ (95\%
probability interval) and that its couplings to electroweak gauge
bosons are very close to those of the SM Higgs boson with the same
mass. This range should be excluded (at 95\%~\cl) at the Tevatron.
Here we extend our analysis to to the case $\mu<0$, reaching
similar conclusions for $\hl^0$ at the Tevatron. We also derive
most probable ranges of several sparticle masses, of the rates of
rare bottom quark processes, and of both spin-independent and
spin-dependent cross sections for dark matter neutralino
scattering off nuclei.

The paper is organized as follows. In Sec.~\ref{sec:analysis} we
outline our theoretical setup. In Sec.~\ref{sec:results} we present our numerical
results for the PS of the CMSSM in terms of the Bayesian statistics
and of the mean quality-of-fit, and  resulting implications
for several observables. We finish
with summary and conclusions in Sec.~\ref{sec:summary}.

\section{The Analysis}\label{sec:analysis}
Our procedure based on MCMC scans and Bayesian analysis has been presented in detail
in~\cite{rtr1}. Here, for completeness, we repeat its main features
following an updated presentation given in~\cite{rrt2}.

\subsection{Theoretical framework}\label{sec:setup}

In the CMSSM the parameters $\mhalf$, $\mzero$ and $\azero$, which are
specified at the GUT scale $\mgut\simeq 2\times 10^{16}\gev$, serve as
boundary conditions for evolving, for a fixed value of $\tanb$, the
MSSM Renormalization Group Equations (RGEs) down to a low energy scale
$\msusy\equiv \sqrt{\mstopone\mstoptwo}$ (where
$m_{\stopone,\stoptwo}$ denote the masses of the scalar partners of
the top quark), chosen so as to minimize higher order loop
corrections. At $\msusy$ the (1-loop corrected) conditions of
electroweak symmetry breaking (EWSB) are imposed and the SUSY spectrum
is computed at $\mz$.

We are interested in delineating high probability regions of the CMSSM
parameters.  We consider separately both signs of $\mu$ and denote
the remaining four free CMSSM parameters by the set
\be \label{indeppars:eq}
\params =  (\mzero, \mhalf, \azero, \tanb ).
\ee
As demonstrated in~\cite{al05,rtr1}, the values of the relevant SM
parameters can strongly influence some of the CMSSM predictions, and,
in contrast to common practice, should not be simply kept fixed at
their central values. We thus introduce a set $\nuis$ of so-called
{\em ``nuisance parameters''} of those SM parameters which are
relevant to our analysis,
\be \label{nuipars:eq} \nuis = ( \mtpole,
\mbmbmsbar, \alphaemmz, \alphas ), \ee
where $\mtpole$ is the pole top quark mass. The other three
parameters:  $\mbmbmsbar$ -- the bottom
quark mass evaluated at $m_b$, $\alphaemmz$ and $\alphas$ -- respectively the
electromagnetic and the strong coupling constants evaluated at the $Z$ pole mass
$M_Z$ -  are all computed in the $\msbar$ scheme.

The set of parameters $\params$ and $\nuis$ form an 8-dimensional set
$\basis$ of our {\em ``basis parameters''} $\basis = (\params,
\nuis)$.\footnote{In~\cite{rtr1} we denoted our basis parameters with a
symbol $\eta$. }
In terms of the basis parameters we compute a number of
collider and cosmological observables, which we call {\em ``derived
variables''} and which we collectively denote by the set $\derived=
(\derived_1, \derived_2, \ldots)$.  The observables, which are listed
below, will be used to compare CMSSM predictions with a set of
experimental data $\data$, which is  available either in the
form of positive measurements or as limits.

In order to map out high probability regions of the CMSSM, we compute
the {\em posterior probability density functions} (pdf's) $p(\basis
|\data)$ for the basis parameters $\basis$ and for several
observables.  The posterior pdf represents our state of knowledge
about the parameters $\basis$ after we have taken the data into
consideration. Using Bayes' theorem, the posterior pdf is given by
\be \label{eq:bayes}
 p(\basis | \data) = \frac{p(\data |
\derived) \pi(\basis)}{p(\data)}. \ee
On the r.h.s. of eq.~\eqref{eq:bayes}, the quantity $p(\data |
\derived)$, taken as a function of $\derived$ for {\em fixed data}
$\data$, is called the {\em likelihood} (where the dependence of
$\derived(\basis)$ is understood).  The likelihood supplies the
information provided by the data and, for the purpose of our analysis,
it is constructed in Sec.~3.1 of ref.~\cite{rtr1}.  The quantity
$\pi(\basis)$ denotes a {\em prior probability density function}
(hereafter called simply {\em a prior}) which encodes our state of
knowledge about the values of the parameters in $\basis$ before we see
the data. The state of knowledge is then updated to the posterior via
the likelihood.  Finally, the quantity in the denominator is called
{\em evidence} or {\em model likelihood}. Here it only serves a
normalization constant, independent of $\basis$, and therefore will be
dropped in the following.  As in ref.~\cite{rtr1}, our posterior pdf's
presented below will be normalized to their maximum values, and {\em
not} in such a way as to give a total probability of 1. Accordingly we
will use the name of a ``relative posterior pdf'', or simply of
``relative probability density''.

The Bayesian approach to parameter inference relies on the
updating of the prior probability to the posterior through the
information provided by the data (via the likelihood). This
requires specification of the prior probabilities for the
parameters of the model, that in our case are taken to be flat
(i.e., constant) over a large range of the CMSSM and SM parameters
given above. If the data are not strongly constraining, the choice
of prior can lead to a significant impact through the effect of
the ``volume'' of the parameter space. Indeed, as we discussed in
ref.~\cite{rtr1}, imagine the situation that there exist a rather
large region of the PS where theoretical predictions match the
data rather well. In addition, let there be a rather small,
possibly fined-tuned, region giving very good match of the data.
The Bayesian posterior probability would give an overwhelming
weight to the larger region, due to the much larger volume it
occupies in parameter space. We notice that this kind of situation
only arises in the ``grey zone'' of insufficient data, since of
course if the data were powerful enough as to rule out such a large
region, then the Bayesian posterior probability would show this by
peaking in correspondence with the best fitting, smaller region.
As done previously in refs.~\cite{rtr1,rrt2}, we therefore
consider also an alternative statistical measure of the mean
quality-of-fit defined in~\cite{rtr1}, which is much more
sensitive to possibly small best-fit regions. Below we will
compare results obtained using the two measures.

\subsection{Constraints}\label{sec:constraints}

\begin{table}
\centering    .

\begin{tabular}{|l | l l | l|}
\hline
SM (nuisance) parameter  &   Mean value  & \multicolumn{1}{c|}{Uncertainty} & ref. \\
 &   $\mu$      & ${\sigma}$ (exper.)  &  \\ \hline
$\mtpole$           &  171.4 GeV    & 2.1 GeV&  \cite{cdf+dzero-mtop-06} \\
$m_b (m_b)^{\overline{MS}}$ &4.20 GeV  & 0.07 GeV &  \cite{pdg06} \\
$\alphas$       &   0.1176   & 0.002 &  \cite{pdg06}\\
$1/\alphaemmz$  & 127.955 & 0.018 &  \cite{pdg06} \\ \hline
\end{tabular}
\caption{Experimental mean $\mu$ and standard deviation $\sigma$
 adopted for the likelihood function for SM (nuisance) parameters,
 assumed to be described by a Gaussian distribution.
\label{tab:meas}}
\end{table}

We perform a scan over very wide ranges of CMSSM
parameters~\cite{rtr1,rrt2}. In particular we take flat priors on
the ranges $50\gev < \mhalf,\mzero < 4 \tev$ (this way including
the focus point region in this kind of
analysis), $|\azero| < 7\tev$ and $2 < \tanb < 62$. For
the SM (nuisance) parameters, we assume flat priors over wide
ranges of their values~\cite{rtr1} and adopt a Gaussian likelihood
with mean and standard deviation as given in table~\ref{tab:meas}.
Note that, with respect to ref.~\cite{rtr1}, we have updated the
values of all the constraints.\footnote{After completing our
numerical scans, a new value of the top mass, $\mtpole=170.9\pm
1.8\gev$, based on Tevatron's Run-II $1\fb^{-1}$ of data was
released~\cite{topmass:mar07}. Including it would not have much
impact on our results, since the shift in the mean value of $M_t$
is very mild if compared to the standard deviation adopted in this
paper.}

\begin{table}
\centering
\begin{tabular}{|l | l l l | l|}
\hline
Observable &   Mean value & \multicolumn{2}{c|}{Uncertainties} & ref. \\
 &   $\mu$      & ${\sigma}$ (exper.)  & $\tau$ (theor.) & \\\hline
 $M_W$     &  $80.392\gev$   & $29\mev$ & $15\mev$ & \cite{lepwwg} \\
$\sineff{}$    &  $0.23153$      & $16\times10^{-5}$
                & $15\times10^{-5}$ &  \cite{lepwwg}  \\
$\deltaamususy \times 10^{10}$       &  28 & 8.1 &  1 & \cite{pdg06}\\
 $\brbsgamma \times 10^{4}$ &
 3.55 & 0.26 & 0.21 & \cite{bsgexp} \\
$\delmbs$     &  $17.33\ps^{-1}$  & $0.12\ps^{-1}$  & $4.8\ps^{-1}$
& \cite{cdf-deltambs} \\
$\abundchi$ &  0.104 & 0.009 & $0.1\,\abundchi$& \cite{wmap3yr} \\\hline
   &  Limit (95\%~\cl)  & \multicolumn{2}{r|}{$\tau$ (theor.)} & ref. \\ \hline
$\brbsmumu$ &  $ <1.0\times 10^{-7}$
& \multicolumn{2}{r|}{14\%}  & \cite{cdf-bsmumu}\\
$\mhl$  & $>114.4\gev$\ ($91.0\gev$)  & \multicolumn{2}{r|}{$3 \gev$}
& \cite{lhwg} \\
$\zetah^2$
& $f(m_h)$ & \multicolumn{2}{r|}{negligible}  & \cite{lhwg} \\
sparticle masses  &  \multicolumn{3}{c|}{See table~4 in ref.~\cite{rtr1}.}  & \\ \hline
\end{tabular}
\caption{Summary of the observables used in the analysis. Upper part:
Observables for which a positive measurement has been
made. $\deltaamususy=\amuexpt-\amusm$ denotes the discrepancy between
the experimental value and the SM prediction of the anomalous magnetic
moment of the muon $\gmtwo$. For central values of the SM input
parameters used here, the SM value of $\brbsgamma$ is
$3.11\times10^{-4}$, while the theoretical error of
$0.21\times10^{-4}$ includes uncertainties other than the parametric
dependence on the SM nuisance parameters, especially on $\mtpole$ and
$\alphas$.  As explained in the text, for each quantity we use a
likelihood function with mean $\mu$ and standard deviation $s =
\sqrt{\sigma^2+ \tau^2}$, where $\sigma$ is the experimental
uncertainty and $\tau$ represents our estimate of the theoretical
uncertainty. Lower part: Observables for which only limits currently
exist.  The likelihood function is given in
ref.~\cite{rtr1}, including in particular a smearing out of
experimental errors and limits to include an appropriate theoretical
uncertainty in the observables. $\mhl$ stands for the light Higgs mass
while $\zetah^2= g^2(\hl ZZ)_{\text{MSSM}}/g^2(\hl ZZ)_{\text{SM}}$,
where $g$ stands for the Higgs coupling to the $Z$ and $W$ gauge boson
pairs.
\label{tab:measderived}}
\end{table}

The experimental values of the collider and cosmological observables
that we apply
(our derived variables) are listed in table~\ref{tab:measderived},
with updates where applicable.  In our treatment of the radiative
corrections to the electroweak observables $M_W$ and
$\sineff$, starting from ref.~\cite{rrt2} we include full two-loop and
known higher order SM corrections as computed in
ref.~\cite{awramik-acfw04}, as well as gluonic two-loop MSSM
corrections obtained in~\cite{dghhjw97}. We further update an experimental
constraint from the anomalous magnetic moment of the muon $\gmtwo$
for which a discrepancy (denoted by $\deltaamususy$) between
measurement and SM predictions (based on $e^+e^-$ data) persists at
the level of $3.5\sigma$~\cite{pdg06}. We note here, however, that the
impact of this (still somewhat uncertain) constraint on our
findings will be rather limited because the corresponding error bar
remains relatively large.\footnote{Although the different evaluations
seem to be converging; \eg, recently $(27.6\pm8.1)\times 10^{-10}$ was obtained
in ref.~\cite{hmnt06}. }

As regards $\brbsgamma$, with the central values of SM input
parameters as given in table~\ref{tab:meas}, for the new SM prediction
we obtain the value of $(3.11\pm 0.21)\times10^{-4}$.\footnote{The
value of $(3.15\pm 0.23)\times10^{-4}$ originally derived in
ref.~\cite{ms06-bsg,mm-prl06} was obtained for slightly different
values of $\mtpole$ and $\alphas$. Note that, in treating the error
bar we have explicitly taken into account the dependence on $\mtpole$
and $\alphas$, which in our approach are treated parametrically. This
has led to a slight reduction of its value.  Note also that even
though the theoretical error is, strictly speaking, not Gaussian, it
can still be approximately treated as such as it represents an
estimate where a larger assumed error of the (dominant) uncertainty
due to non-perturbative effects is assigned lower probability -
M.~Misiak, private communication.}  We compute SUSY contribution to
$\brbsgamma$ following the procedure outlined in
refs.~\cite{dgg00,gm01} which were extended in
refs.~\cite{or1+2,for1+2} to the case of general flavor mixing. In
addition to full leading order corrections, we include large
$\tanb$-enhanced terms arising from corrections coming from beyond the
leading order and further include (subdominant) electroweak
corrections.

Regarding cosmological constraints, we use the determination of the
relic abundance of cold DM based on the 3-year data from
WMAP~\cite{wmap3yr} to constrain the relic abundance $\abundchi$ of
the lightest neutralino which we compute with high precision,
including all resonance and coannihilation effects, as explained in
ref.~\cite{rtr1}, and solve the Boltzmann equation numerically as in
ref.~\cite{darksusy}.  In order to remain on a conservative side, we
impose the following Gaussian distribution
\bea
\abundchi= 0.104\pm \sqrt{(0.009)^2+ (0.1\,\abundchi)^2}
=0.104\pm 0.009\,\sqrt{1+ 1.335\,(\abundchi/0.104)^2}.
\label{eq:abundchitot}
\eea
Note that our estimated theoretical uncertainty is of the same order as the
uncertainty from current cosmological determinations of
$\abundcdm$.

We further include in our likelihood function an improved 95\%~\cl\
limit on $\brbsmumu$ and a recent value of $\bbbarmix$ mixing,
$\delmbs$, which has recently been precisely measured at the Tevatron
by the CDF~Collaboration~\cite{cdf-deltambs}. In both cases we use
expressions from ref.~\cite{for1+2} which include dominant large
$\tanb$-enhanced beyond-LO SUSY contributions from Higgs penguin
diagrams.  Unfortunately, theoretical uncertainties, especially in
lattice evaluations of $f_{B_s}$ are still very large (as reflected in
table~\ref{tab:measderived} in the estimated theoretical error for
$\delmbs$), which makes the impact of this precise measurement on
constraining the CMSSM parameter space rather limited.\footnote{On
the other hand, in the MSSM with general flavor mixing, even with the
current theoretical uncertainties, the bound from
$\delmbs$ is in many cases much more constraining than from other rare
processes~\cite{for3+4}.}

For the quantities for which positive measurements have been made (as
listed in the upper part of table~\ref{tab:measderived}), we assume a
Gaussian likelihood function with a variance given by the sum of the
theoretical and experimental variances, as motivated by eq.~(3.3) in
ref.~\cite{rtr1}. For the observables for which only lower or upper
limits are available (as listed in the bottom part of
table~\ref{tab:measderived}) we use a smoothed-out version of the
likelihood function that accounts for the theoretical error in the
computation of the observable, see eq.~(3.5) and fig.~1
in~\cite{rtr1}. In applying the Higgs boson $\hl^0$ lower mass bounds
from LEP-II we take into account its dependence on its coupling to the
$Z$ boson pairs $\zetah^2 \equiv g^2(\hl ZZ)_{\text{MSSM}}/g^2(\hl
ZZ)_{\text{SM}}$, as described in detail in ref.~\cite{rrt2}. When
$\zetah^2\simeq 1$, the LEP-II lower bound of $114.4\gev$
(95\%~\cl)~\cite{lhwg} is applicable. For arbitrary values of
$\zetah$, we apply the LEP-II 95\%~\cl\ bounds on $\mhl$ and $\mha$,
which we translate into the corresponding 95\%~\cl\ bound in the
$(\mhl, \zetah^2)$ plane. We then add a theoretical uncertainty
$\tau(\mhl) = 3\gev$, following eq.~(3.5) in
ref.~\cite{rtr1}. However, a posteriori we find $\zeta_h^2 \simeq 1$
which means that the CMSSM light Higgs boson is invariably SM-like.
This procedure results in a conservative likelihood function for
$\mhl$, which does not simply cut away points below the 95\%~\cl\
limit of LEP-II, but instead assigns to them a lower probability that
gradually goes to zero for lower masses.

Finally, points that do not fulfil the conditions of radiative
EWSB and/or give non-physical (tachyonic) solutions are discarded.
We adopt the same convergence and mixing criteria as described in
appendix~A2 of ref.~\cite{rtr1}, while our sampling procedure is
described in appendix~A1 of ref.~\cite{rtr1}. We have the total of
$N=10$ MC chains, with a merged number of samples $2\times10^5$,
and an acceptance rate of about 1.5\%. We adopt the Gelman \&
Rubin mixing criterion, with the inter-chain variance divided by
the intra-chain variance (the $R-1$ parameter) being less then
0.1 along all directions in parameter space. More details of our
numerical MCMC scan can be found in~\cite{rtr1}.

\section{Results}\label{sec:results}

We will now explore the implications of the above constraints on the
CMSSM parameters, paying some more attention to the impact of
$\brbsgamma$. We will compare the posterior probability distributions
of the Bayesian language with the ranges favored by the mean
quality-of-fit.  Next, we will discuss implications for Higgs and
superpartner masses and for direct detection of the neutralino dark
matter. In computing the Higgs (and SUSY) mass spectrum we employ the code
SOFTSUSY~v2.08~\cite{softsusy}.

\subsection{Implications for the CMSSM parameters}\label{sec:cmssmps}

\begin{figure}[tbh!]
\begin{center}
\begin{tabular}{c c c}
  \includegraphics[width=0.3\textwidth]{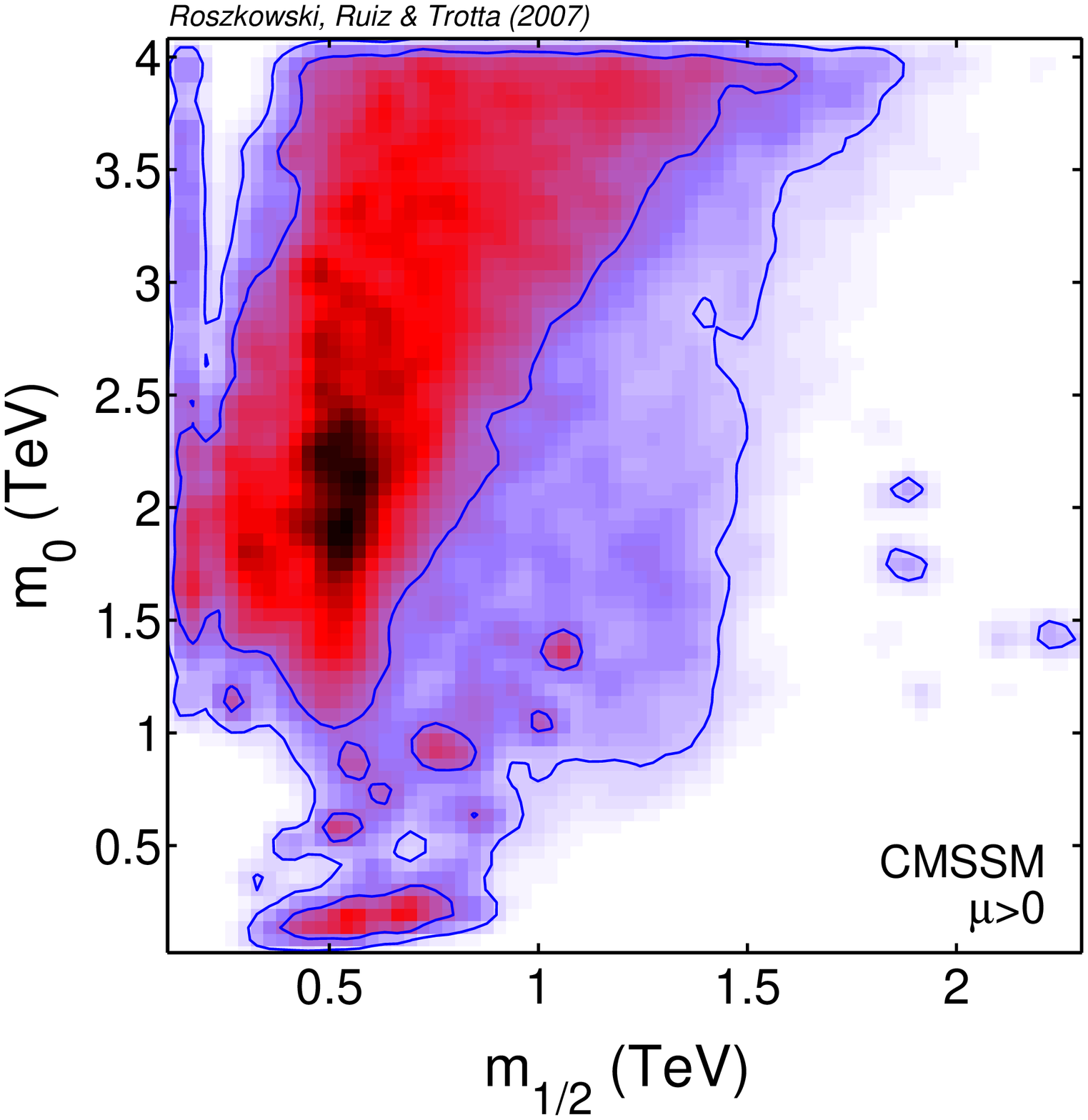}
& \includegraphics[width=0.3\textwidth]{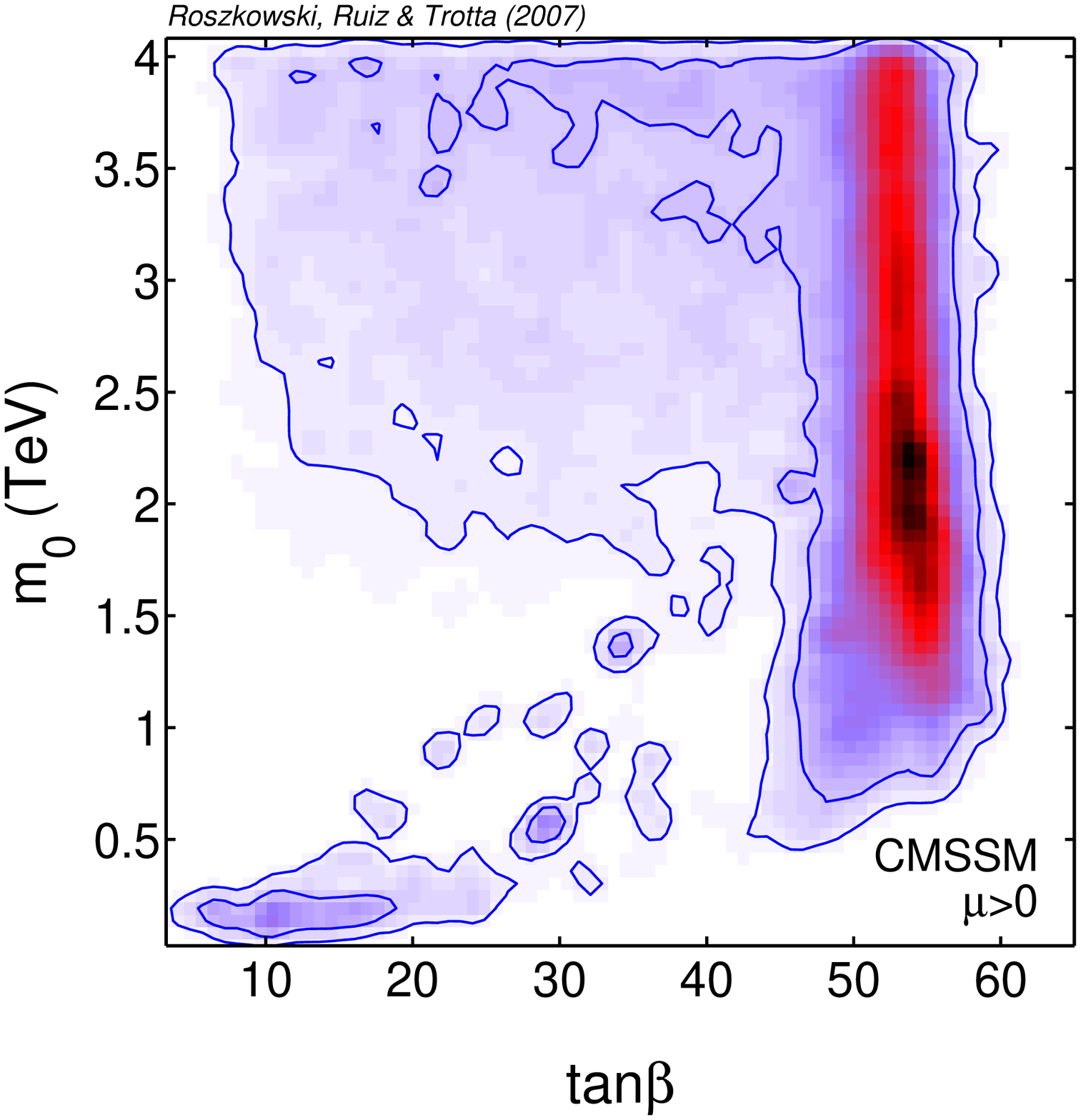}
& \includegraphics[width=0.3\textwidth]{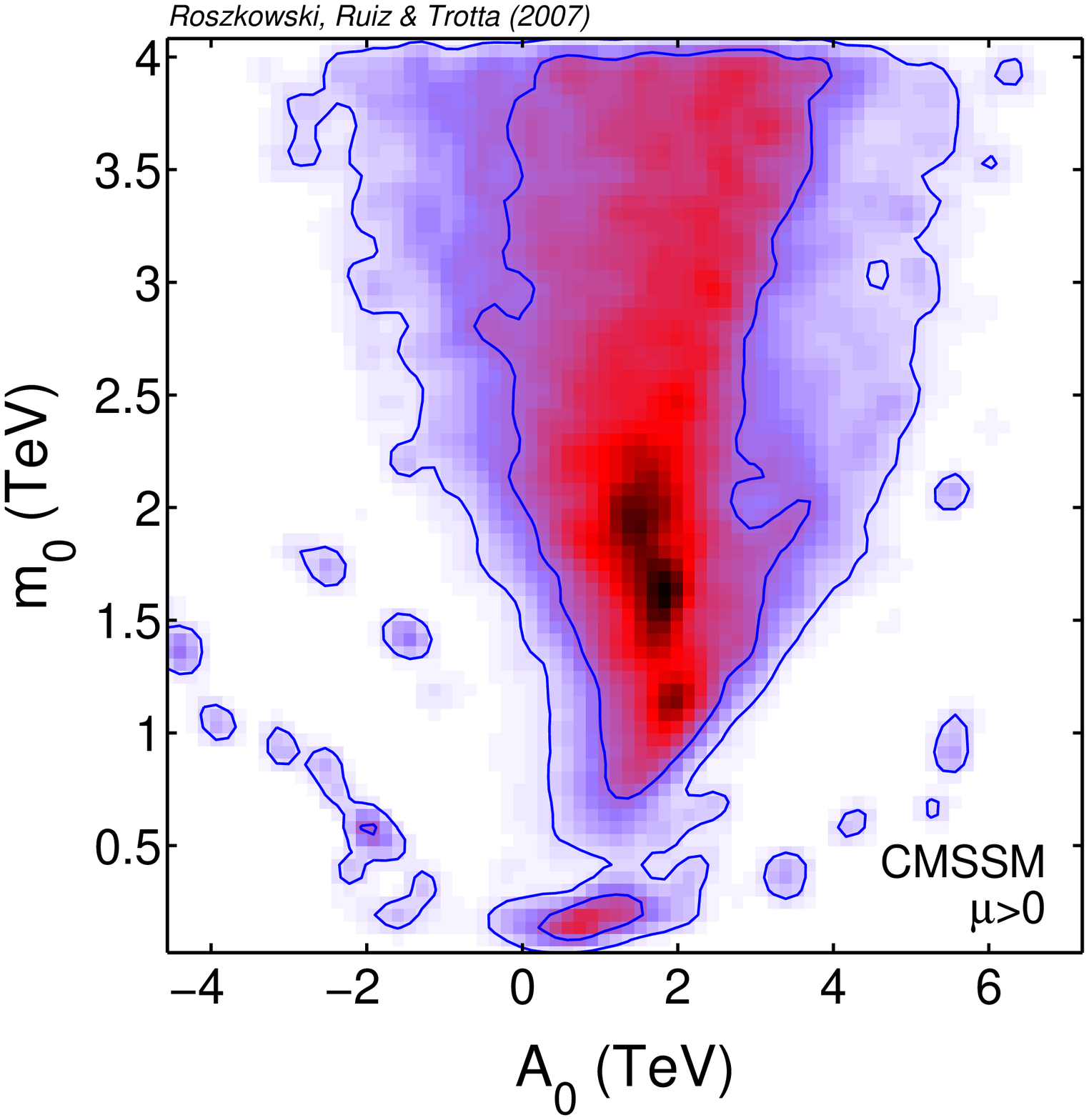}\\
  \includegraphics[width=0.3\textwidth]{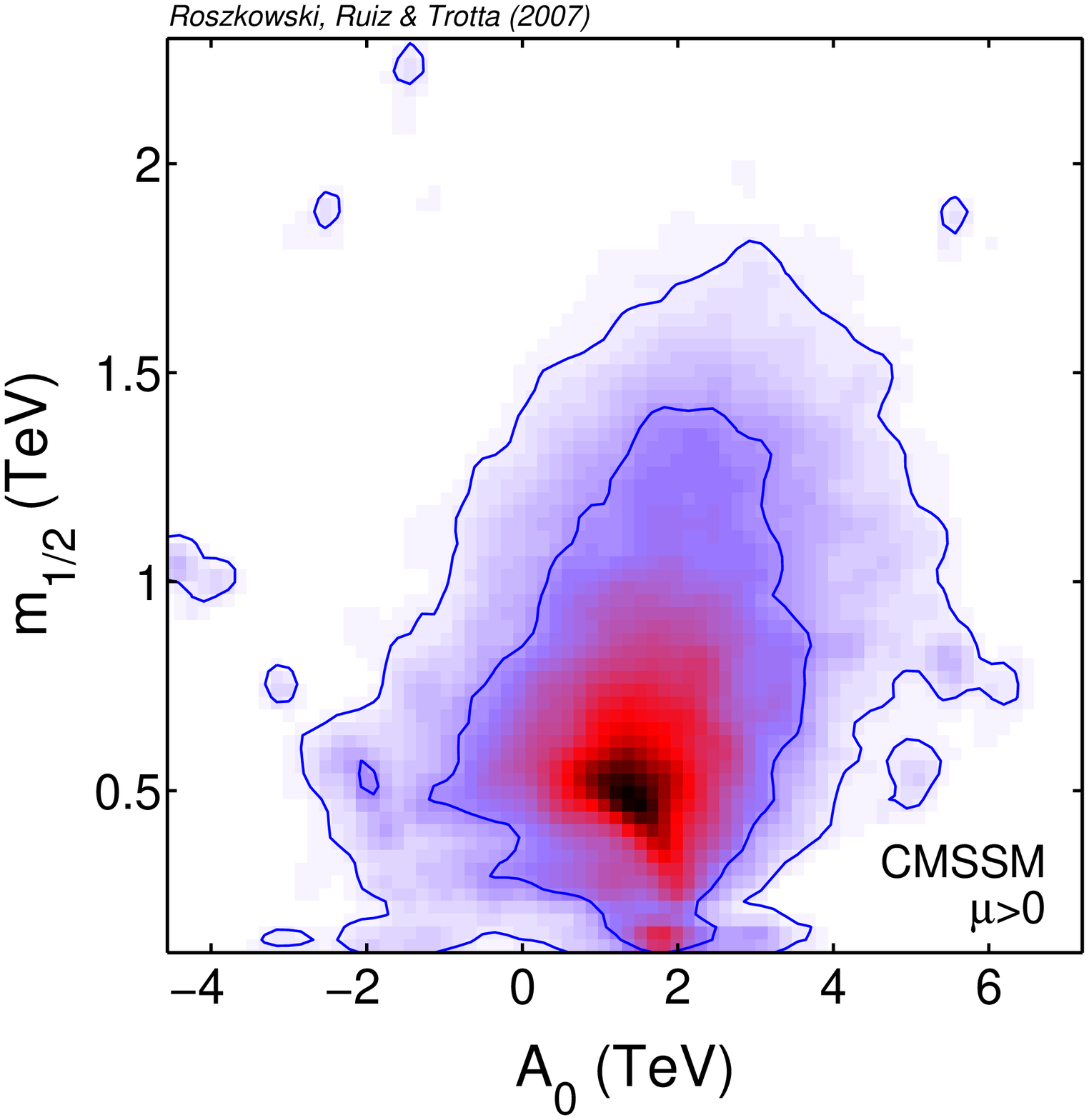}
& \includegraphics[width=0.3\textwidth]{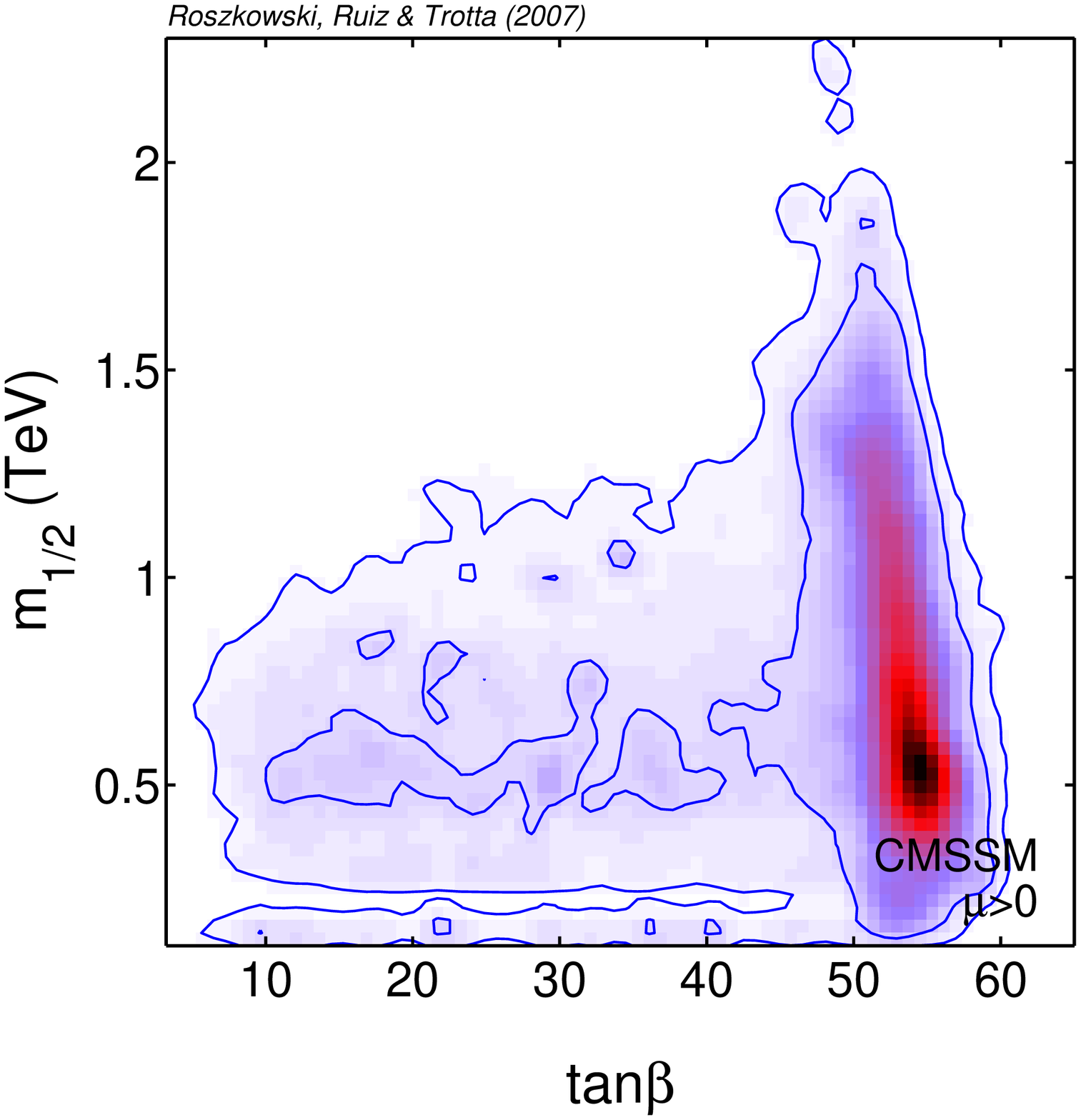}
& \includegraphics[width=0.3\textwidth]{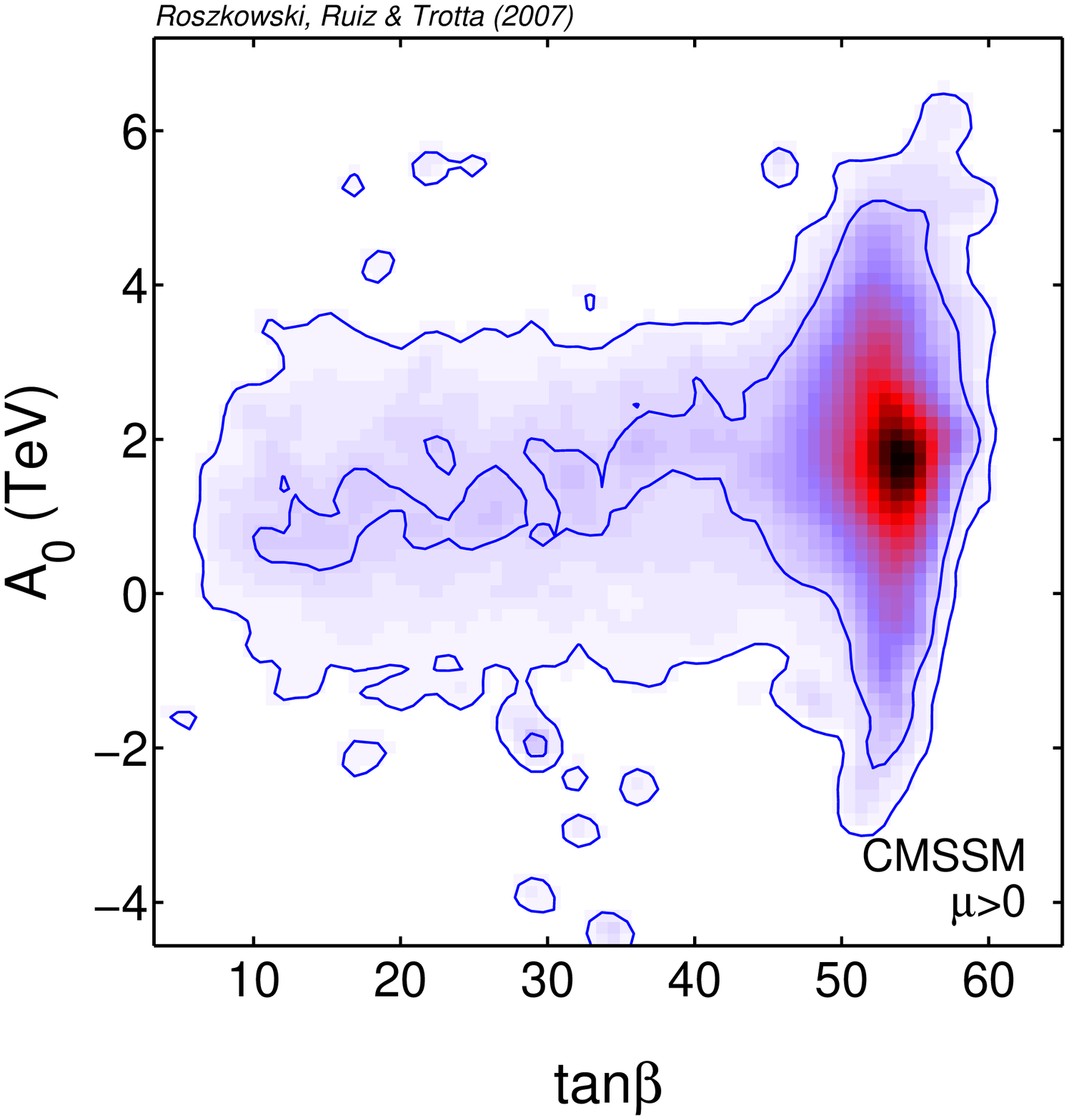}\\
\end{tabular}
  \includegraphics[width=0.3\textwidth]{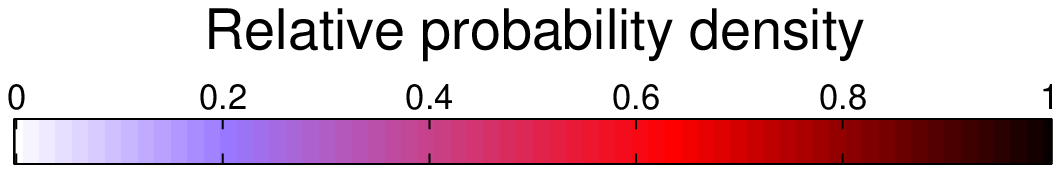}
\end{center}
\caption{\label{fig:cmssm2dcontoursmup} {The 2-dim relative
probability density functions in the planes spanned by the CMSSM
parameters: $\mhalf$, $\mzero$, $\azero$ and $\tanb$ for $\mu>0$. The
pdf's are normalized to unity at their peak. The inner (outer) blue
solid contours delimit regions encompassing 68\% and 95\% of the total
probability, respectively. All other basis parameters, both CMSSM and
SM ones, in each plane have been marginalized over. 
This figure should be compared with figure~2 in ref.~\protect\cite{rtr1}.
}}
\end{figure}
\begin{figure}[tbh!]
\begin{center}
\begin{tabular}{c c c}
  \includegraphics[width=0.3\textwidth]{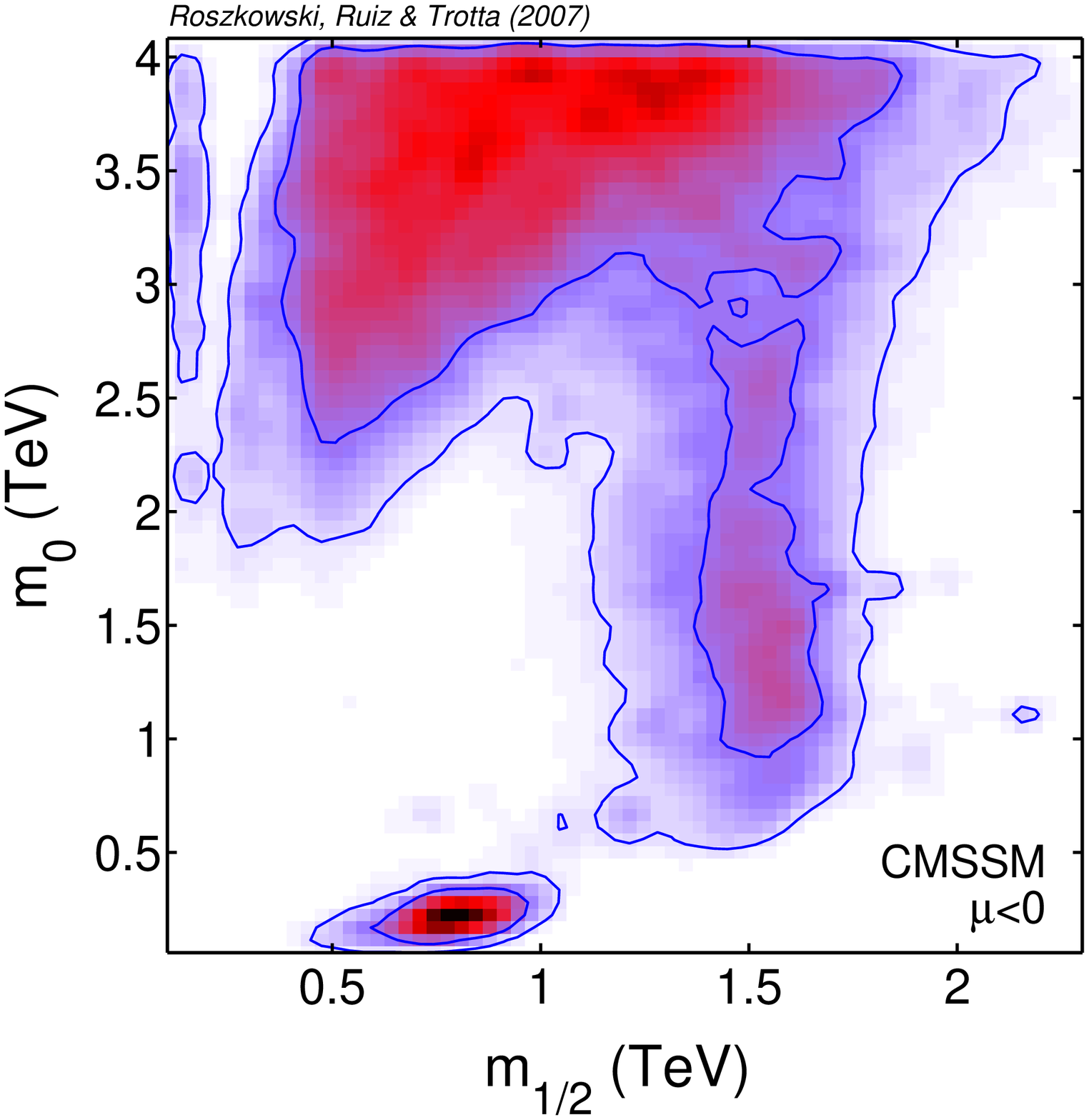}
& \includegraphics[width=0.3\textwidth]{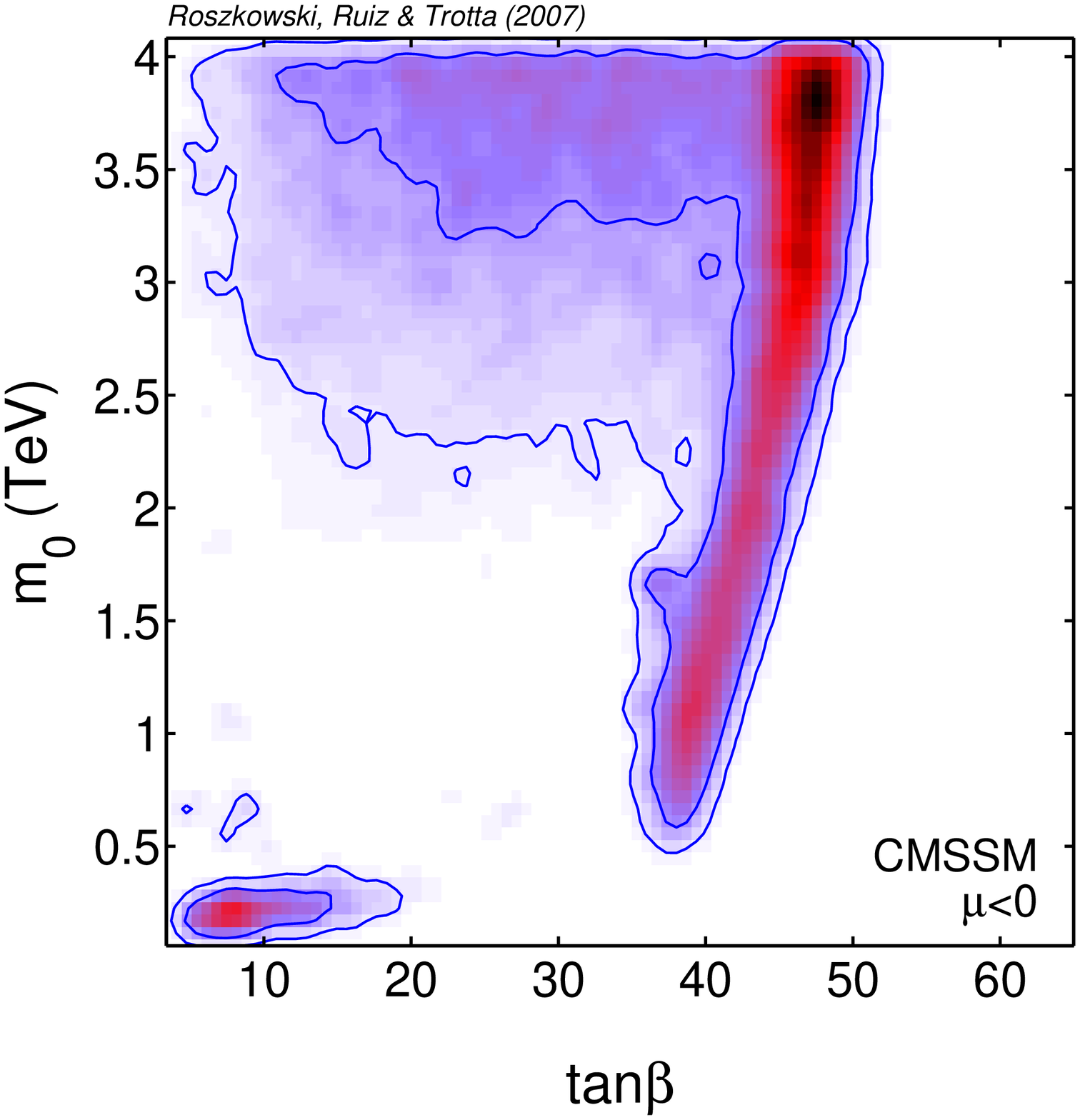}
& \includegraphics[width=0.3\textwidth]{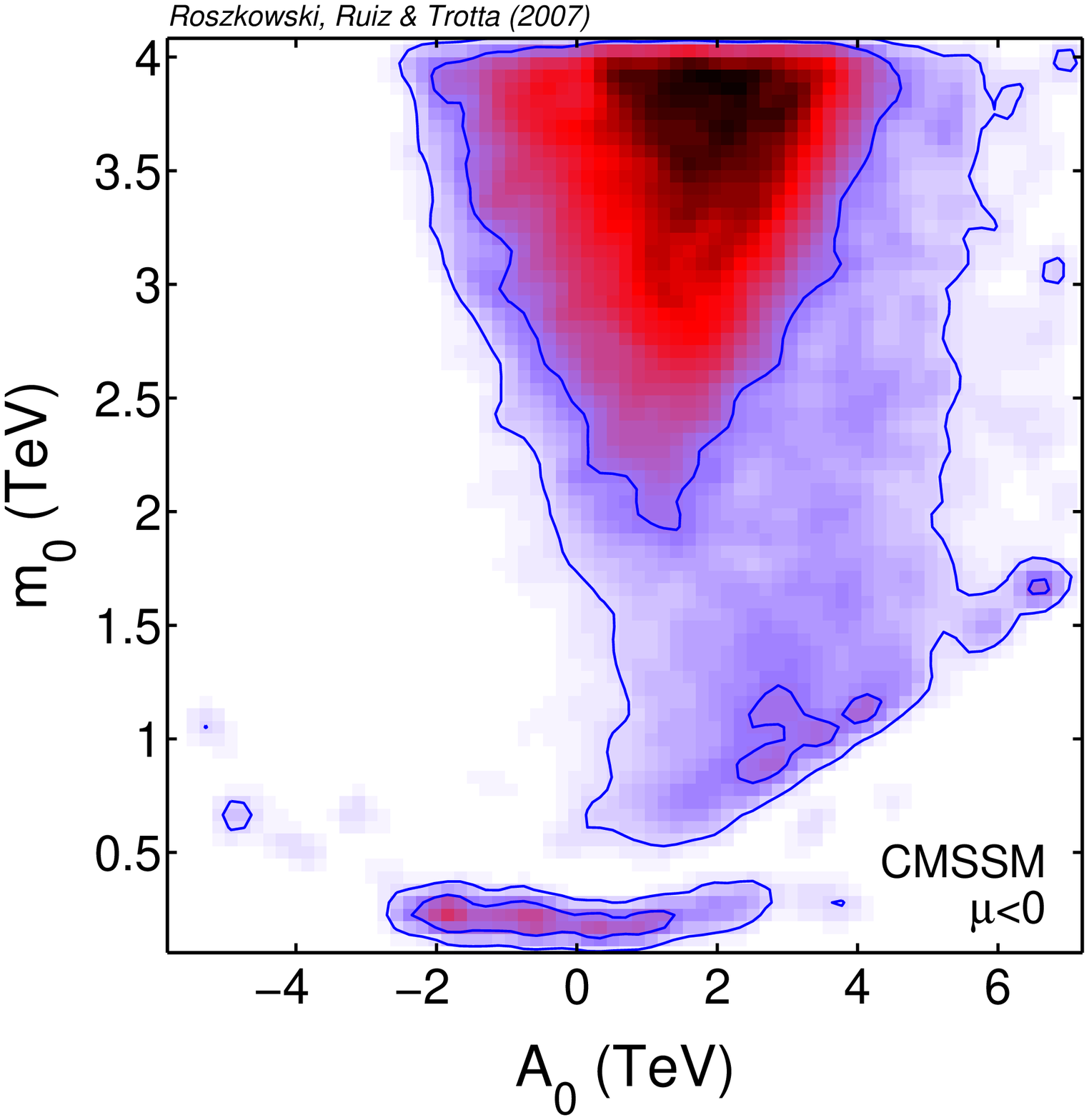}\\
  \includegraphics[width=0.3\textwidth]{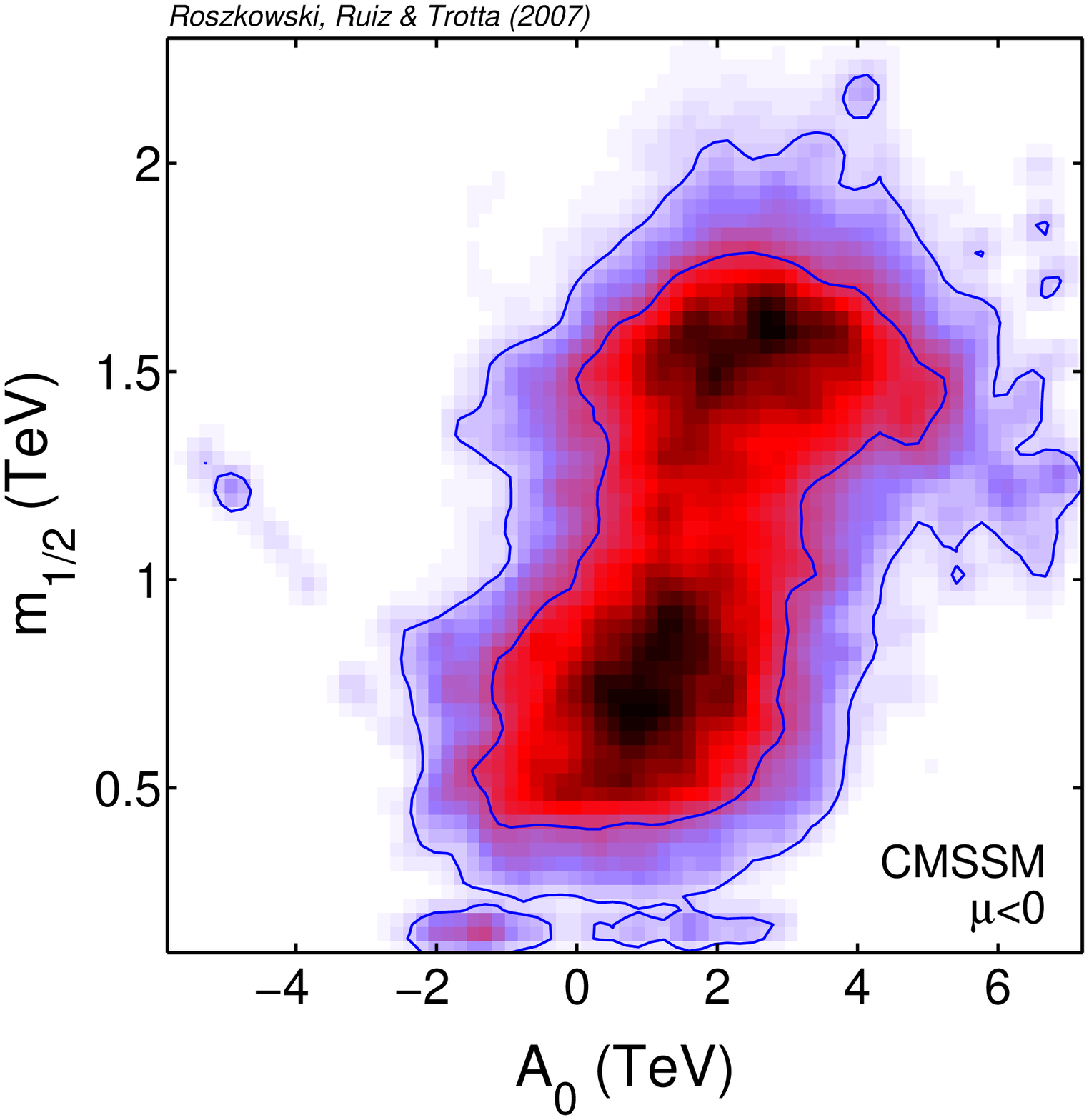}
& \includegraphics[width=0.3\textwidth]{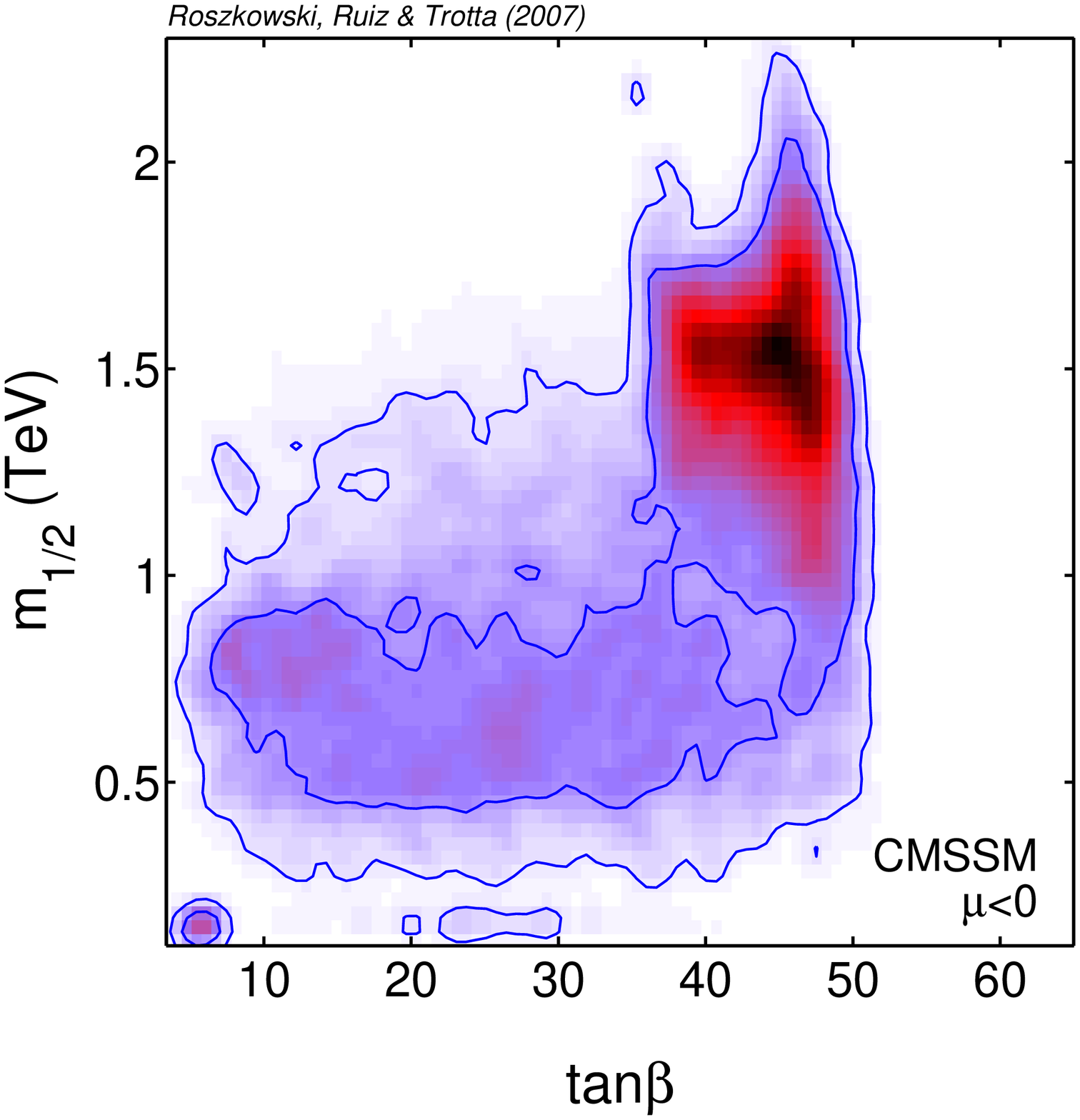}
& \includegraphics[width=0.3\textwidth]{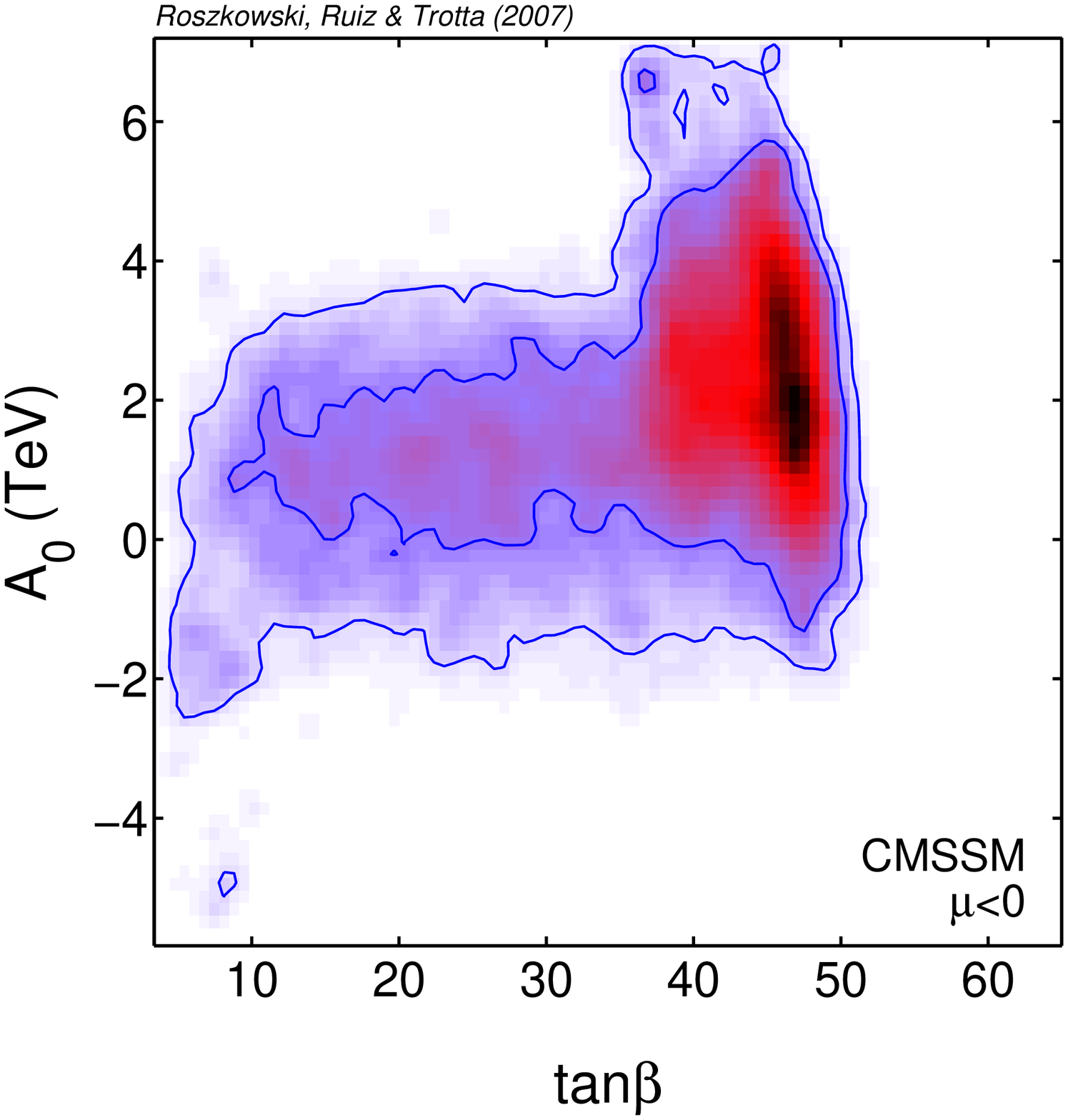}
\end{tabular}
  \includegraphics[width=0.3\textwidth]{rrt3-colorbar.ps}
\end{center}
\caption{\label{fig:cmssm2dcontoursmun} {The same as in
  fig.~\protect\ref{fig:cmssm2dcontoursmup} but for $\mu<0$.
}}
\end{figure}
\begin{figure}[tbh!]
\begin{center}
\begin{tabular}{c c}
  \includegraphics[width=0.45\textwidth]{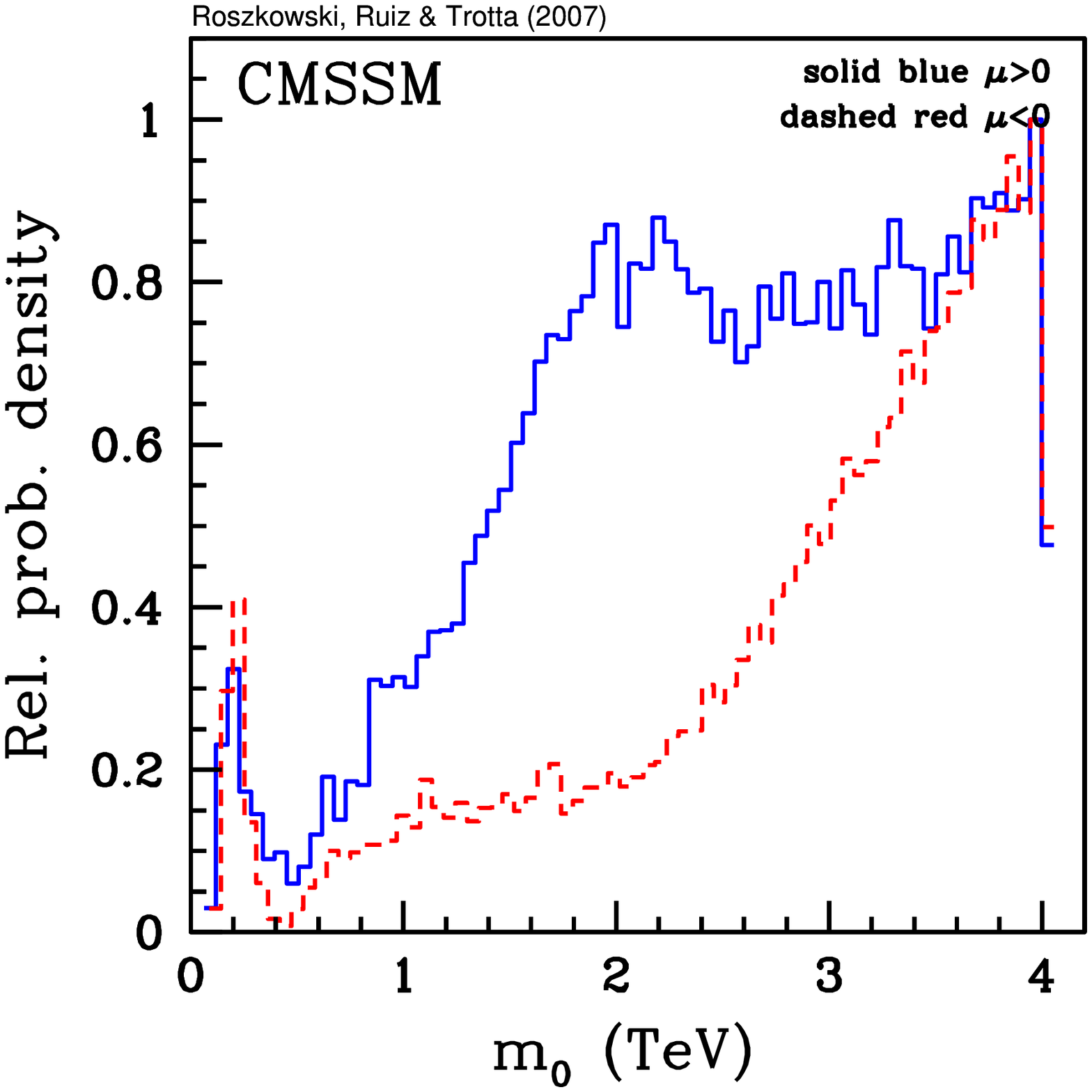}
& \includegraphics[width=0.45\textwidth]{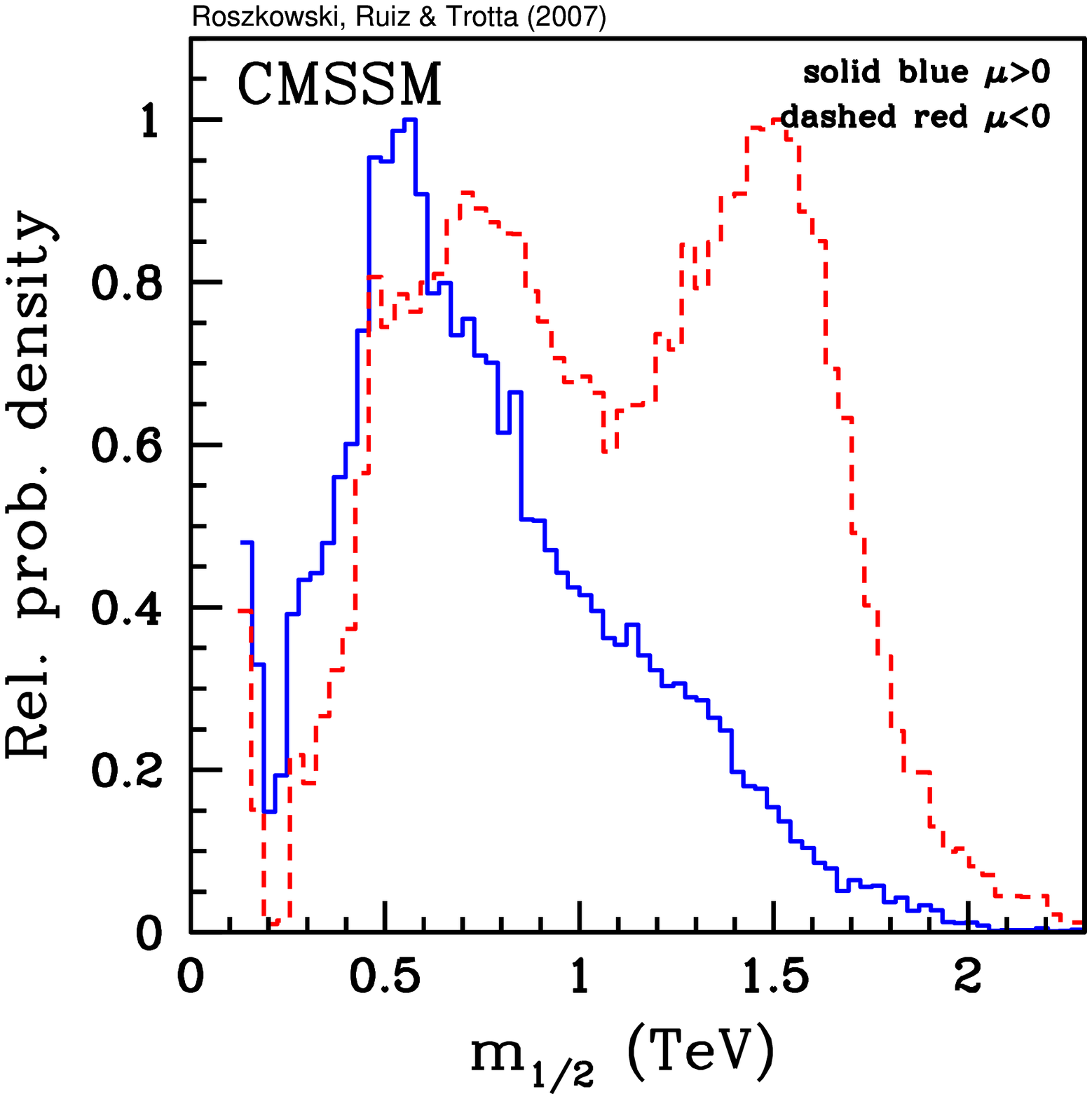}\\
  \includegraphics[width=0.45\textwidth]{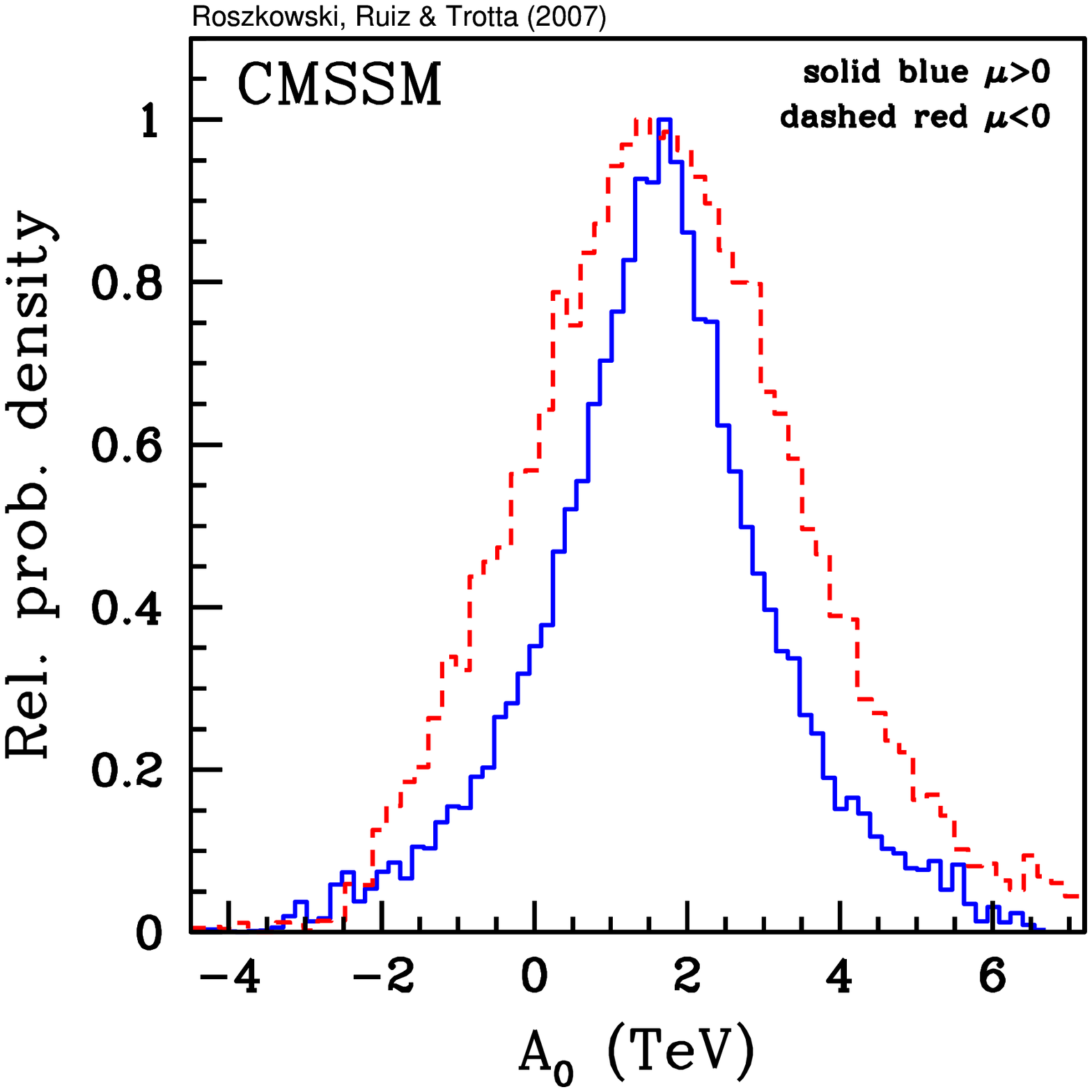}
& \includegraphics[width=0.45\textwidth]{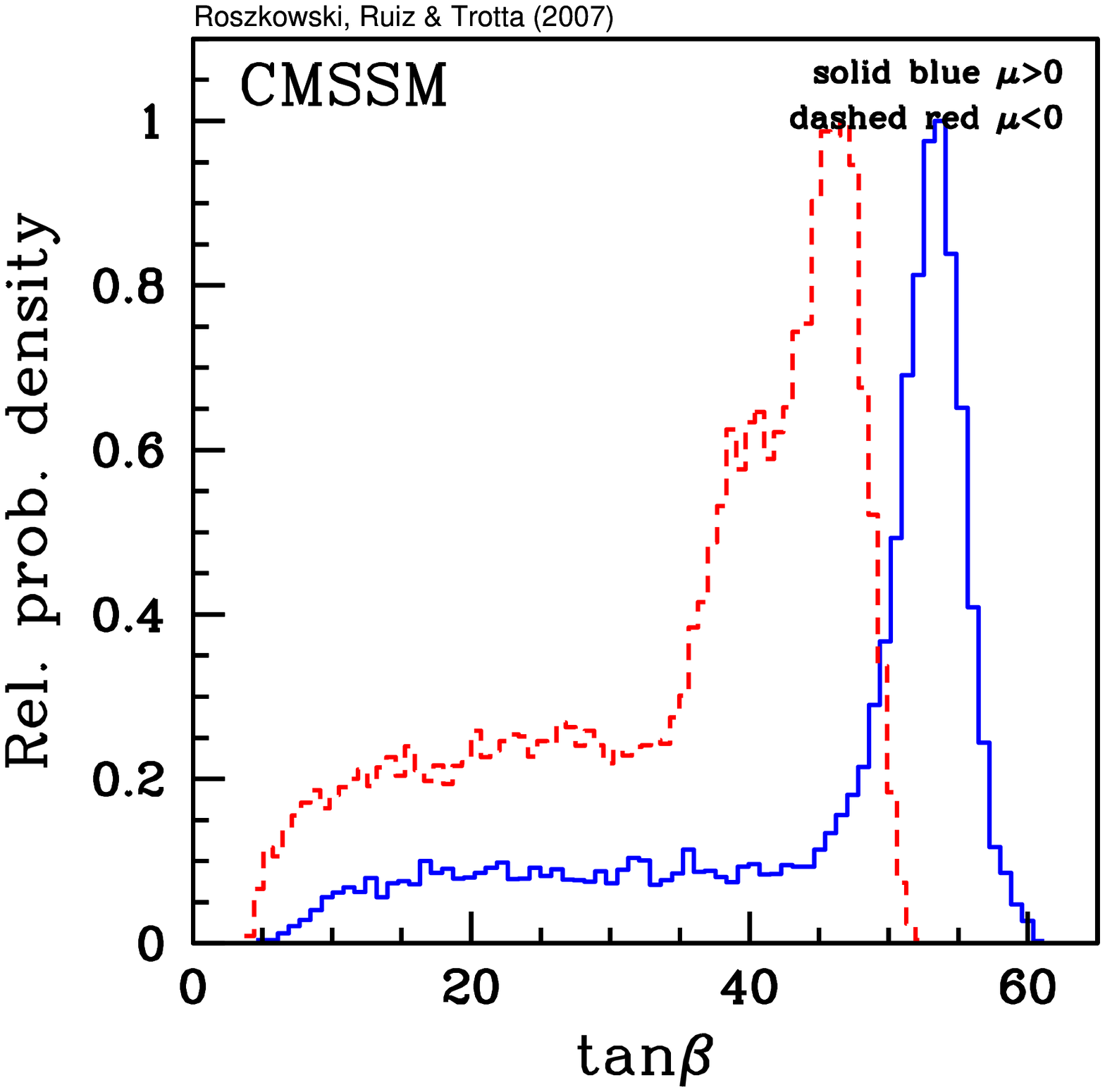}
\end{tabular}
\end{center}
\caption{\label{fig:cmssm1dpdf} {The 1-dim relative probability
densities for the CMSSM parameters $\mzero$, $\mhalf$, $\azero$ and
$\tanb$. All other parameters have been marginalized over. Solid blue
(dashed red) curves correspond to $\mu>0$ ($\mu<0$).  
This figure should be compared with figure~4 (black solid lines) in
ref.~\protect\cite{rtr1}.}}
\end{figure}

We first show in fig.~\ref{fig:cmssm2dcontoursmup} the 2-dim relative
probability density functions in the planes spanned by the CMSSM
parameters: $\mhalf$, $\mzero$, $\tanb$, $\azero$, and assuming
$\mu>0$, while in fig.~\ref{fig:cmssm2dcontoursmun} the same is shown
for $\mu<0$. In each panel all other basis parameters have been
marginalized over. Redder (darker) regions correspond to higher
probability density. Inner and outer blue (dark) solid contours
delimit regions of 68\% and 95\% of the total probability,
respectively, and remain well within the assumed priors, except for
$\mzero$.
In all the 2-dim plots, the MC samples have been divided into
$70\times 70$ bins, with a mild smoothing across adjacent bins to
improve the quality of the presentation (this has not impact on
our statistical conclusions). Jagged contours are a result of a
finite resolution of the MC chains.

In the case of $\mu>0$ (fig.~\ref{fig:cmssm2dcontoursmup}) we can see
a strong preference for large $\mzero\gsim1\tev$. On the other hand,
the peak of probability for $\mhalf$ is around $0.5\tev$, although the
68\% range of total probability is rather wide, increases with
$\mzero$ and exceeds $1.5\tev$ for $\mzero\simeq4\tev$. Additionally,
at smaller $\mzero\lsim 1\tev$ there are a few confined 68\% total
probability regions.

The strong preference for large $\mzero\gg\mhalf$ is primarily the
result of the sizable shift in the SM value of $\brbsgamma$, as can
be seen by comparing fig.~\ref{fig:cmssm2dcontoursmup} with fig.~2 in
ref.~\cite{rtr1} (or fig.~8 of ref.~\cite{alw06}) where the previous
value of $\brbsgamma$ has been used. (While the other CMSSM parameters
also experience some shift in their most probable values, it is not as
dramatic as that of $\mzero$ towards larger values.) The underlying
reason is that, at fairly small $\mhalf$ the charged Higgs mass
remains relatively light, in the few hundred GeV range, and, via a
loop exchange with the top quark, it adds substantially to the SM
value of $\brbsgamma$, towards the experimentally allowed range. (In
fact, for $\mhm\simeq650\gev$, the contribution is sufficient to fill
the gap between the SM and the experimental central values of
$\brbsgamma$~\cite{ms06-bsg}.) At smaller $\mhalf$ and/or $\mzero$, the
(negative, for $\mu>0$) chargino-stop contribution is too large and
needs to be compensated by the $\hm$-top contribution. In fact, we do
find some small ``islands'' of 68\% total probability at
$\mhalf\lsim1\tev$ and $\mzero\lsim1.3\tev$ (in particular, notably,
an interesting case of $\mhalf\simeq0.5\tev$ and $\mzero\simeq
0.2\tev$) but the bulk of high probability region corresponds to
$\mzero\gsim1\tev$.

At $95\%$ total probability level the available parameter space opens
widens considerably, and also some new features arise. In
particular, at $\mhalf\simeq 0.2\tev$ we can see a narrow
high-probability funnel induced by the light Higgs boson
resonance~\cite{al05,rtr1}. Also, $\tanb$ becomes less confined to its
most preferred range of large values between some 50 and 60, while
$\azero$ remains on a positive side.

In the case of $\mu<0$ (fig.~\ref{fig:cmssm2dcontoursmun}), one
can see a strong preference for even larger $\mzero$. Also, the
68\% total probability
region of $\mhalf$ shifts towards larger values, although still
remains basically below $2\tev$. (Although we again find an
interesting isolated high probability region at
$\mhalf\simeq0.75\tev$ and $\mzero\simeq 0.2\tev$.) This shift
towards larger $\mzero$ and/or $\mhalf$ is again caused mostly by
the $\brbsgamma$ constraint. At large $\mzero$ (and not too large
$\mhalf$) the charged Higgs mass {\em decreases} and its
contribution tends to be on a high side, while the chargino-stop
one becomes too small. A similar effect is observed at large
$\mhalf\sim 1.5\tev$ and $\mzero\gsim1\tev$ where the chargino and
stop masses become too large to contribute much as well.

In fig.~\ref{fig:cmssm1dpdf} the 1-dim marginalized probability
distributions for the CMSSM parameters are compared for both
$\mu>0$ and $\mu<0$. In each panel all the other CMSSM parameters
and all SM (nuisance) parameters have been marginalized over. It
is clear that non-negligible probability ranges of the CMSSM
parameters, other than $\mzero$, are confined well within their
assumed priors. Again, we can see strong preference for large
$\mzero\gg1\tev$ (even stronger for $\mu<0$ than for $\mu>0$).
Larger values of $\mhalf$ are also favored for $\mu<0$ although,
for both signs, this parameter is well confined within $2\tev$.
The preferred range of $\azero$ is fairly uncorrelated with the
other parameters~\cite{rtr1}, and it is symmetrically peaked
(basically independently of the sign of $\mu$) at some $1.5\tev$.
This value is however basically twice as large as for the previous
SM value of $\brbsgamma$~\cite{rtr1}. On the other hand, $\tanb$
is well-peaked at some $53$ for $\mu>0$ and some $48$ for $\mu<0$.
In both cases, there remains a sizable tail of much smaller values
which remain allowed at large $\mzero$.

What most strongly contributes to confining $\mhalf$ well within its
prior (for both signs of $\mu$) is the relic abundance $\abundchi$
which becomes too large. On the one hand, at large $\tanb$ it becomes
easier to satisfy the constraint from $\abundchi$ due to the increased
role of the neutralino annihilation via the pseudoscalar Higgs effect
and/or the coannihilation effect. On the other, as explained in
ref.~\cite{rtr1,rrt2}, as $\tanb$ becomes very large, $\gsim60$ for
$\mu>0$ ($\gsim50$ for $\mu<0$), it becomes very difficult to find
self-consistent solutions of the RGE's.

\begin{figure}[tbh!]
\begin{center}
\begin{tabular}{c c c}
    \includegraphics[width=0.3\textwidth]{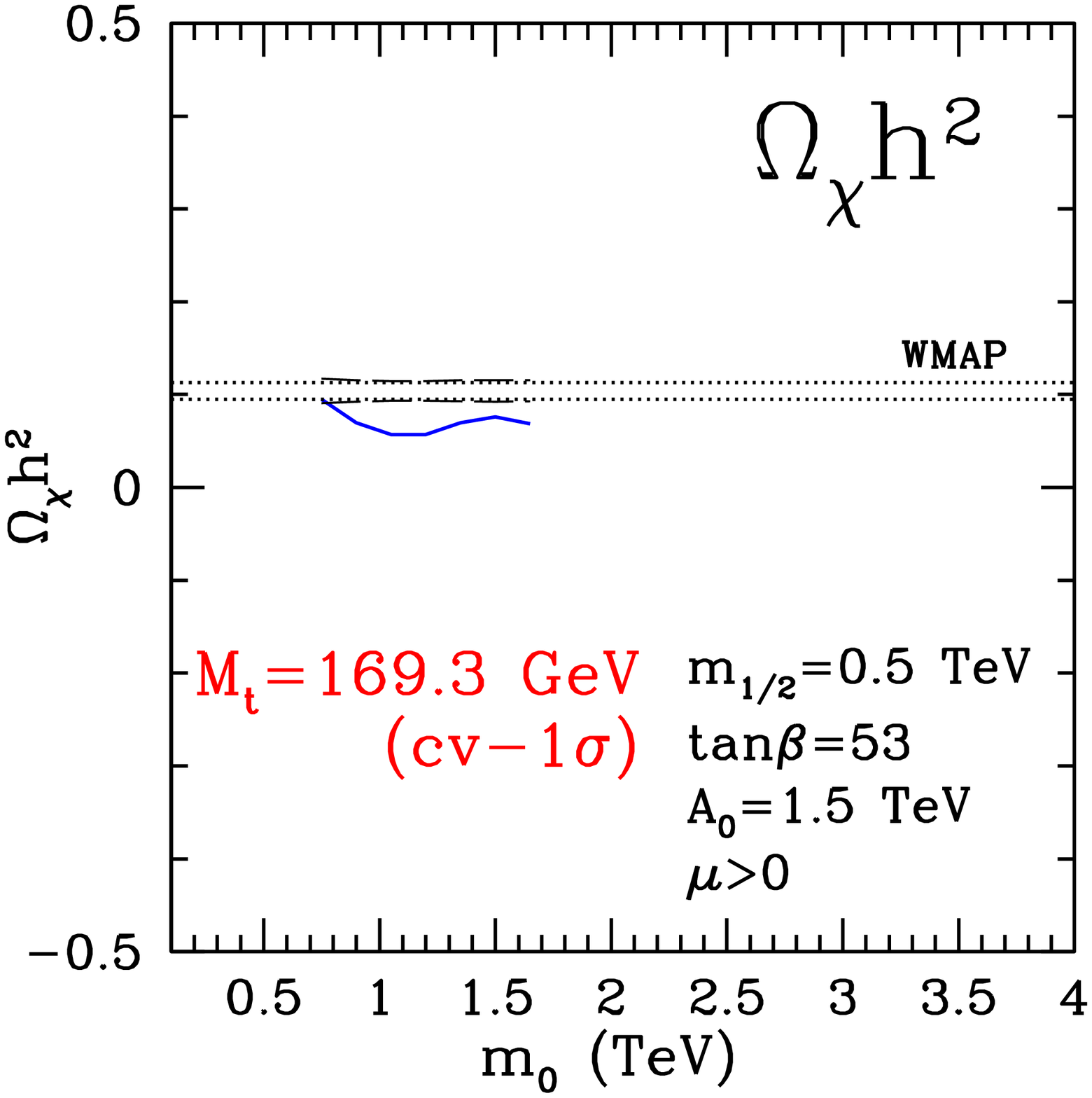}
&   \includegraphics[width=0.3\textwidth]{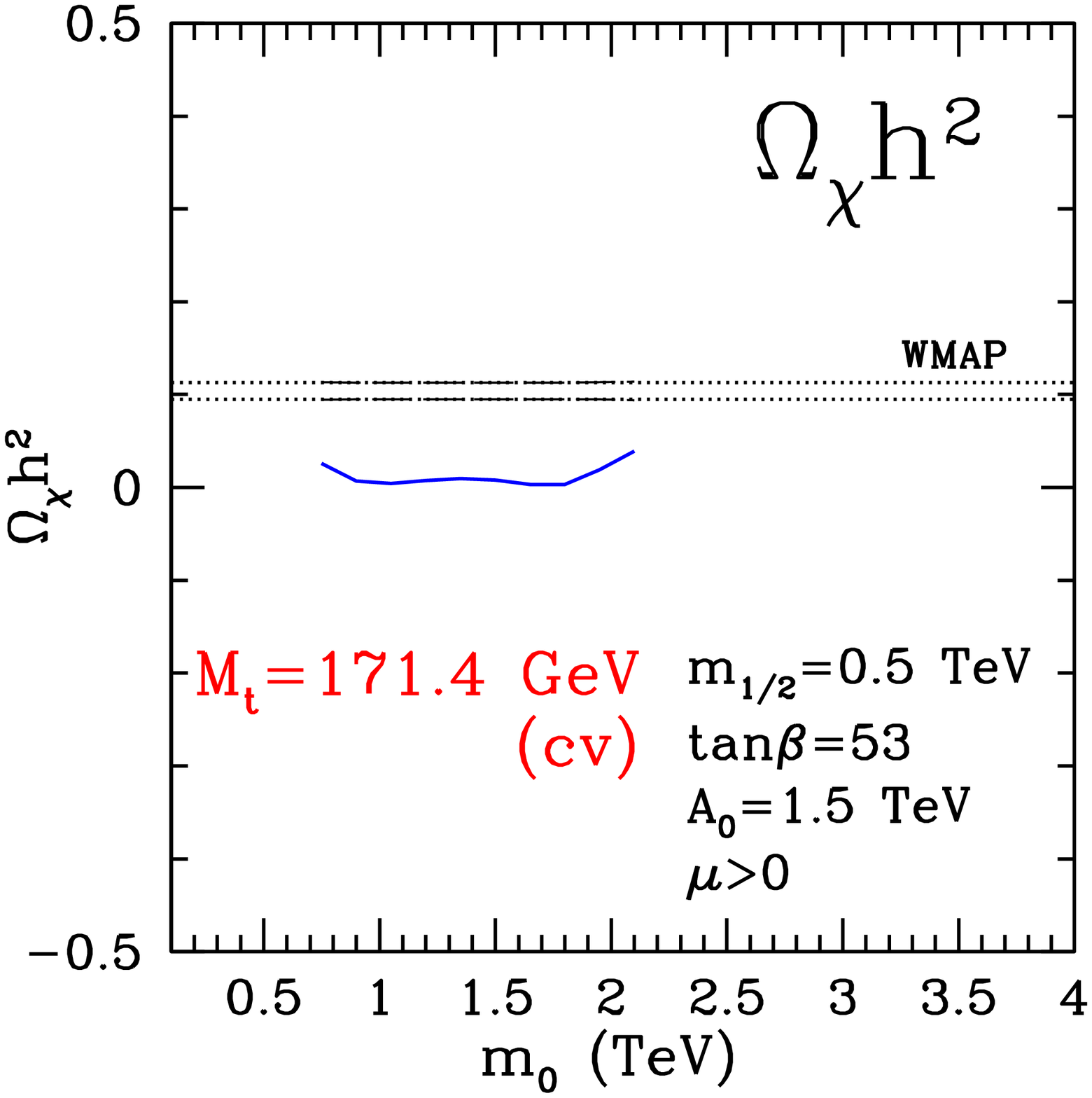}
&   \includegraphics[width=0.3\textwidth]{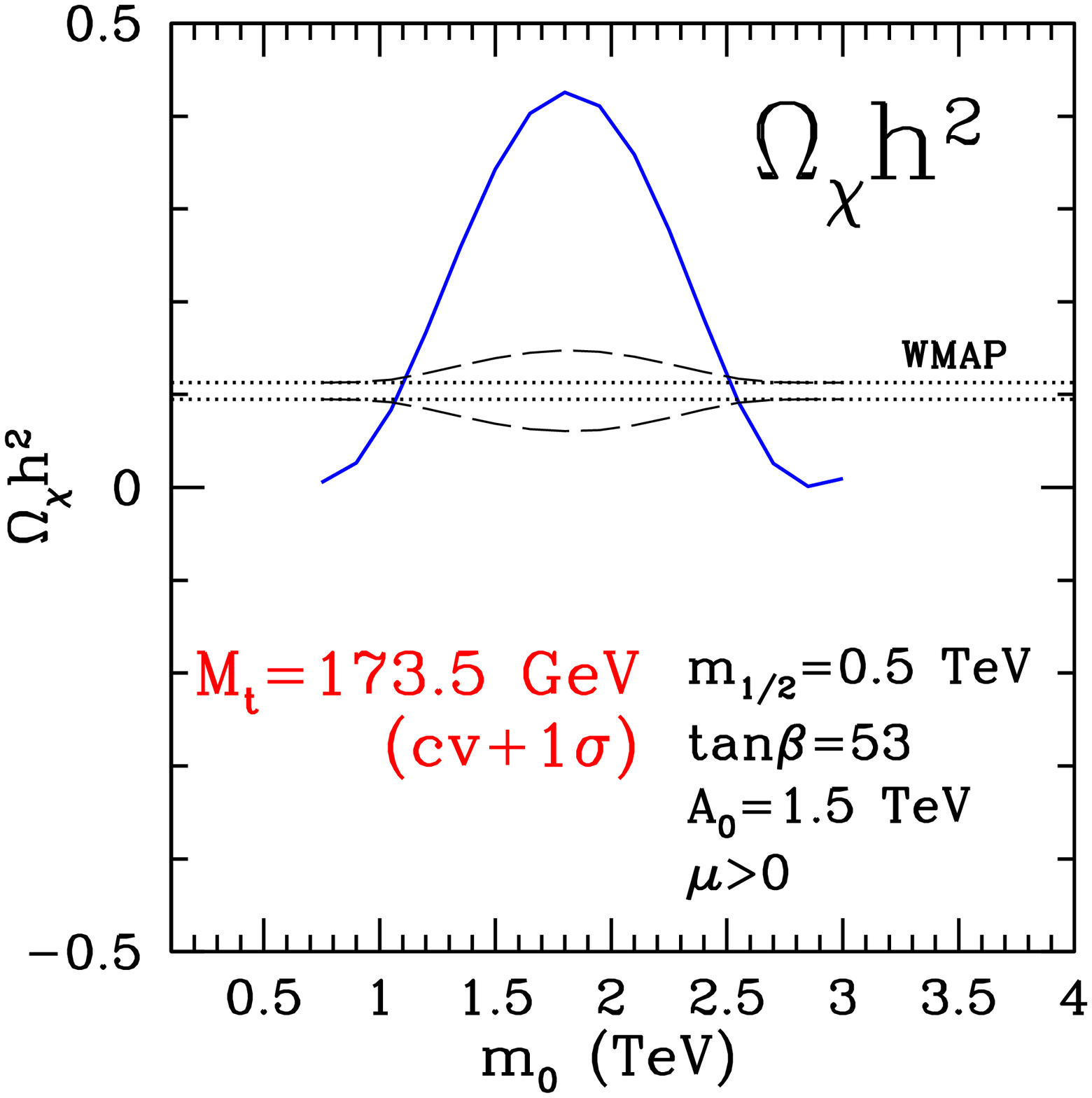}\\
\end{tabular}
\end{center}
\caption{An illustration of the sensitivity of the cosmologically
  favored range of $\mzero$ to the top mass $\mtpole$ in the focus
  point region. Other CMSSM parameters have been fixed close to their
  most preferred values for $\mu>0$ (see
  fig.~\protect\ref{fig:cmssm2dcontoursmup}).  We show $\abundchi$
  (blue solid line) for the central value (cv) of $\mtpole$ (middle
  panel) and for the values decreased and increased by $1\sigma$ (left
  and right panel). The parallel 
  dotted lines denote the very narrow $2\sigma$ range, as
  determined from the 3-year data of WMAP~\cite{wmap3yr}. The
  long-dashed line denotes the combined theoretical plus experimental
  error, as described in the text. Note that, as $\mtpole$ is varied
  between its cv minus $1\sigma$ to its cv plus $1\sigma$, the range
  of $\mzero$, for which consistent solutions can be found, roughly
  doubles.
\label{fig:oh2-m0vsmtop}
}
\end{figure}

The feature that very large values of $\mzero$ (the FP region), up to
the assumed prior of $4\tev$, remain allowed (actually, even
preferred), is unfortunate but is a consequence of the fact that
current data are not constraining enough. Normally, as all the
superpartner masses (including the LSP) increase, bino-like neutralino
annihilation in the early Universe becomes suppressed and it becomes
harder to satisfy the WMAP constraint on $\abundchi$. This is why
$\mhalf$ is well confined within some $2\tev$, as described
above. Unfortunately, in the FP region the behavior of $\abundchi$ is
much more sensitive to input parameters.  We illustrate this in
fig.~\ref{fig:oh2-m0vsmtop} where we plot $\abundchi$ vs. $\mzero$ for
$\mu>0$ and a choice of the other CMSSM parameters close to their
highest probability values
(fig.~\ref{fig:cmssm2dcontoursmup}). Clearly, as $\mtpole$ is varied
within $1\sigma$ around its central value (cv), $\abundchi$ changes
quite substantially. In the presented example, by fixing $\mtpole$ at
its central value~\cite{ehow} (as it is normally done in fixed-grid
scans) one would find no cosmologically allowed $\mzero$. On the other
hand, by reducing (increasing) $\mtpole$ by $1\sigma$ we can find one
(two widely disconnected ) narrow region(s) of $\mzero$ where
$\abundchi$ in the WMAP range. Furthermore, in a probabilistic
approach, even the case at the central value of $\mtpole$ is not
excluded but only less favored by $\abundchi$.

Another feature that is evident in fig.~\ref{fig:oh2-m0vsmtop} is
that, as $\mtpole$ is varied by $1\sigma$ around its central value,
the range of $\mzero$ where self-consistent solutions of the RGE's and
the conditions of EWSB can be found changes by as much as a factor of
two. It is therefore clear that, if one includes the FP region, it is
basically impossible to locate the cosmologically favored range of
$\mzero$. In particular, it would be misleading to simply fix the top
mass at its central value (and likewise with the bottom mass at large
$\tanb$).

\begin{figure}[!tbh]
\begin{center}
\begin{tabular}{c}
    \includegraphics[width=0.45\textwidth]{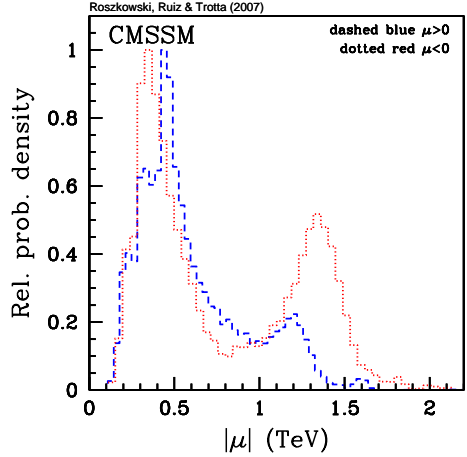}
\end{tabular}
\end{center}
\caption{The 1-dim relative probability density of $|\mu|$. All other
parameters have been marginalized over. Dashed blue (dotted red)
curves correspond to $\mu>0$ ($\mu<0$).
\label{fig:mu} }
\end{figure}

The rather special properties of the FP region are to a large extent
related to the behavior of the parameter $\mu$. As $\mzero$ increases
(along, say, fixed $\mhalf$), $\mu^2$ (which is determined the the
conditions of EWSB) decreases rather quickly, thus increasing the
higgsino component of the neutralino. This in turn reduces $\abundchi$
to an acceptable range for some narrow range of $\mzero$. At slightly
larger $\mzero$, $\mu^2$ drops below zero, thus delimiting the zone
where consistent, non-tachyonic solutions can be found.  Thus
generally in the FP region $\mu$ is rather small relative to
$\mzero$. We can see this feature in fig.~\ref{fig:mu} where we plot
the 1-dim relative probability density for the parameter. We can see a
clear peak in the few hundred~GeV region for both signs of $\mu$. In
this region $|\mu|\simeq \mhalf/3$.  Additionally, for $\mu<0$ there
is a second well-pronounced peak around some $1.4\tev$ which
corresponds to the band of higher relative probability around
$\mhalf\simeq1.5\tev$ and $\mzero\gsim1\tev$ in
fig.~\ref{fig:cmssm2dcontoursmun}, outside of the FP region.

\begin{table}
 \centering
 \begin{tabular}{|l |c c| c c|}
  \hline
            & \multicolumn{2}{|c|}{$\mu<0$} &
 \multicolumn{2}{|c|}{$\mu>0$}\\
  Parameter & 68\% region    & 95\% region            & 68\% region    & 95\%
            region\\\hline
  $\mzero$~(TeV)  &$< 3.51$ &$< 3.93$               &$< 3.1$     &$< 3.87 $  \\
  $\mhalf$~(TeV)  &$(0.57, 1.55)$  &$(0.87, 1.86)$  & $(0.4, 1.16)$  &$(0.145,
            1.6)$  \\
  $\azero$~(TeV) &$(-0.19, 3.42)$  &$(-1.79, 5.49)$  &$(0.11, 2.94)$ &$(-1.86, 4.84)$   \\
  $\tanb $   &$(18.6, 46.08)$  &$(7.51, 48.9)$  &$(26.38, 54.18)$  &$(11.17,56.78)$  \\

\hline
  \end{tabular}
 \caption{CMSSM parameter ranges corresponding to 68\% and 95\% of
 posterior probability (with all other parameters marginalized over) for
 both signs of $\mu$.
} \label{table:CMSSMtable}
 \end{table}

In order to summarize the above discussion, in
table~\ref{table:CMSSMtable} we give the 68\% and 95\% total
probability ranges of the CMSSM parameters for both signs of $\mu$.

We emphasize that the above results do not at all imply that it is
equally easy to fit all the data for both signs of $\mu$. (In
figs.~\ref{fig:cmssm2dcontoursmup}--~\ref {fig:cmssm1dpdf} the
posterior probabilities are normalized {\em relative} to the
respective highest values.) Actually, for negative $\mu$ the fit
is considerably poorer. In fact, we find that the best-fit
$\chi^2$ for the $\mu>0$ case is $6.3$, while for the $\mu<0$ case
it more than doubles to $14.4$. It is difficult to attach a
precise statistical significance to this result, as clearly the
distributional properties of the parameter space are far from
being Gaussian (and hence the $\chi^2$ is not chi-square
distributed). A proper evaluation of the significance of this
difference in the goodness-of-fit would require Monte Carlo
simulations of the measurements for both the $\mu>0$ and $\mu<0$
cases, which is beyond the scope of this work. However, this is an
indication of the fact that the $\mu<0$ case is at greater tension
with the data than $\mu>0$, although it cannot be conclusively
ruled out yet.

A fully Bayesian approach would consider computing the Bayes factor
among the two possibilities for $\mu$, along the lines of what has
been done in ref.~\cite{alw06}. This procedure is, however,
computationally demanding, and the result is potentially strongly
dependent on volume effects deriving from the choice of priors (see,
eg., ref.~\cite{trotta07}). An interesting alternative is
to use the procedure outlined in ref.~\cite{gt07} which employs
Bayesian calibrated p-values to obtain 
an upper limit on the Bayes factor regardless of the prior for the
alternative hypothesis. In the present case, the application of this
procedure would require Monte Carlo simulation to obtain the p-value
corresponding to the observed $\chi$-square difference. However, if we
take again the result of $\Delta \chi^2 = 7$ at face value, assuming
that it is indeed $\chi$-square distributed (which is probably a very
poor approximation, as argued above), then the corresponding upper
limit on the Bayes factor is $10:1$. (This rough estimate is actually in
surprising good agreement with the result found in ref.~\cite{alw06}
using numerical integrations) This would mean that the minimum
probability of $\mu<0$ is 0.1, which certainly does not constitute
strong evidence against $\mu<0$.  The above considerations only
highlight the difficulty of translating our result into a precise
statement about the relative probability of $\mu>0$ vs $\mu<0$. 

\begin{figure}[tbh!]
\begin{center}
\begin{tabular}{c c c}
  \includegraphics[width=0.3\textwidth]{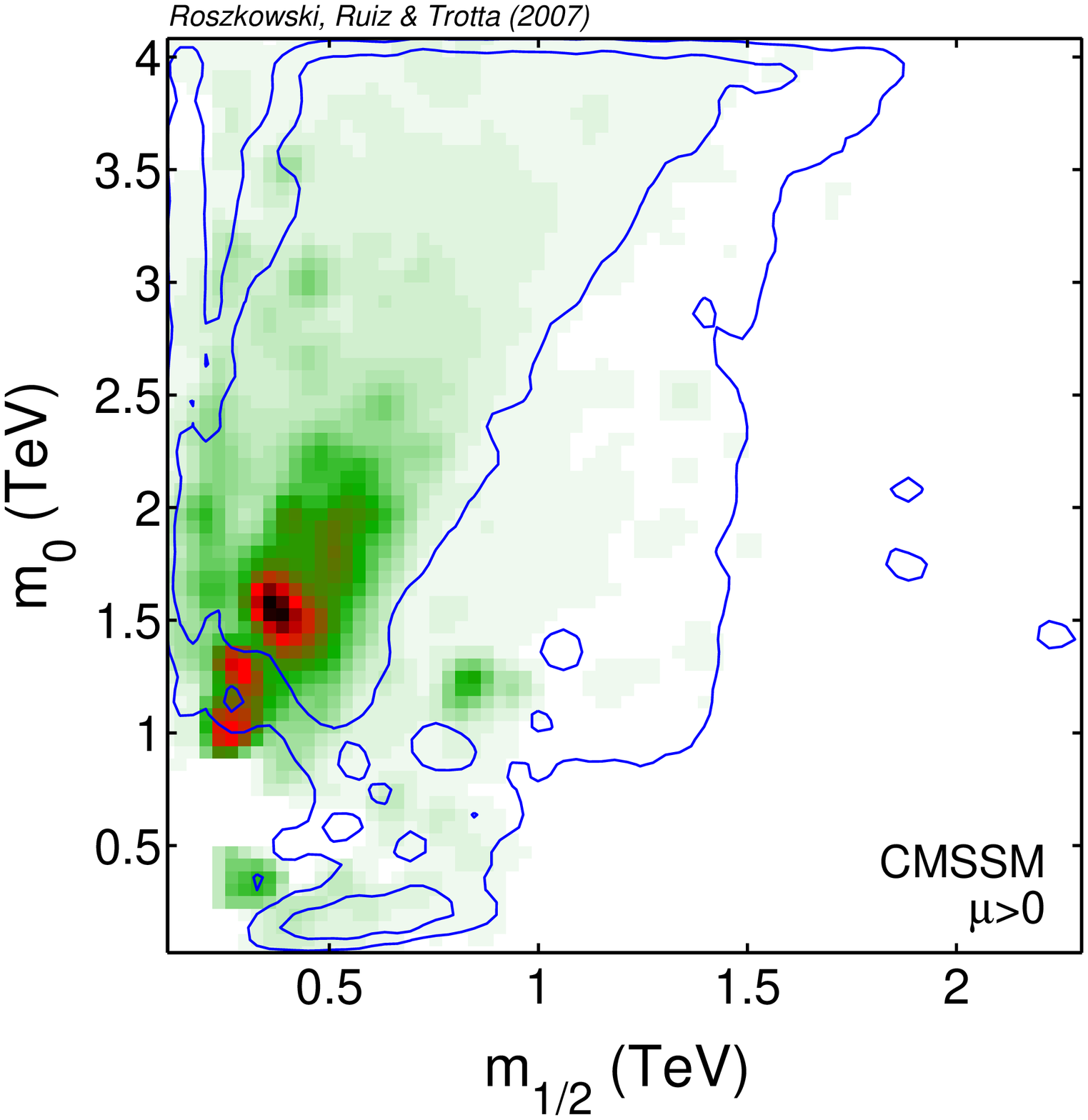}
& \includegraphics[width=0.3\textwidth]{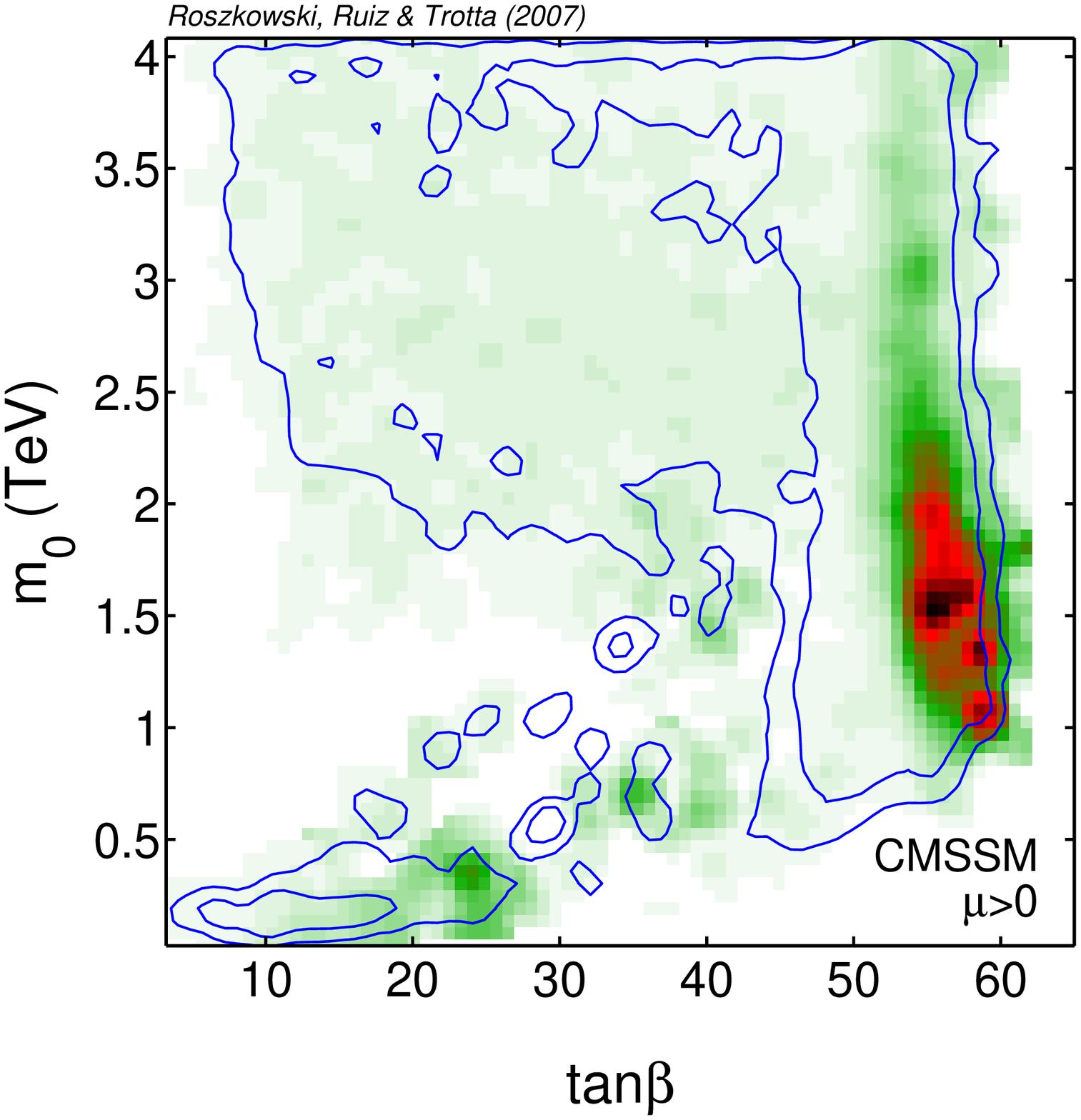}
& \includegraphics[width=0.3\textwidth]{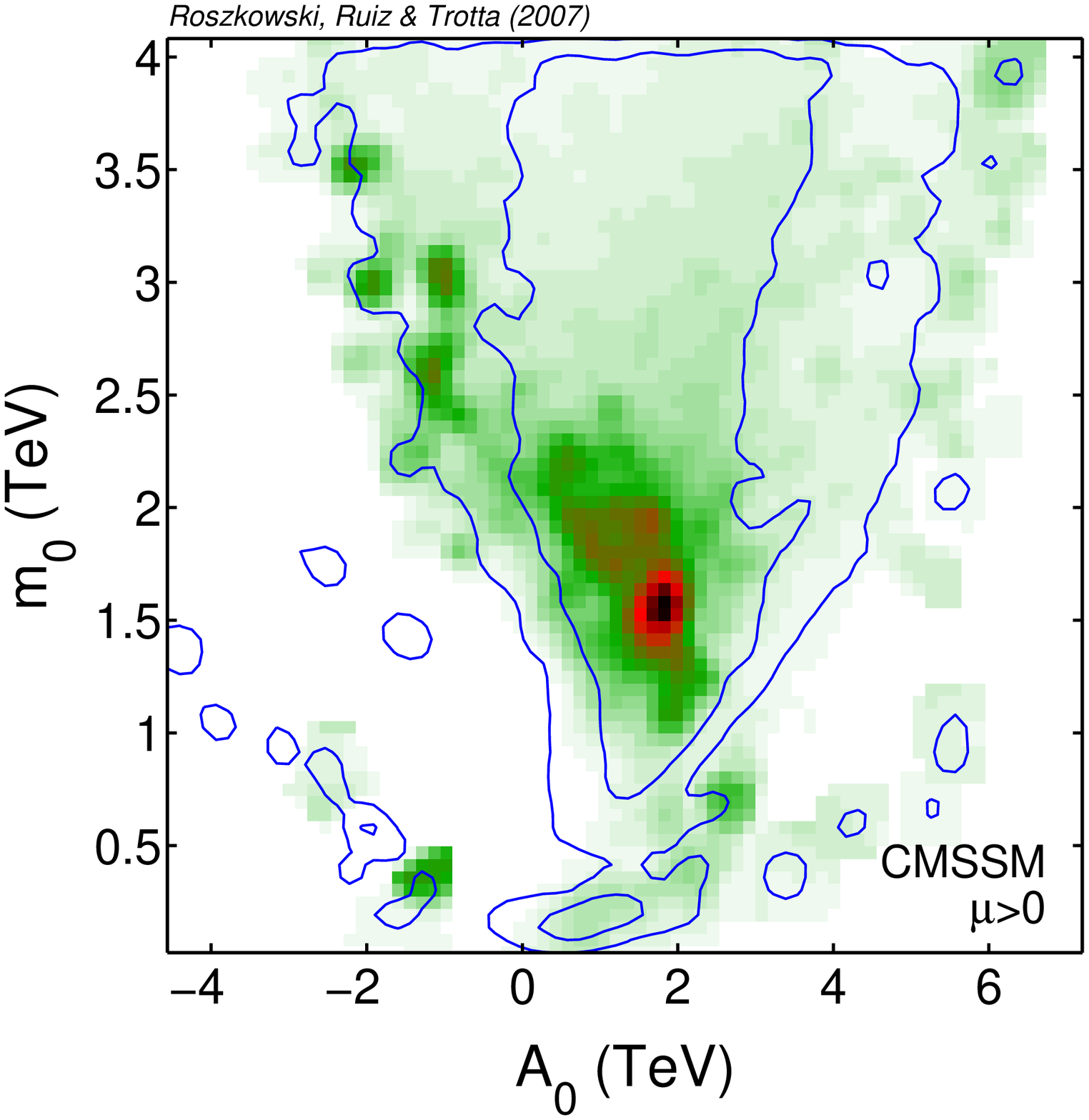}\\
  \includegraphics[width=0.3\textwidth]{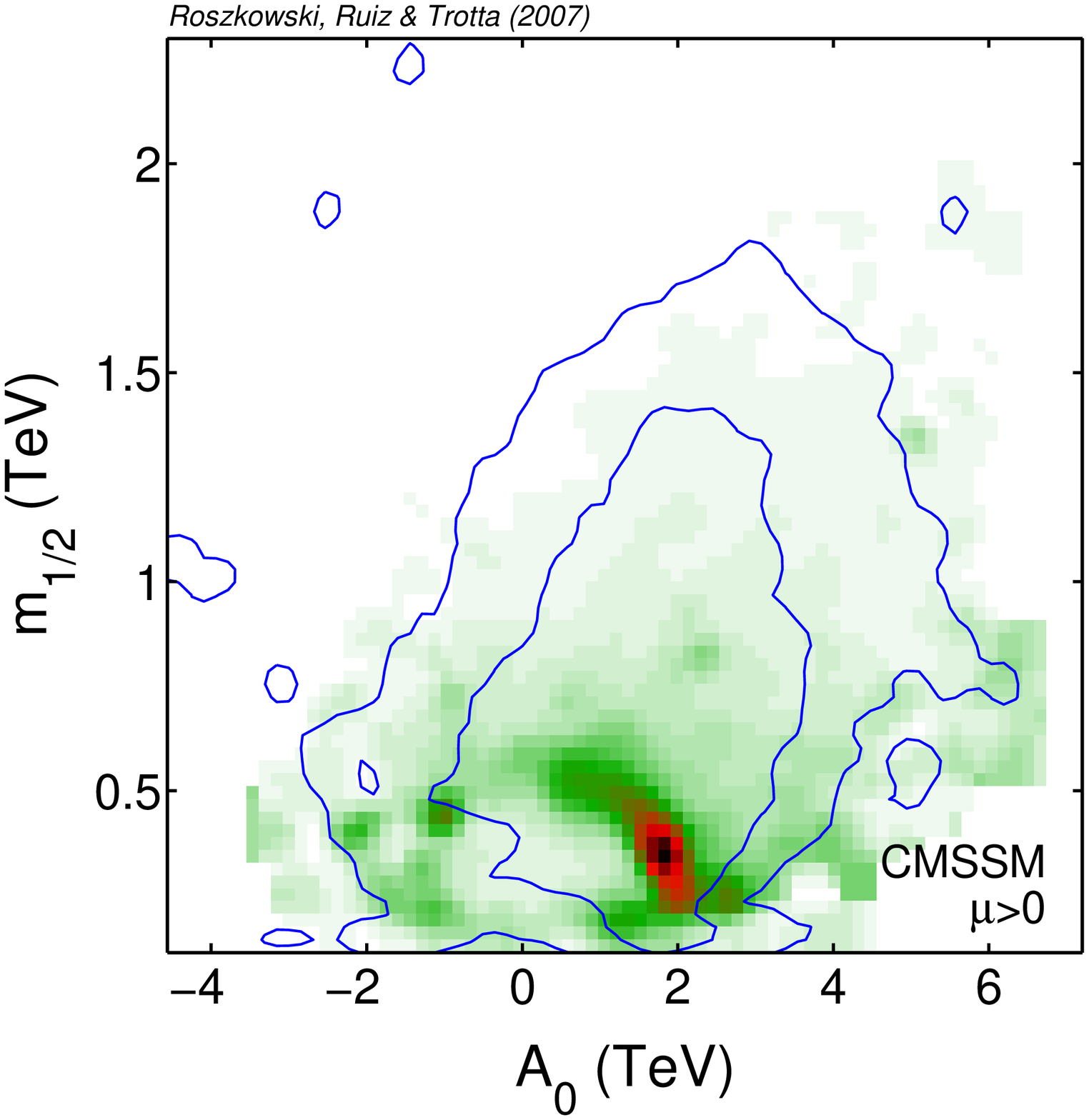}
& \includegraphics[width=0.3\textwidth]{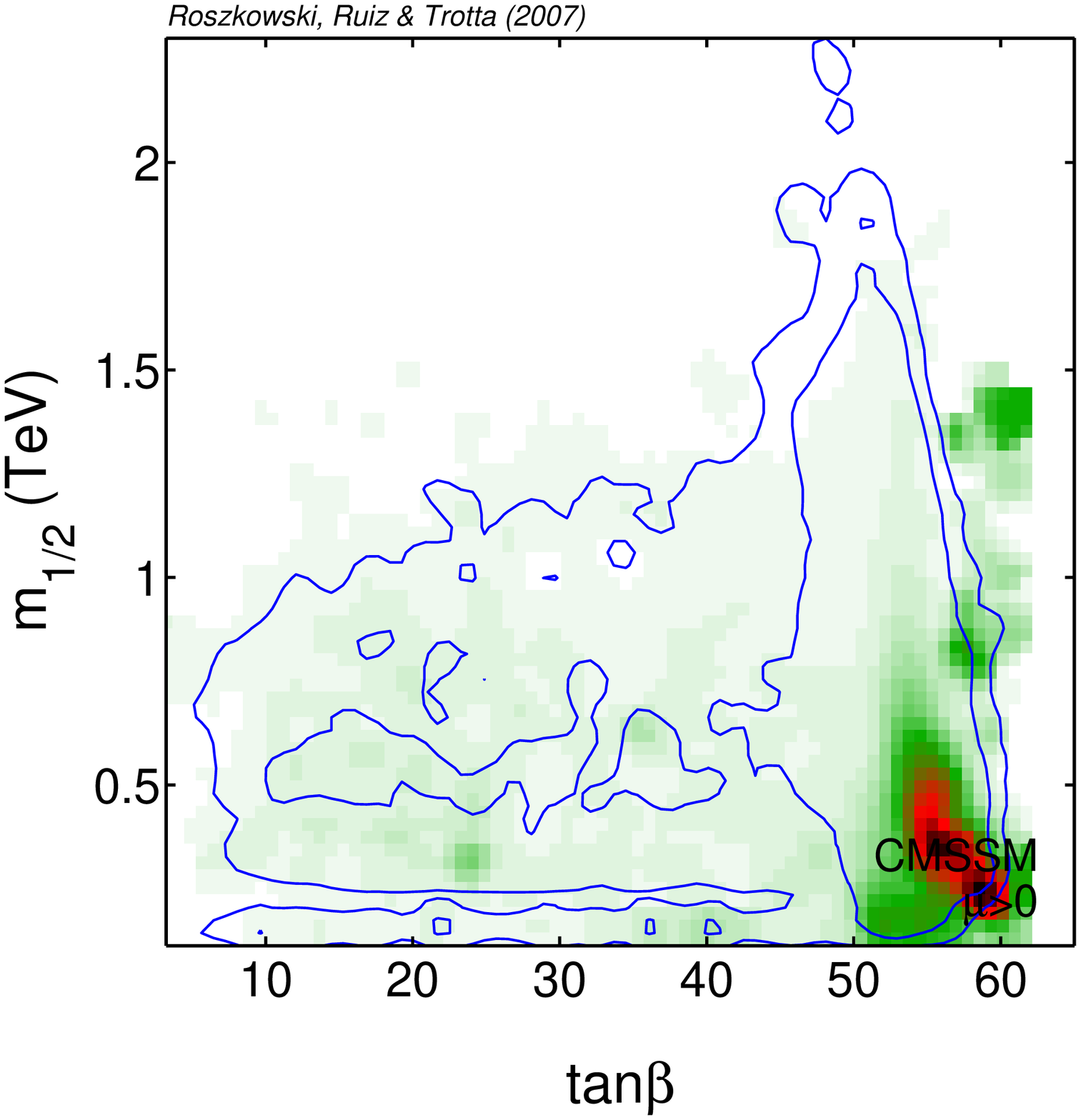}
& \includegraphics[width=0.3\textwidth]{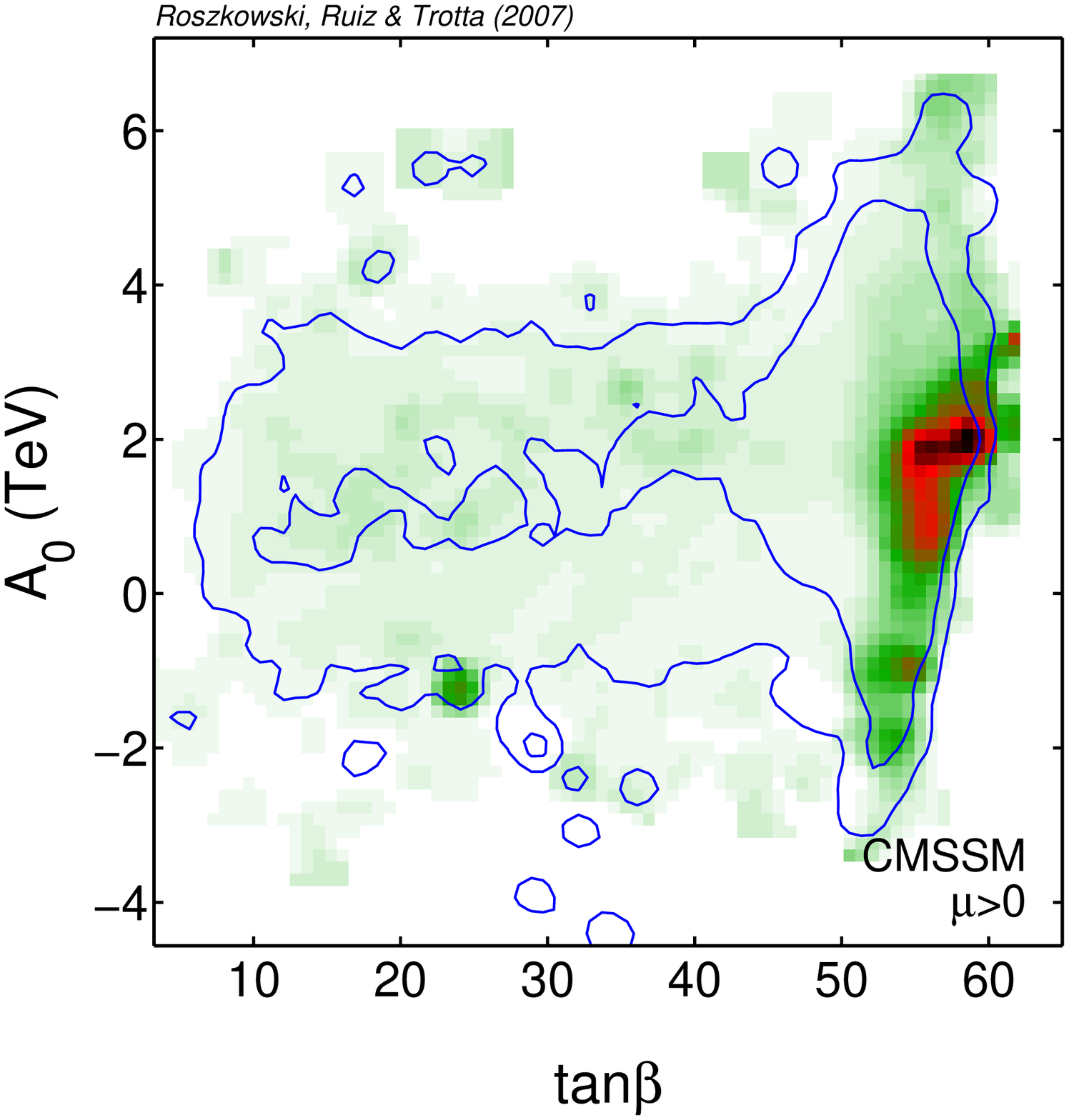}\\
\end{tabular}
  \includegraphics[width=0.3\textwidth]{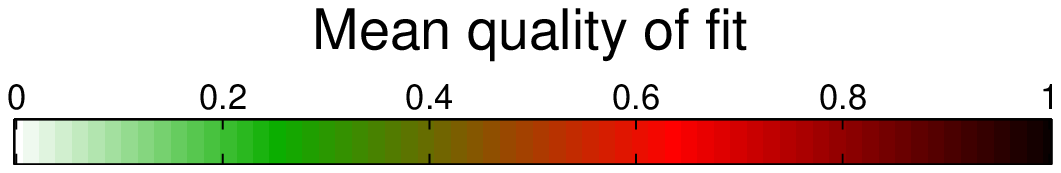}
\end{center}
\caption{\label{fig:cmssm2dcontoursmup-like} {The mean quality-of-fit
    in the planes spanned by the CMSSM parameters: $\mhalf$, $\mzero$,
    $\azero$ and $\tanb$ for $\mu>0$. For comparison, (blue solid)
    total posterior probability contours of 68\% and 95\% from
    fig.~\protect\ref{fig:cmssm2dcontoursmup} have been added. 
This figure should be compared with figure~11 in ref.~\protect\cite{rtr1}.
}}
\end{figure}

\begin{figure}[tbh!]
\begin{center}
\begin{tabular}{c c c}
  \includegraphics[width=0.3\textwidth]{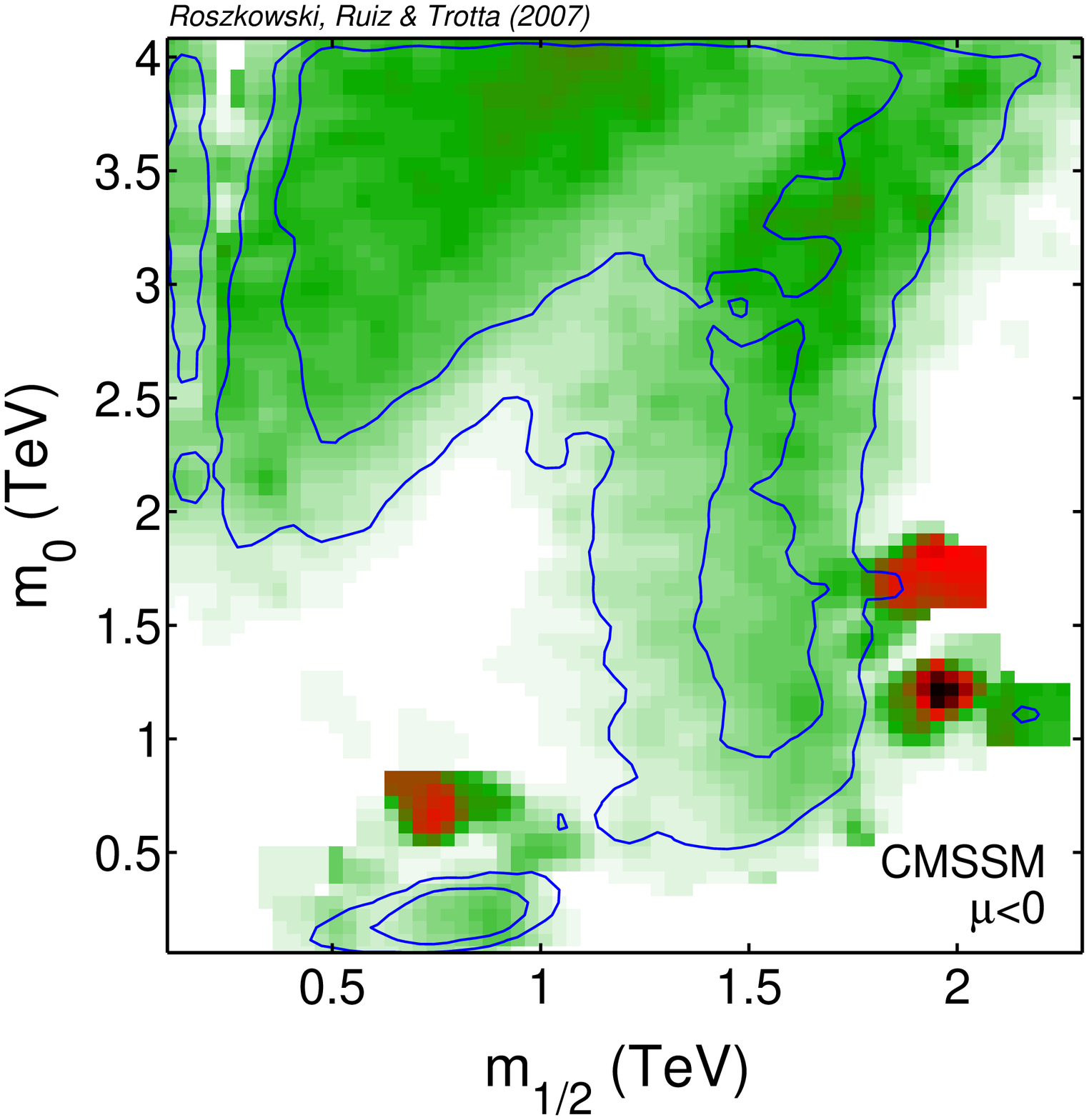}
& \includegraphics[width=0.3\textwidth]{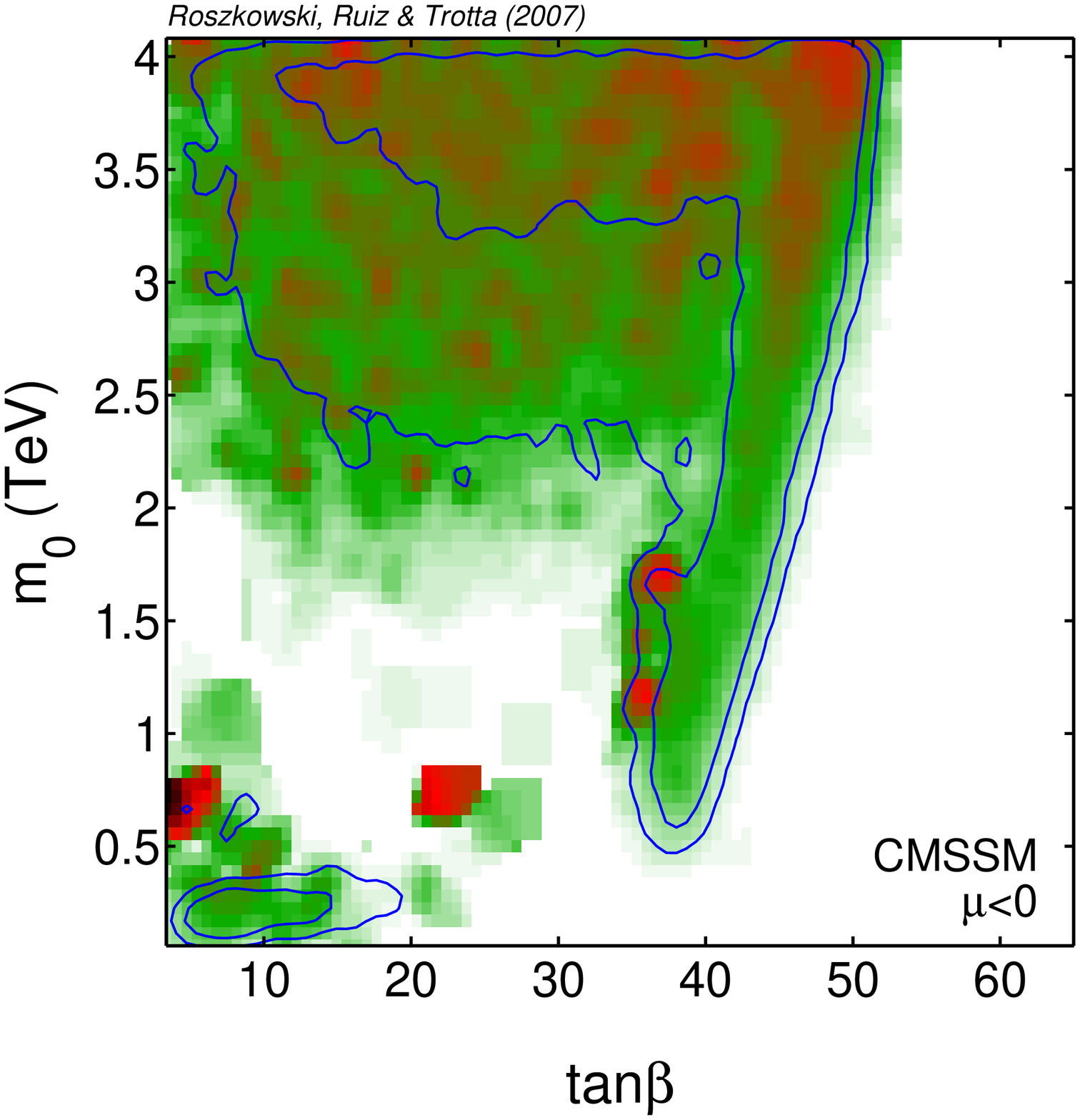}
& \includegraphics[width=0.3\textwidth]{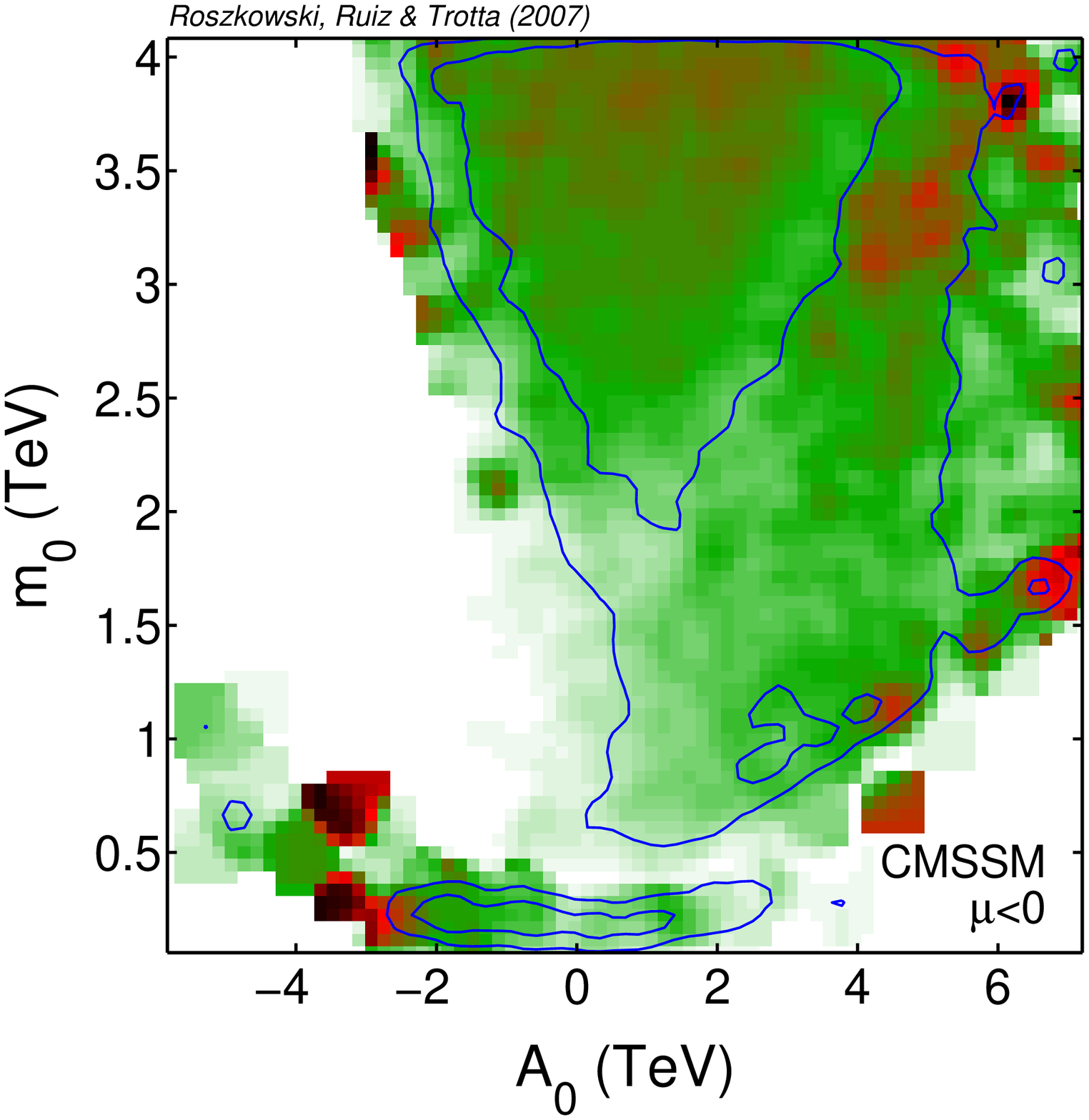}\\
  \includegraphics[width=0.3\textwidth]{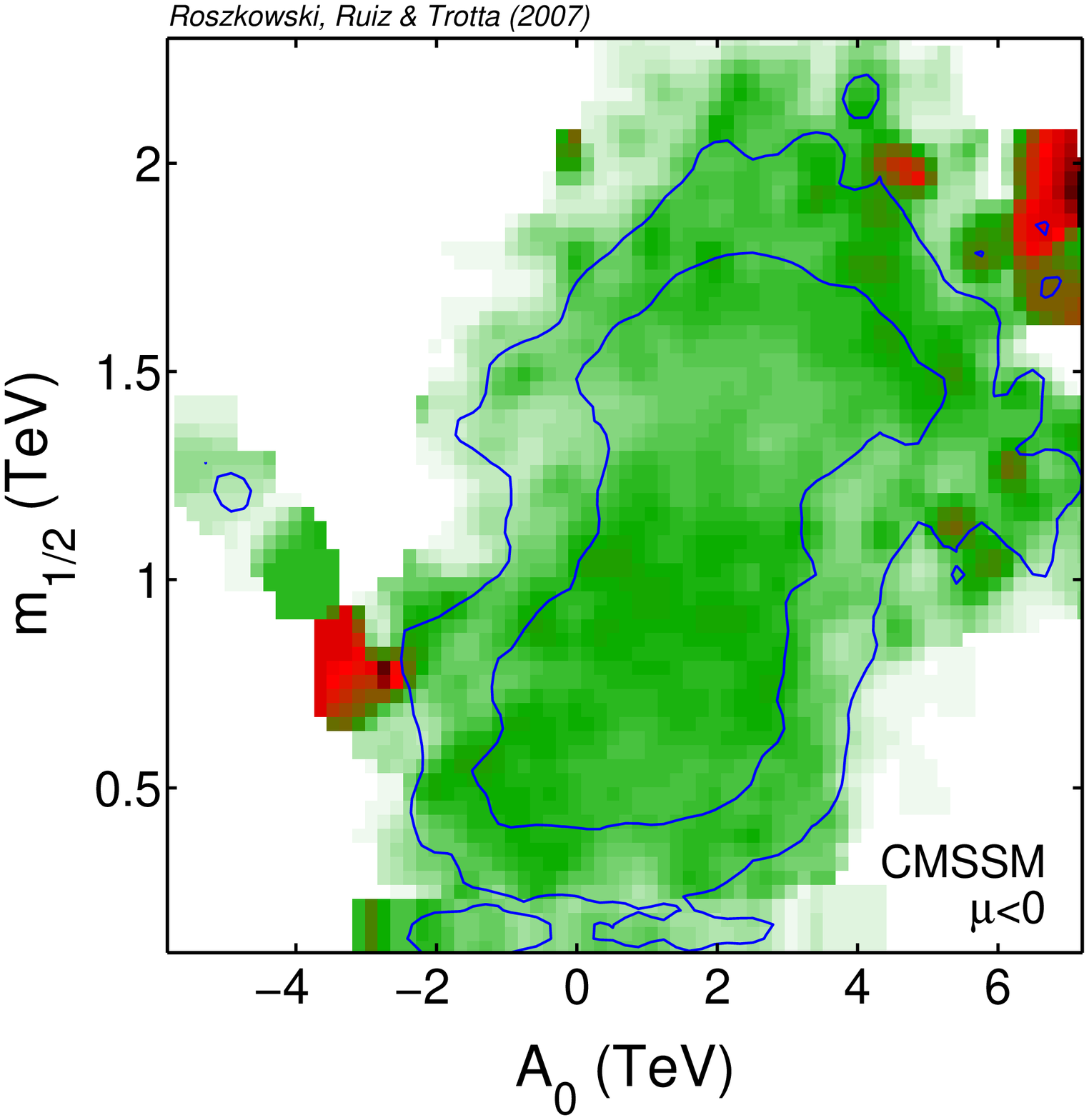}
& \includegraphics[width=0.3\textwidth]{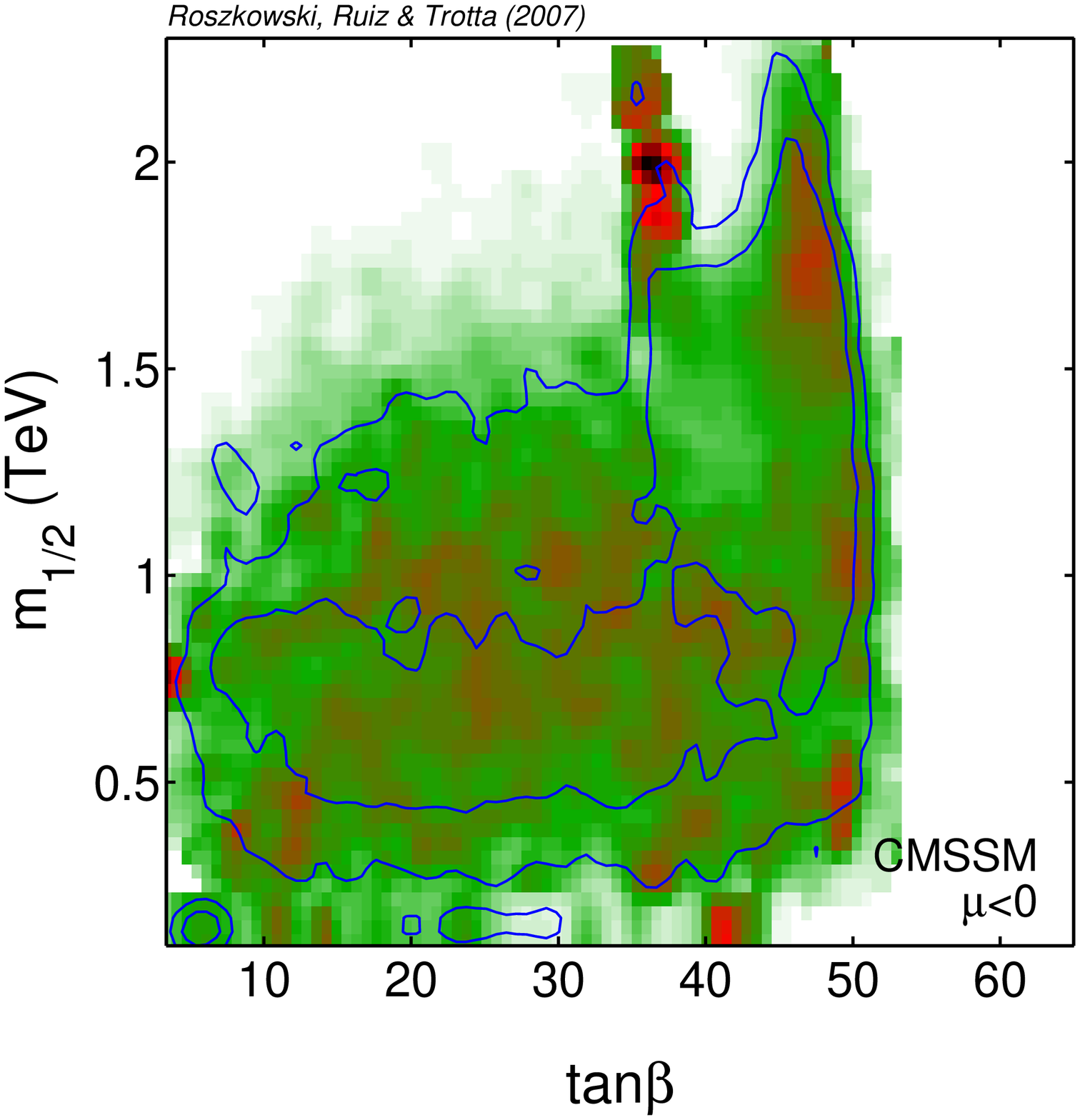}
& \includegraphics[width=0.3\textwidth]{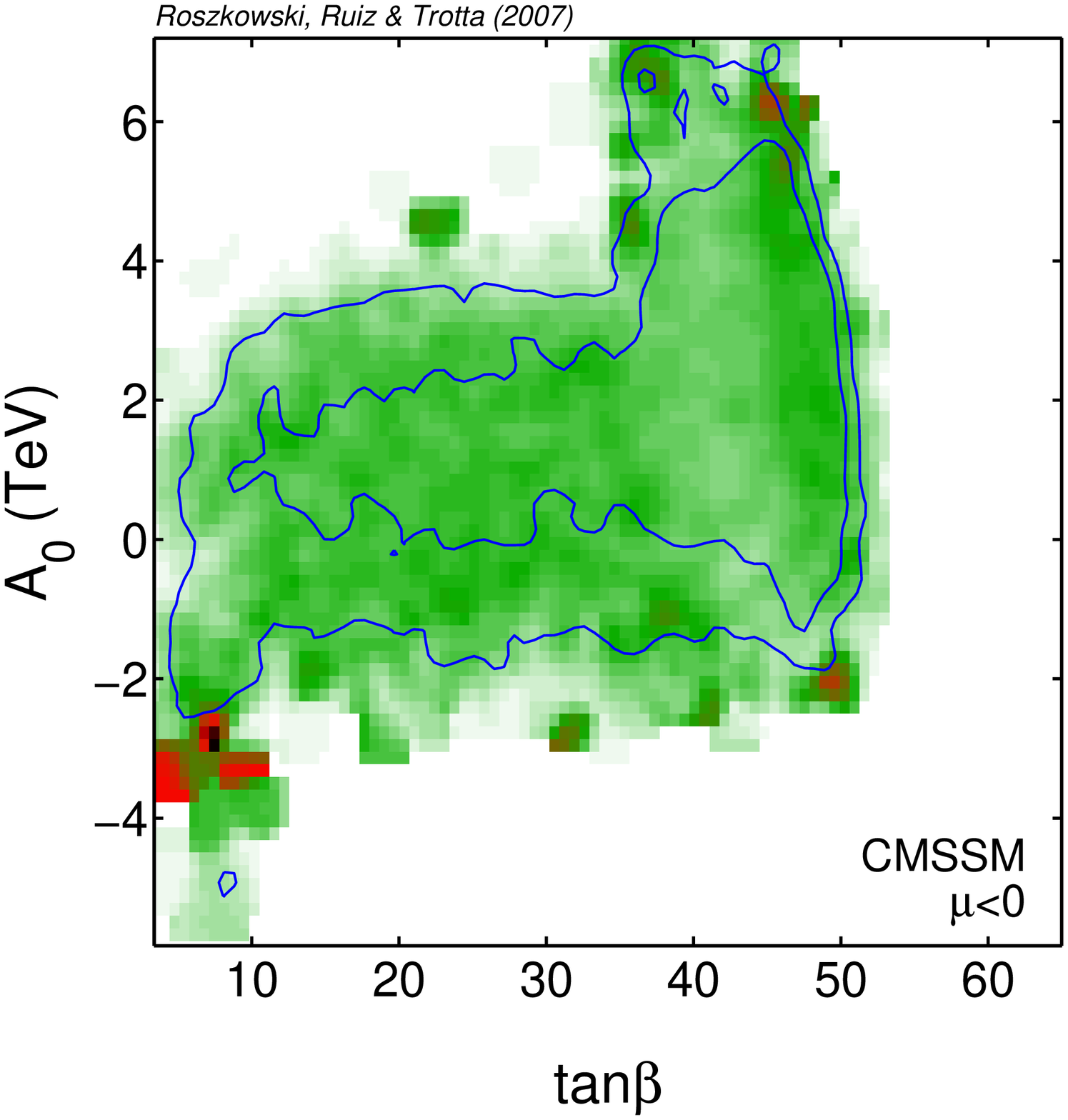}\\
\end{tabular}
  \includegraphics[width=0.3\textwidth]{rrt3-colorbar-like.ps}
\end{center}
\caption{\label{fig:cmssm2dcontoursmun-like} {The same as in
fig.~\protect\ref{fig:cmssm2dcontoursmup-like} but for $\mu<0$. For
comparison, (blue solid) total posterior probability contours of 68\%
and 95\% from fig.~\protect\ref{fig:cmssm2dcontoursmun} have been
added.}}
\end{figure}

A related (although somewhat different) issue is to identify
regions of the CMSSM PS where the fit to the data is much better
than elsewhere.  As stated above, if such regions occupy a small
volume of parameter space (given our choice of priors), the
posterior pdf will consequently weight them down, although they
might exhibit a higher goodness-of-fit. 
To address this point, we consider an alternative measure of the
mean quality-of-fit, which is defined as the average of the effective
$\chi^2$ under the posterior distribution, i.e.,
\begin{equation}
\langle \chi^2 \rangle = \int  d\,\basis\, 
[-2 \ln{p(\data |\basis)}] p(\basis|\data),
\label{eq:mqofdef}
 \end{equation}
which is a quantity that is largely insensitive to the choice of
priors (as long as the best-fitting points are explored by the MCMC
scan).  Its distribution for the CMSSM parameters is plotted in
figs.~\ref{fig:cmssm2dcontoursmup-like}
and~\ref{fig:cmssm2dcontoursmun-like} for $\mu>0$ and $\mu<0$,
respectively, in each case they are normalized to the respective
best-fit value.
We can see that, for $\mu>0$, there indeed exists at
least one well-localized region around $\mhalf\simeq 0.4\tev$ and
$\mzero\simeq 1.5\tev$ (not far from the location of the highest
relative pdf, and in any case within the 68\% total probability
contour), with another one at somewhat smaller values of both
$\mhalf$ and $\mzero$. (Below we will show however that such
best-fit regions may be in conflict with dark matter search
limits, which have not been applied as constraints at this stage.)
On the other hand, at $\mu<0$ it is generally more difficult to
find a good fit to the data, as indicated by the larger value of
the $\chi^2$ given above. We also notice that many of the
best-fitting regions for the case $\mu<0$ in
fig.~\ref{fig:cmssm2dcontoursmun-like} lie outside the $95\%$
posterior probability contour, further indicating a strong tension
with the data.

\begin{figure}[!tbh]
\begin{center}
\begin{tabular}{c c c}
    \includegraphics[width=0.3\textwidth]{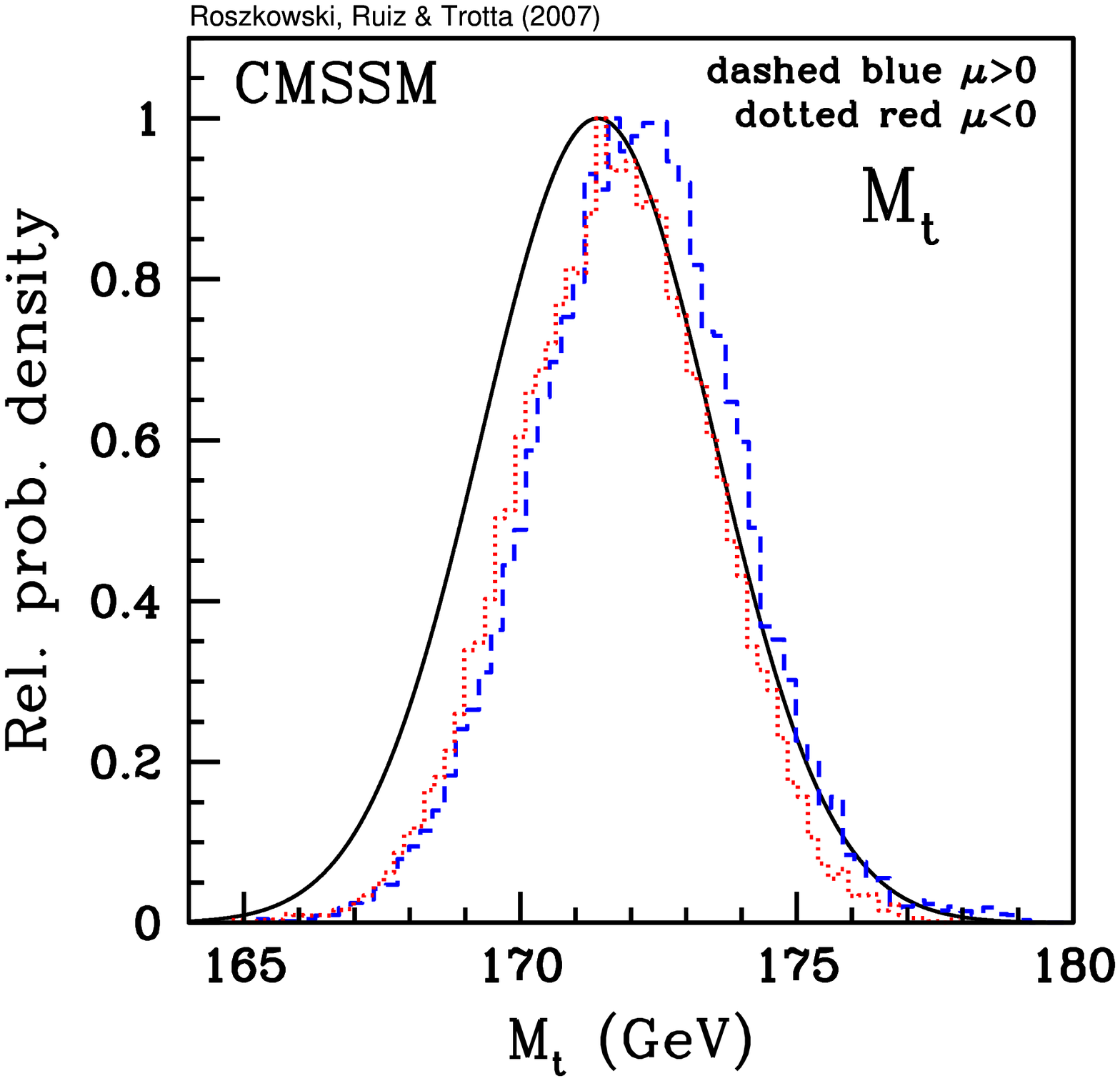}
&   \includegraphics[width=0.3\textwidth]{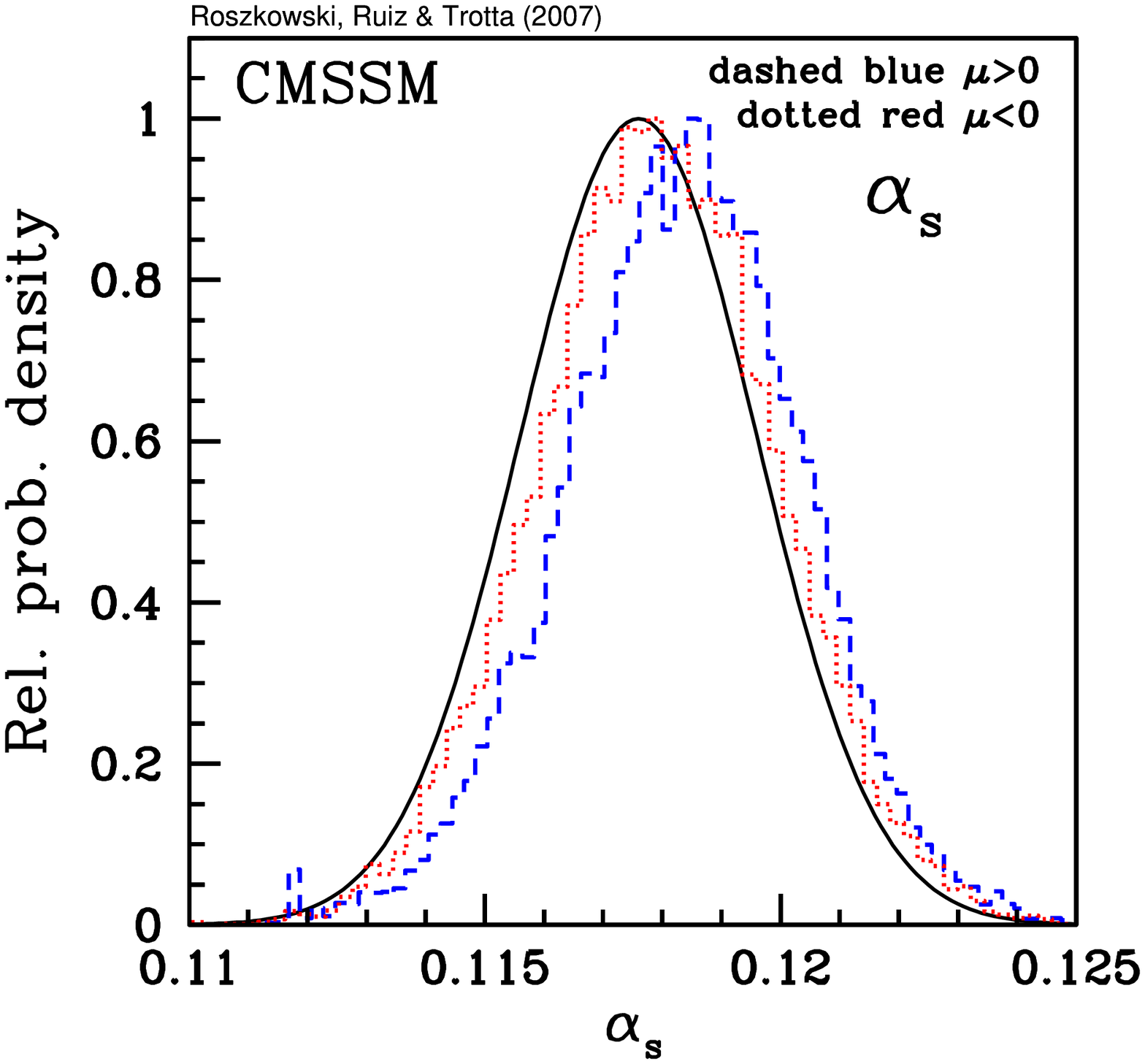}
&   \includegraphics[width=0.3\textwidth]{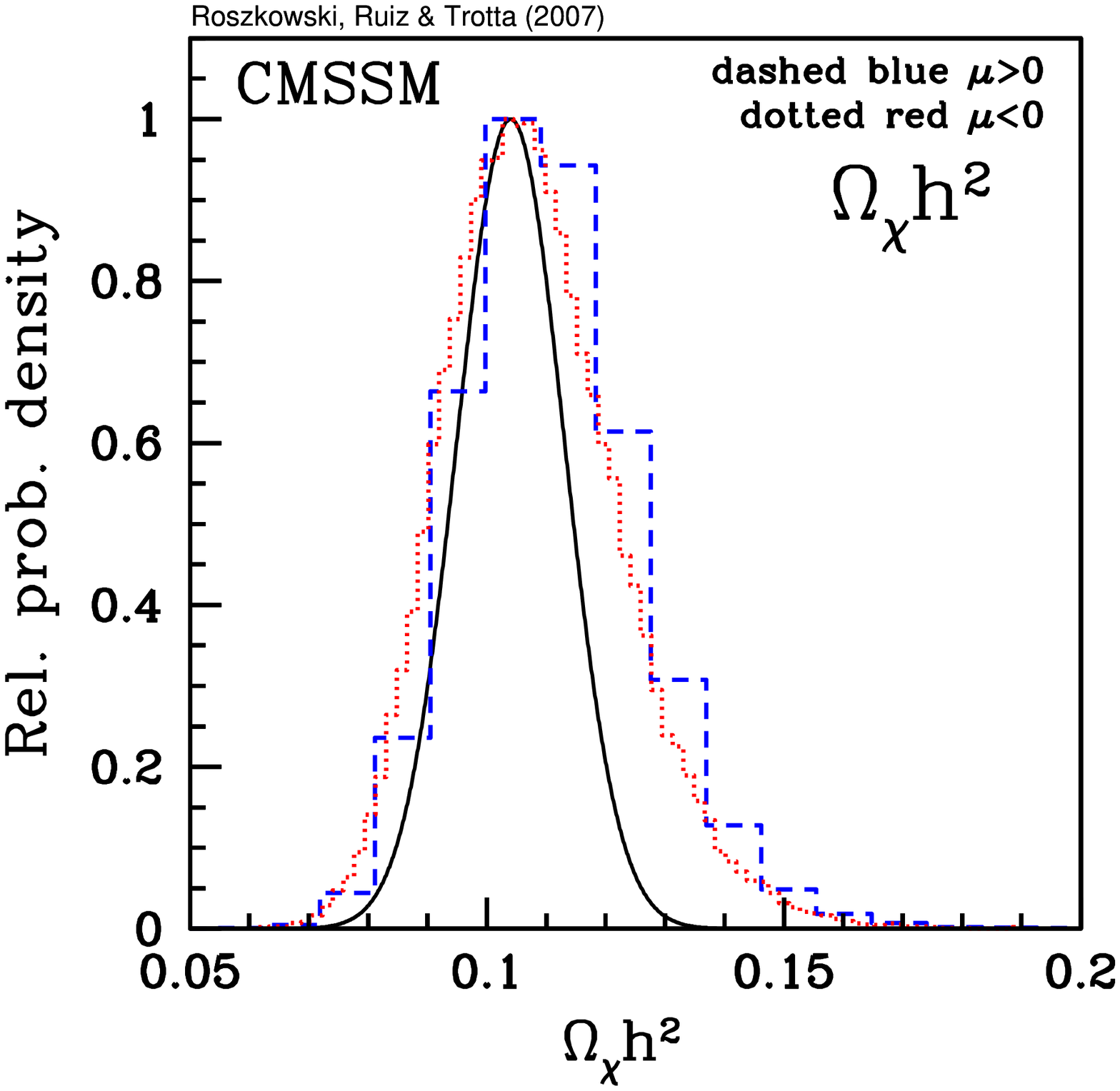}\\
    \includegraphics[width=0.3\textwidth]{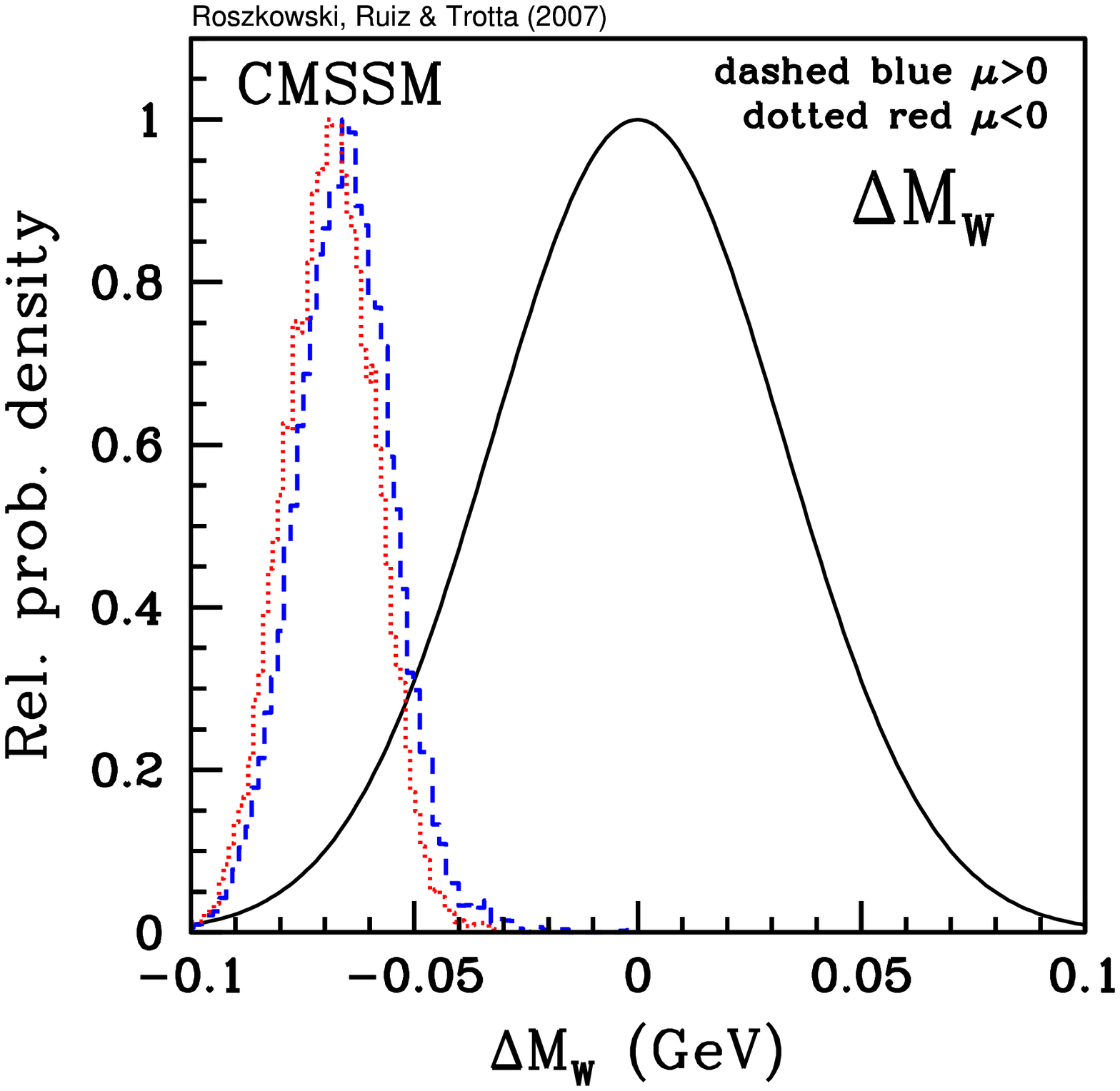}
&   \includegraphics[width=0.3\textwidth]{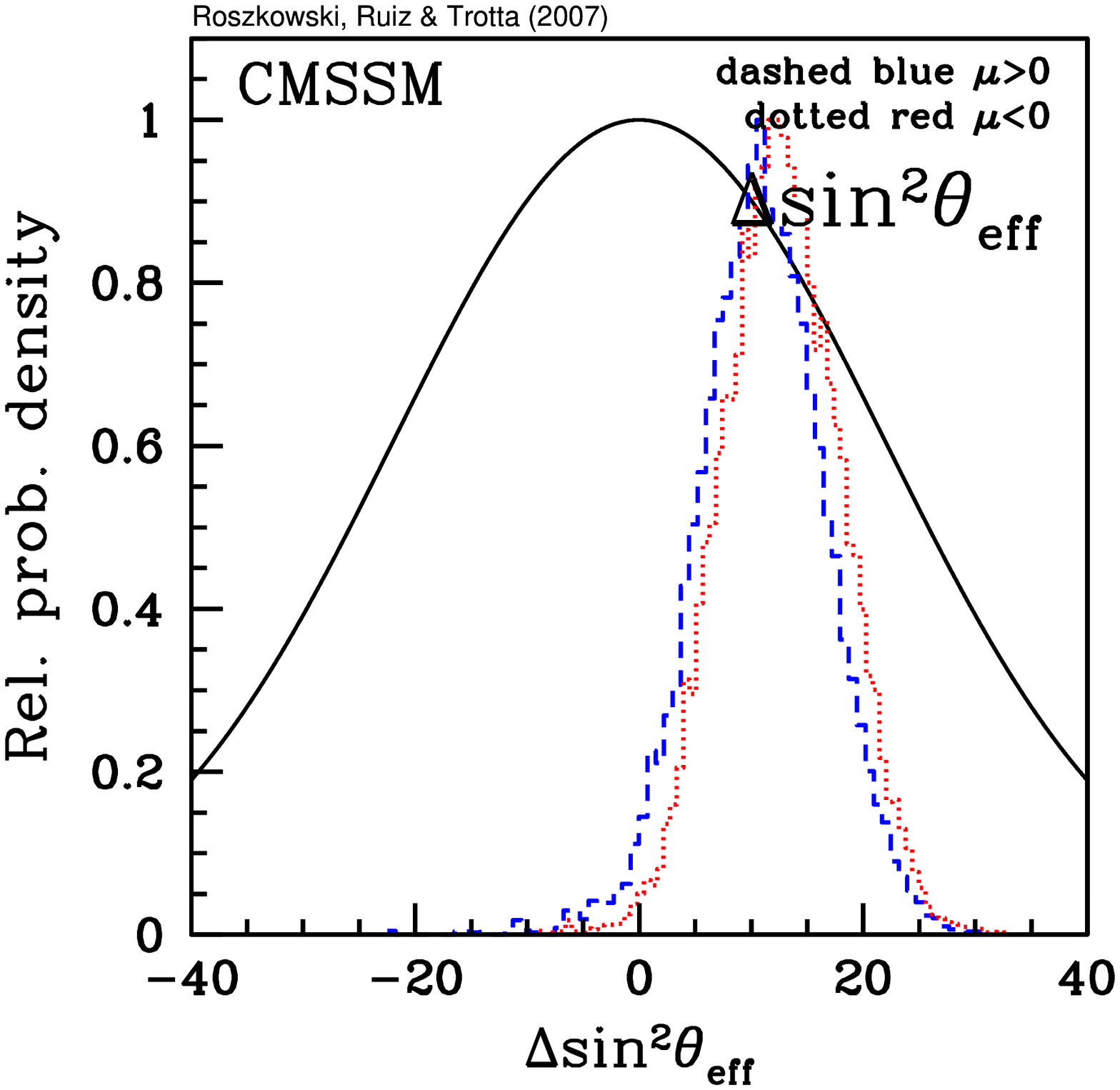}
&   \includegraphics[width=0.3\textwidth]{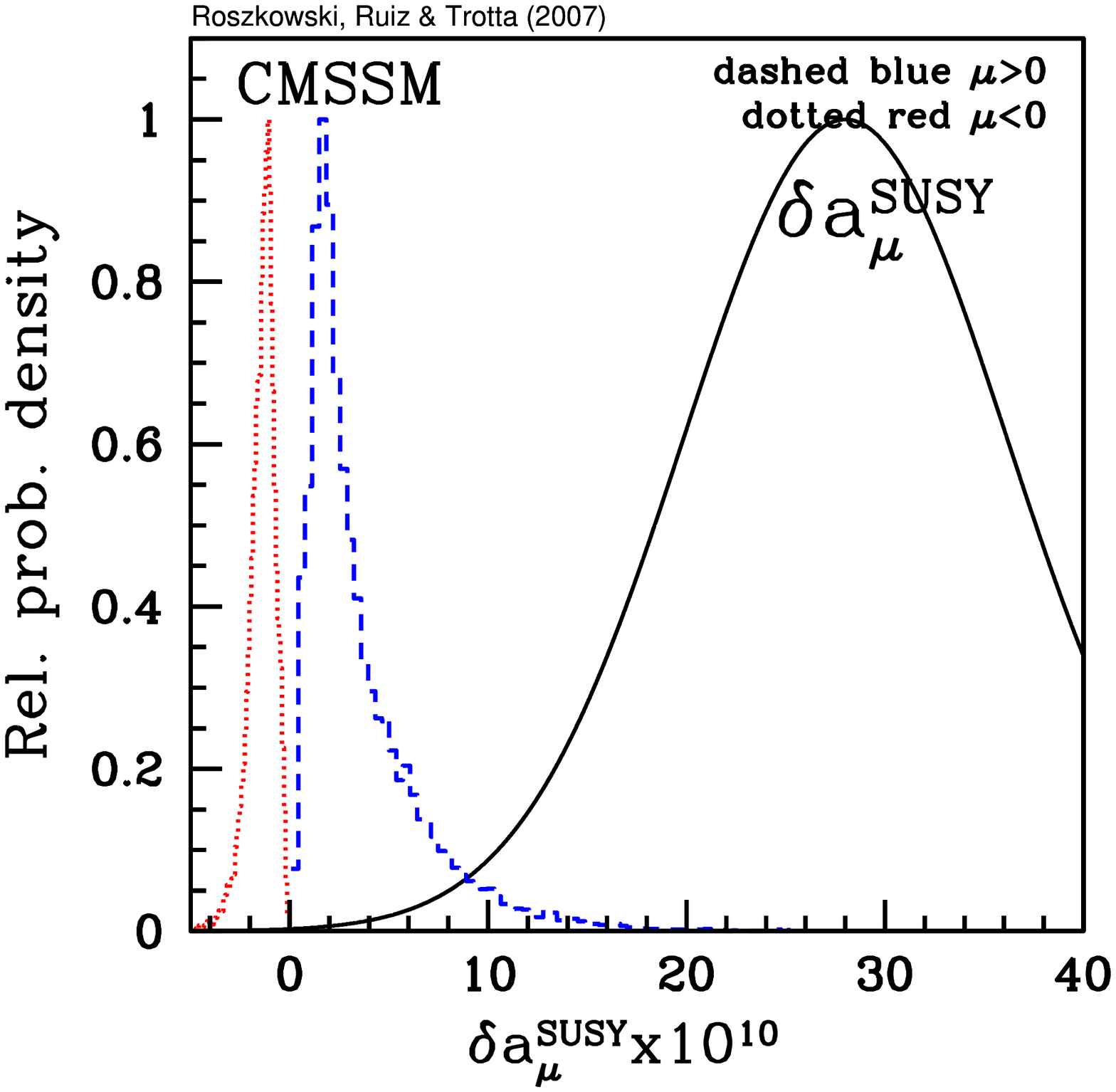}\\
    \includegraphics[width=0.3\textwidth]{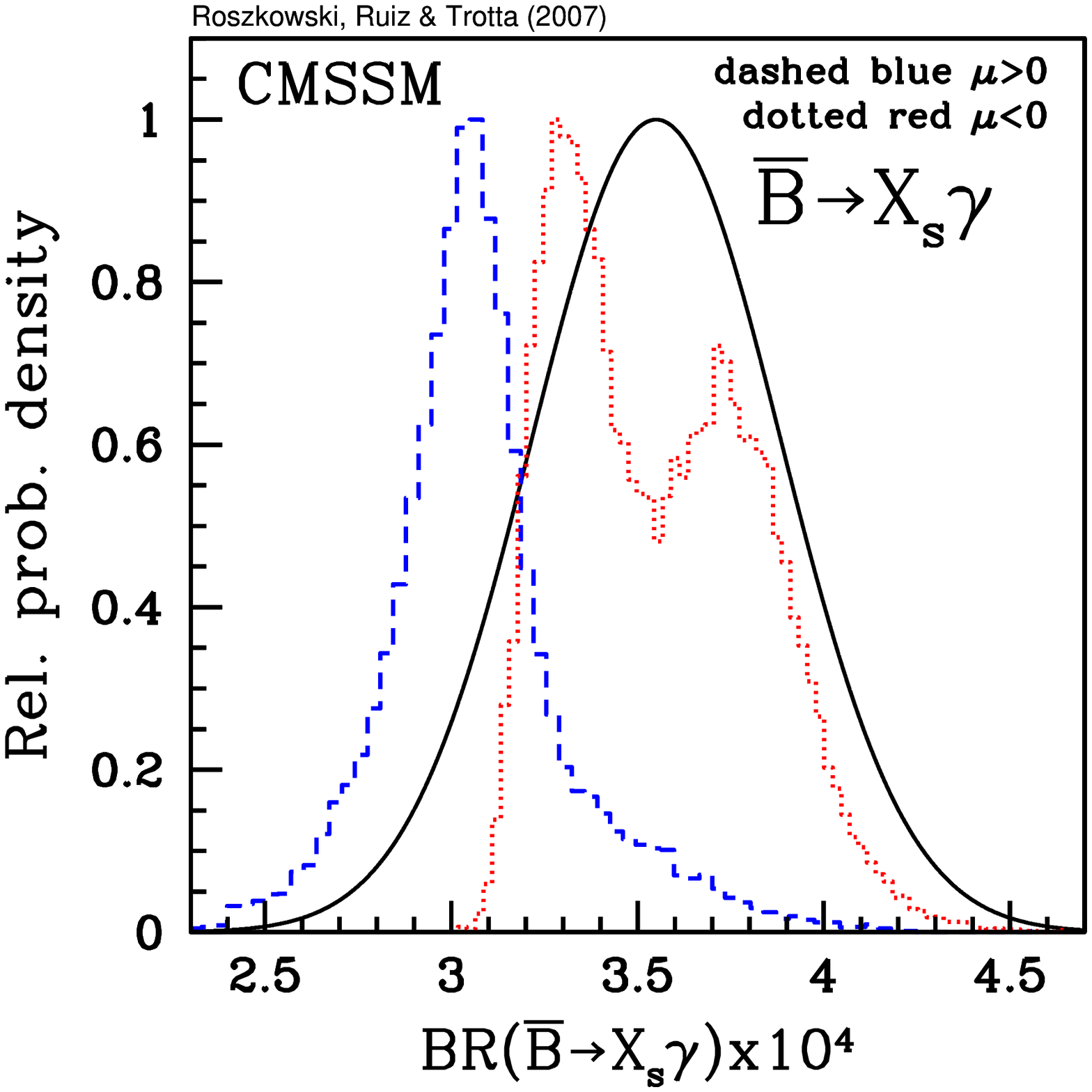}
&   \includegraphics[width=0.3\textwidth]{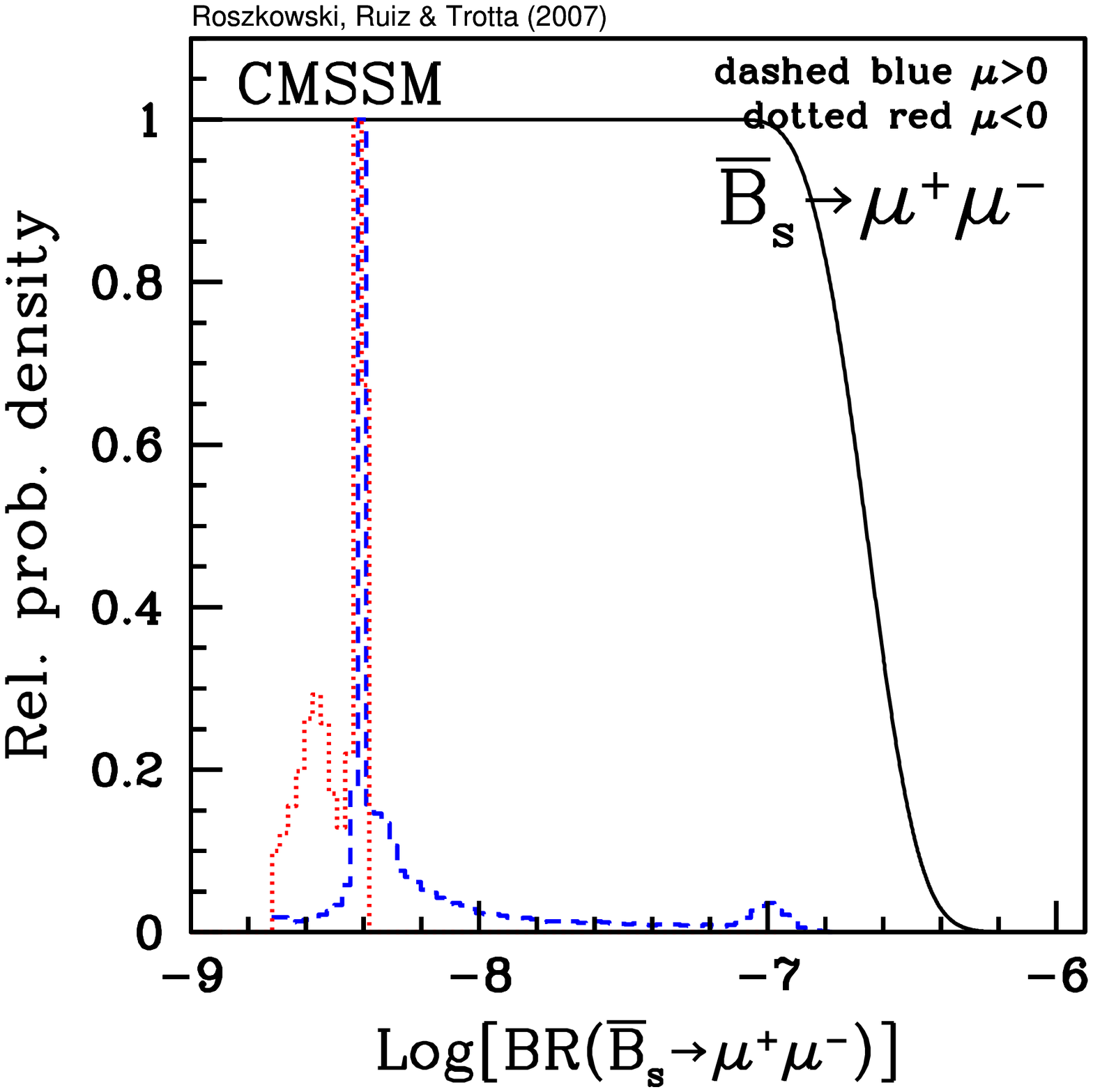}
&   \includegraphics[width=0.3\textwidth]{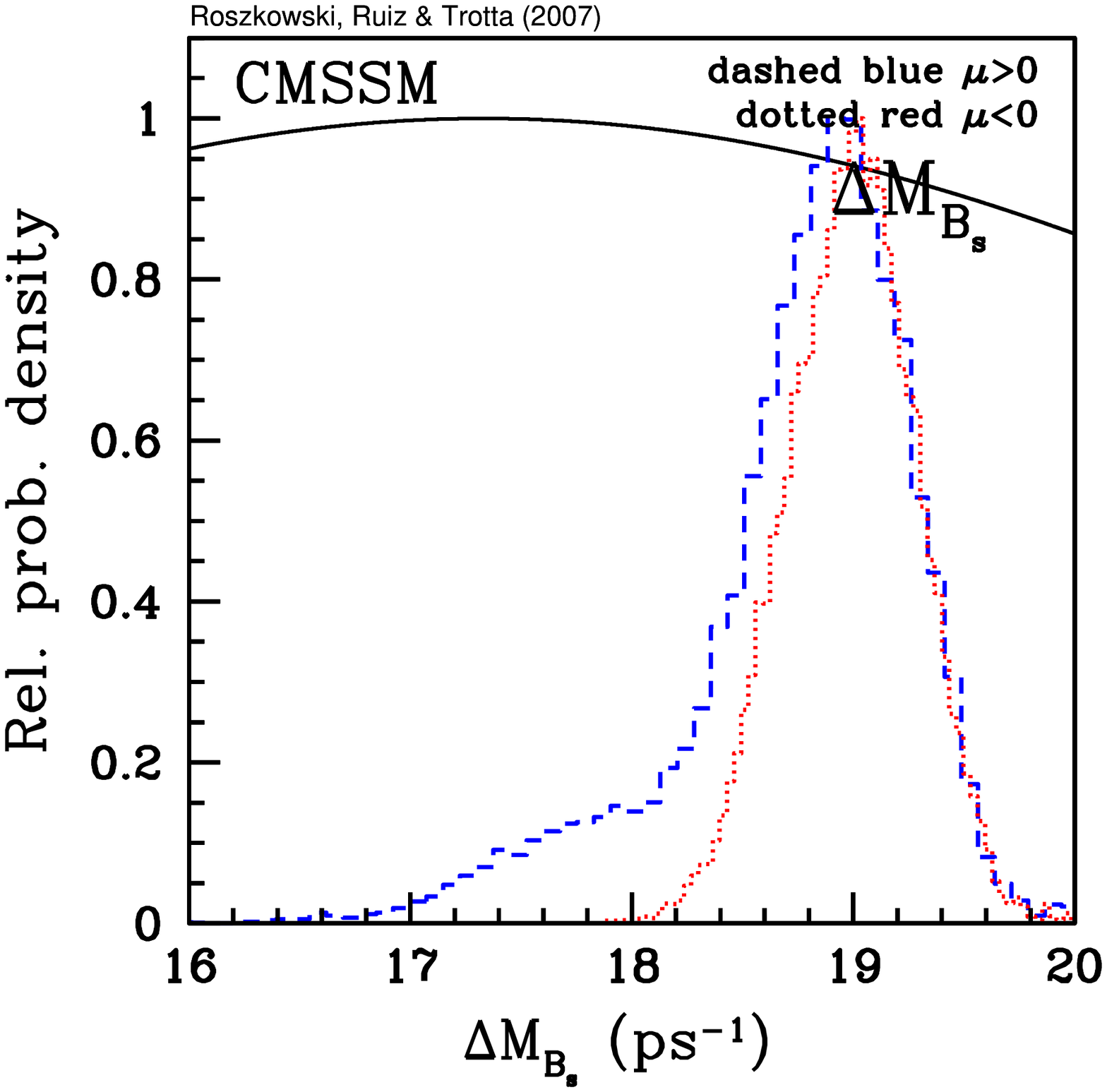}
\end{tabular}
\end{center}
\caption{The 1-dim relative probability density functions for several variables:
  $\mtpole$, $\alphas$, $\abundchi$, $\Delta\mW=\mW-80.392\gev$,
  $\Delta\sineff{}=\sineff{}-0.23153$, $\brbsgamma$,
  $\deltaamususy$, $\brbsmumu$, $\delmbs$.  In each panel the dashed blue
  (dotted red) curves correspond to $\mu>0$ ($\mu<0$) while the black
  solid line represents the data as encoded in the likelihood function.
This figure should be compared with figure~10 in ref.~\protect\cite{rtr1}.
\label{fig:tensionfig}
}
\end{figure}

Different experimental observables may constrain or favor
different regions of the CMSSM parameters; they may ``pull'' in
different directions. We display this in fig.~\ref{fig:tensionfig}
where we plot the 1-dim posterior pdf's for several variables for
both signs of $\mu$.  For comparison, we also plot the
corresponding Gaussian likelihood functions representing the data
used in the fit. If there was no tension among different
observables then, in the absence of strong correlations among
them, the relative probability curves should overlap with the
data. This is basically the case for $\mtpole$, $\alphas$ and
$\abundchi$. (In the last case the slightly skewed shape of the
pdf's is a result of our treatment of the theoretical uncertainty
which is larger for larger $\abundchi$.)  On the other hand, the
electroweak observables $\mW$ and $\sineff{}$ show some pull away
from their expected values, in general agreement with
ref.~\cite{alw06} where the two variables were computed with a
similar precision. On the other hand, the tension is insufficient
to provide convincing preference for low $\msusy$, in apparent
contrast to the findings of ref.~\cite{ehow}.

The biggest tension between best-fit values and experiment is
displayed, unsurprisingly, in the SUSY contribution to the anomalous
magnetic moment of the muon. For both signs of $\mu$ the peaks of the
relative probability are far below the central experimental value
(about $3.2\sigma$ for $\mu>0$ versus about $3.7\sigma$ for $\mu<0$),
and close to each other~\cite{alw06}. We conclude that it is not
justified to use this sole observable to select the positive sign of
$\mu$ -- one has to perform a global fit in all of the variables and
judge the two cases by this criterion. Furthermore, the new
$\brbsgamma$ actually seems to agree with the data slightly better for
$\mu<0$ than for the other sign. Generally, for $\mu>0$ the total
$\brbsgamma$ remains peaked around the SM central value, while for
$\mu<0$ it is somewhat above it. Finally, $\brbsmumu$ and $\delmbs$
are peaked at their SM values, somewhat more so than
before~\cite{rtr1}.

\subsection{Implications for collider searches}\label{sec:susyobs}

In our Bayesian formalisms, high probability ranges of the CMSSM
parameters can easily be translated into analogous ranges for
Higgs and superpartner masses, and for other observables,
including indirect processes and dark matter detection cross
sections. We discuss them in turn below.

\begin{figure}[!tbh]
\begin{center}
\begin{tabular}{c c}
    \includegraphics[width=0.45\textwidth]{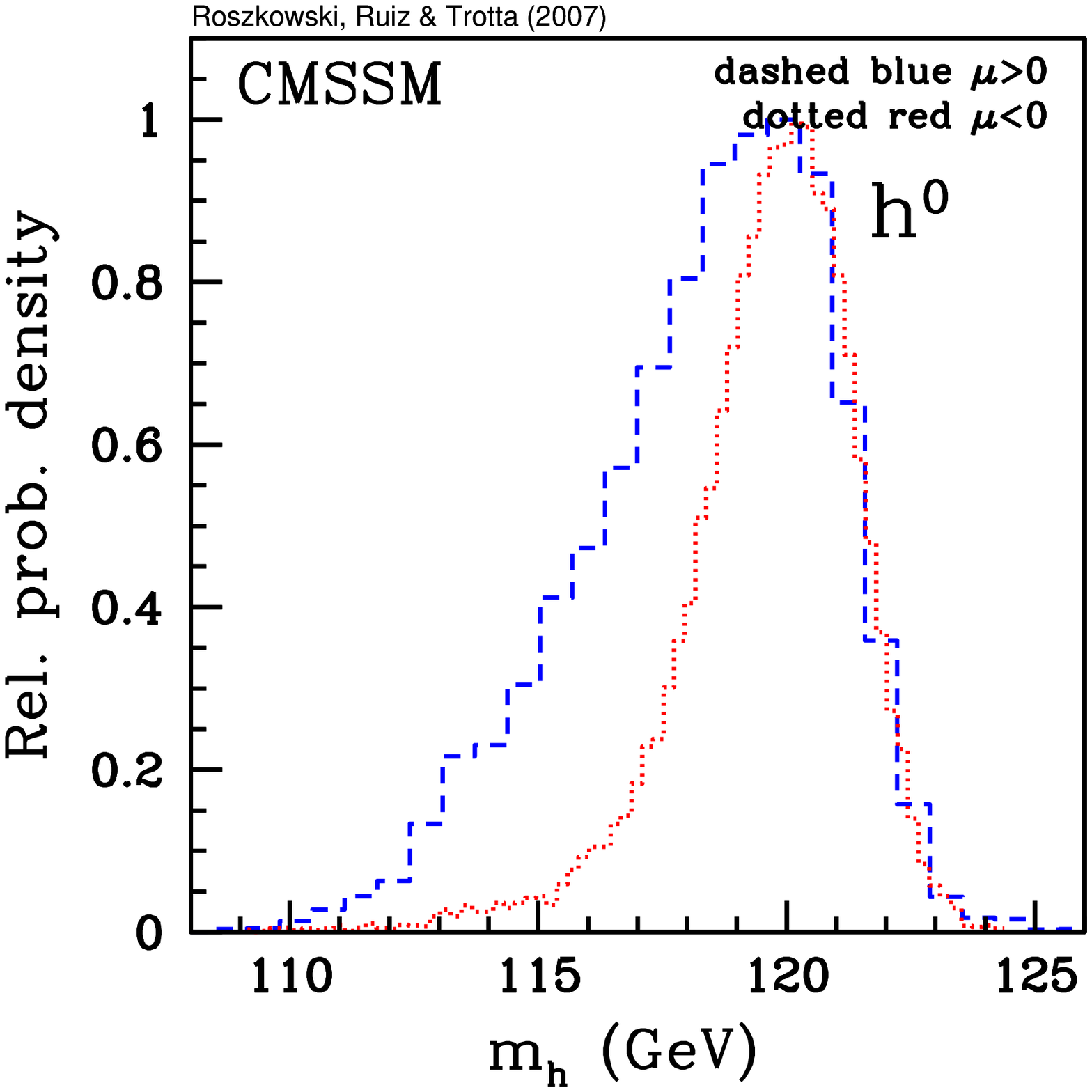}
  & \includegraphics[width=0.45\textwidth]{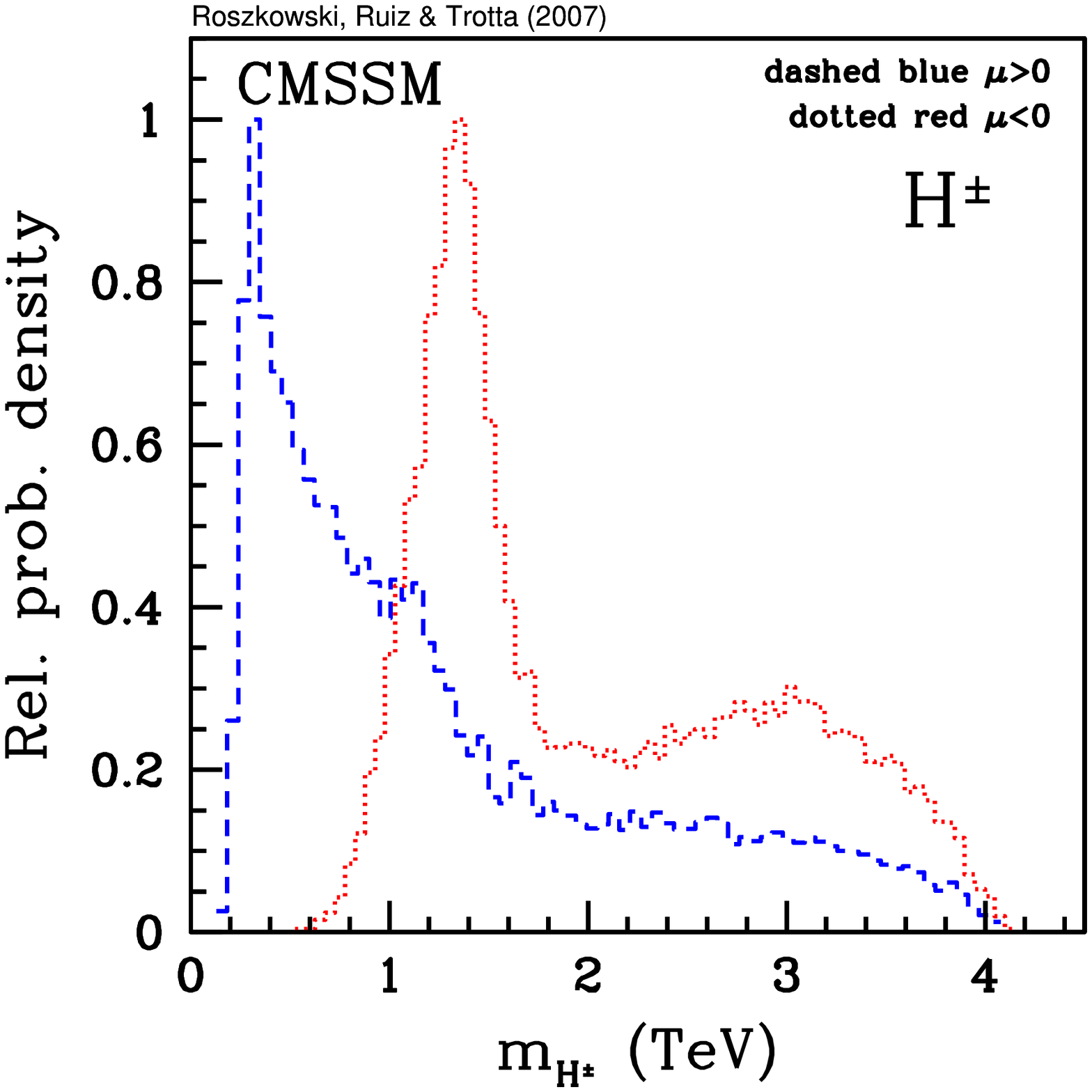}
\end{tabular}
\end{center}
\caption{The 1-dim relative probability density functions for the
masses of $\hl^0$ (left panel) and $\hpm$ (right panel). Dashed blue
(dotted red) curves correspond to $\mu>0$ ($\mu<0$).  Analogous and
more detailed figures for the case $\mu>0$ were presented in fig.~7 of
ref.~\cite{rrt2}.
\label{fig:hmass-like} }
\end{figure}

We start with the Higgs bosons. In fig.~\ref{fig:hmass-like} we
plot the Bayesian relative probability density distributions of
the mass of the lightest Higgs boson $\hl^0$ and of the charged
Higgs boson $\hpm$ for both signs of $\mu$.\footnote{The case of
$\mu>0$ was already presented in fig.~7 of ref.~\cite{rrt2}. We
include it here for comparison.} (The other Higgs bosons are
basically degenerate in mass with $\hpm$.) In both cases we can
see a clear peak in $\mhl$ close to $120\gev$, and a sharp
drop-off for larger values of the mass.  The other Higgs are
typically considerably heavier for negative $\mu$ than for the
other choice. This may provide one way of an experimental
determination of the sign of $\mu$.  The 68\% and 95\% total
probability ranges of the Higgs masses are given in
table~\ref{table:massesTable} below. We should also note that the
alternative measure of the mean quality-of-fit favors lower ranges
of the masses of all the Higgs bosons in the case of
$\mu>0$~\cite{rrt2}. For the case $\mu<0$, the mean quality-of-fit
distribution of $\mhl$ is roughly similar to that of the pdf but
with the peak shifted to the right by about $1\gev$ (plus some
moderate preference for smaller values, around the current LEP-II
limit), while the masses of all the other Higgs bosons are
preferably very heavy, in the TeV regime. This is a reflection of
the absence of regions giving good fit to the data in the case of
$\mu<0$ (compare fig.~\ref{fig:cmssm2dcontoursmun-like}).

In ref.~\cite{rrt2} we have investigated in detail light Higgs masses
and couplings for the case of $\mu>0$.  In particular we showed that,
throughout the whole CMSSM parameter space, the couplings of the
lightest Higgs boson $\hl^0$ to the gauge bosons $Z$ and $W$ are very
close to those of the SM Higgs boson with the same mass, while its
couplings to bottom quark and tau lepton pairs show some variation. We
concluded that, at the Tevatron, with about $2\fb^{-1}$ of integrated
luminosity per experiment (already on tape), it should be possible to
set a 95\%~\cl\ exclusion limit for the whole 95\% posterior
probability range of $\mhl$.  Based on fig.~\ref{fig:hmass-like} and
table~\ref{table:massesTable} we extend this conclusion to the case of
$\mu<0$. On the optimistic side, should a Higgs signal be found, in
order to be able to claim a $3\sigma$ evidence, at least about
$4\fb^{-1}$ will be needed, independently of the sign of $\mu$. The
Tevatron's ultimate goal is to collect about $8\fb^{-1}$ per
experiment.

One should remember that the above conclusions do depend on the
assumed prior range of $\mzero<4\tev$, as well as on the choice of
adopting flat priors in the CMSSM variables of
eq.~\ref{indeppars:eq}. For instance, adopting a much more generous
upper limit $\mzero<8\tev$ would lead to changing the ranges for
$\mu>0$ to roughly $120.4\gev \lsim \mhl \lsim 124.4\gev$ (68\%~\cl)
and $115.4\gev \lsim \mhl \lsim 125.6 \gev$ (95\%~\cl), the latter of
which could be excluded at 95\%~\cl\ with about $3\fb^{-1}$ of
integrated luminosity per experiment~\cite{rrt2}. Still, should no
Higgs signal be found at the Tevatron, large ranges of $\mzero$ will
become excluded at high~\cl, with the specific value depending on the
accumulated luminosity.

\begin{figure}[tbh!]
\begin{center}
\begin{tabular}{c c c}
    \includegraphics[width=0.3\textwidth]{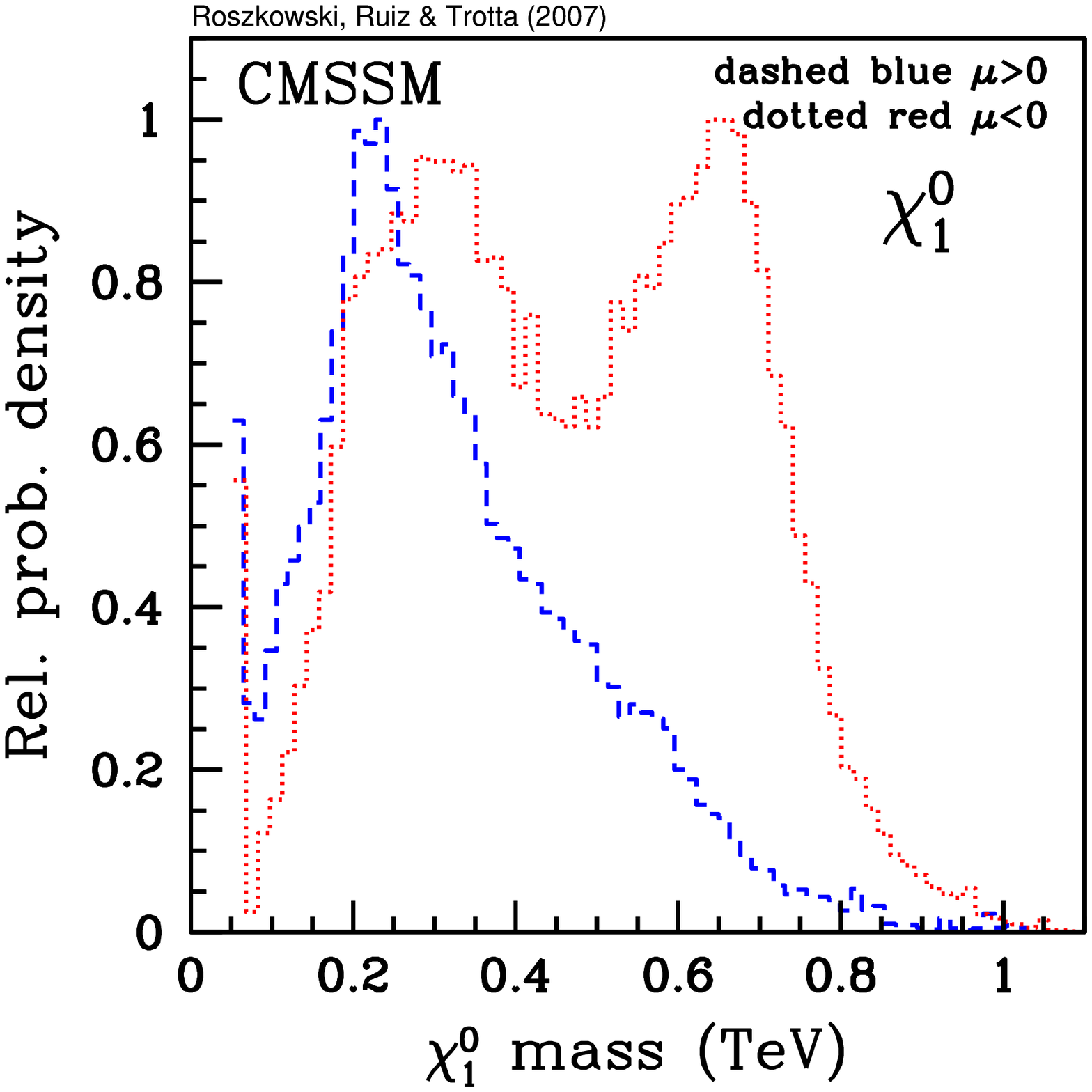}
&   \includegraphics[width=0.3\textwidth]{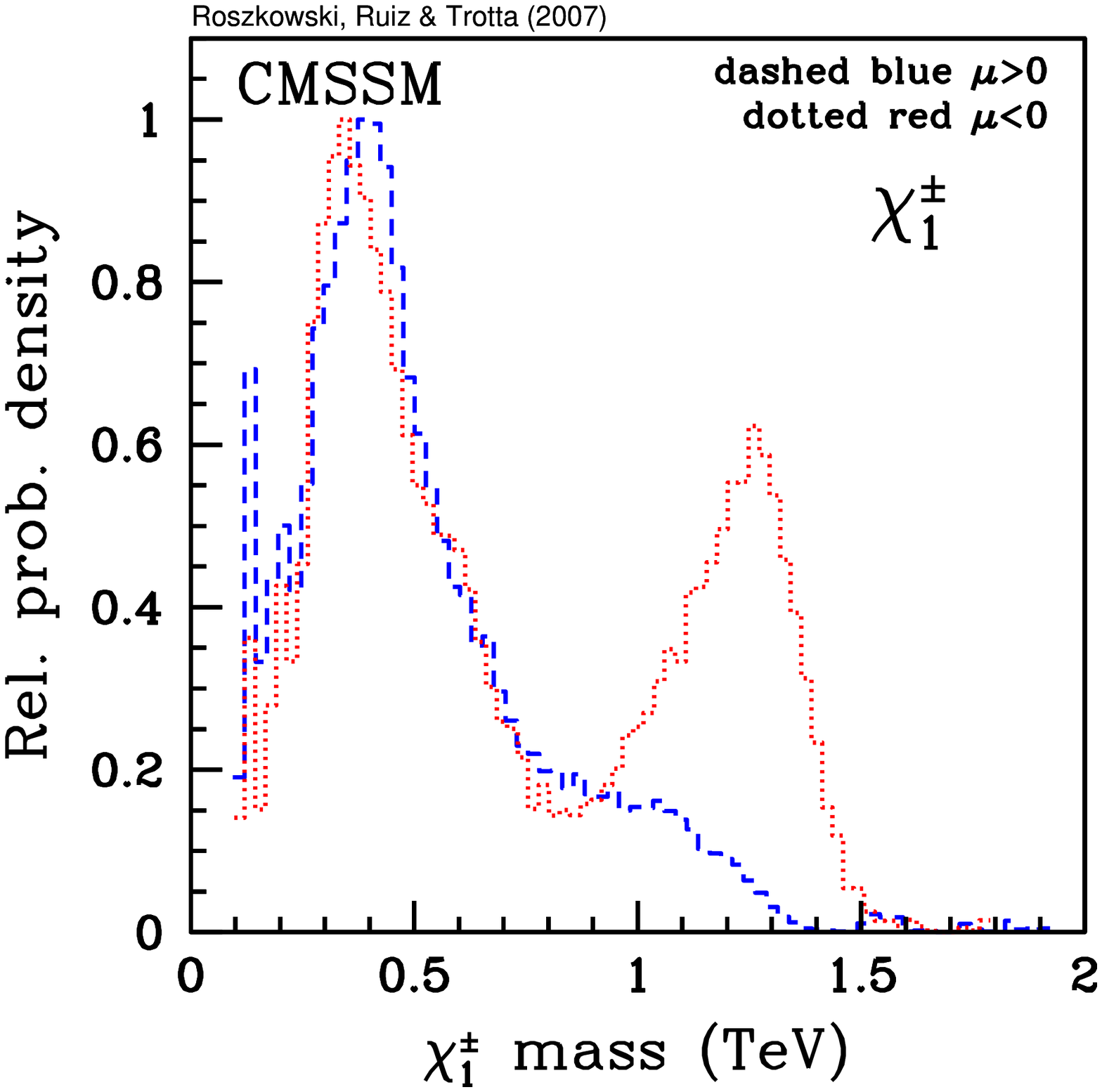}
&   \includegraphics[width=0.3\textwidth]{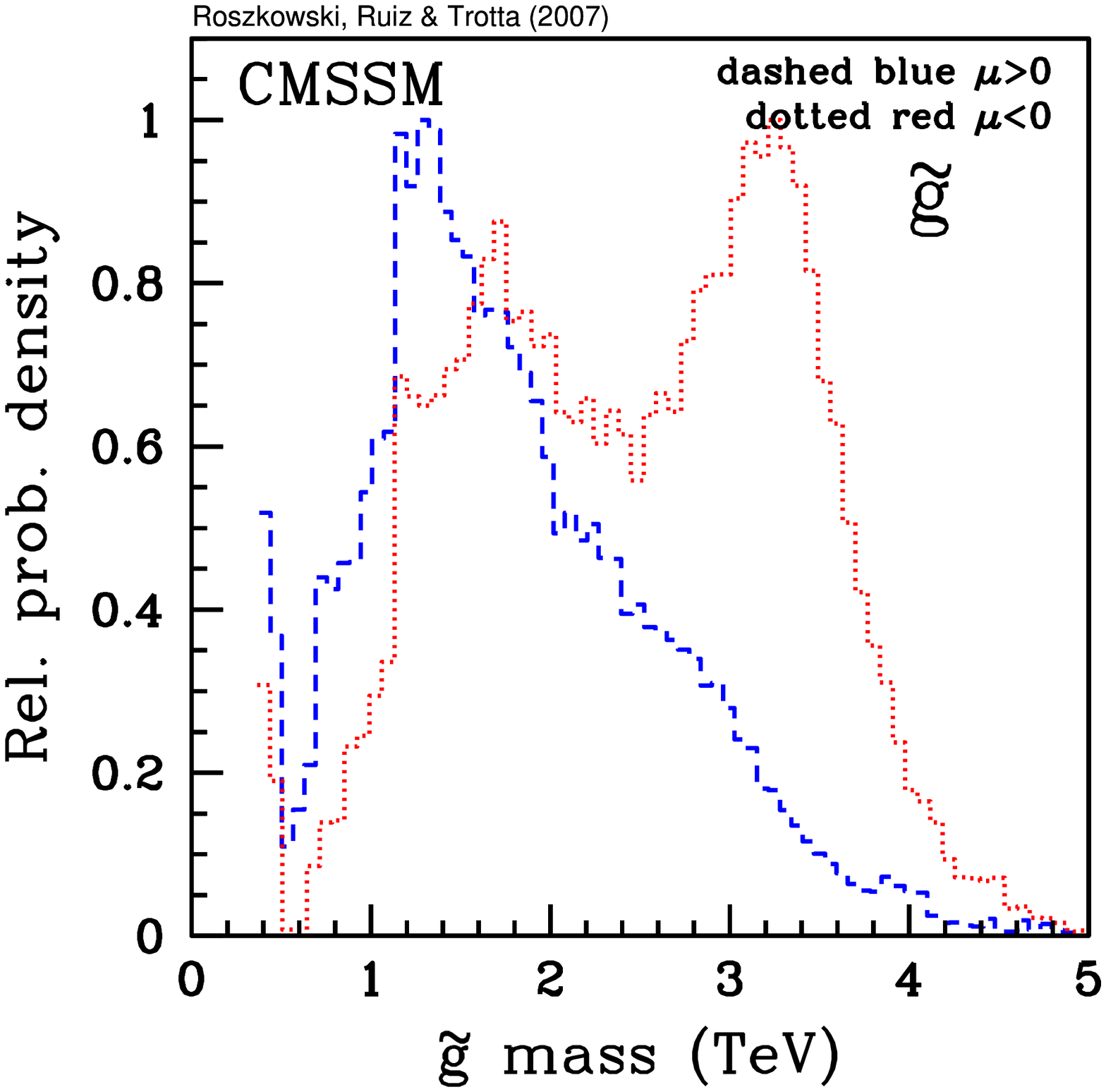}\\
    \includegraphics[width=0.3\textwidth]{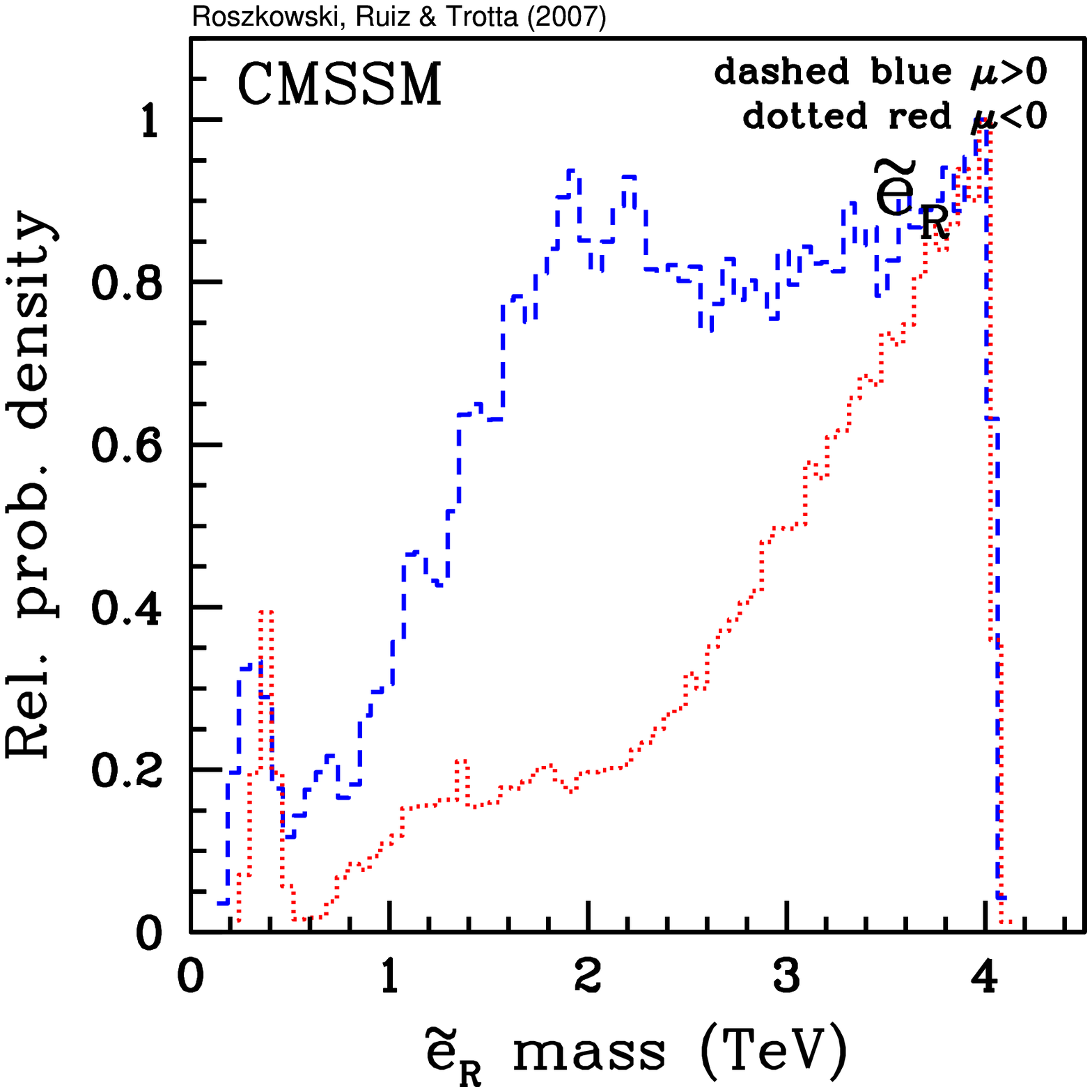}
&   \includegraphics[width=0.3\textwidth]{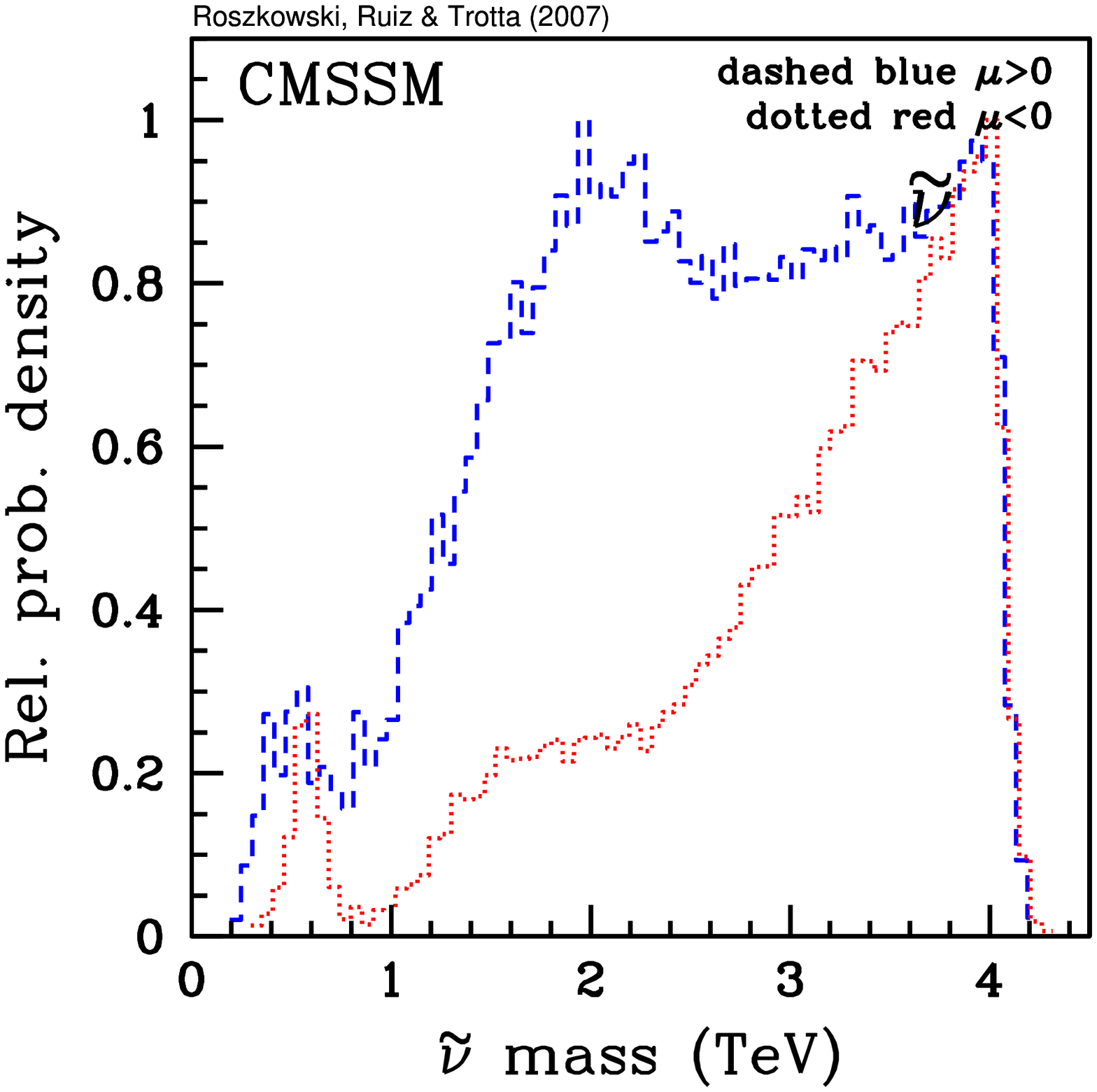}
&   \includegraphics[width=0.3\textwidth]{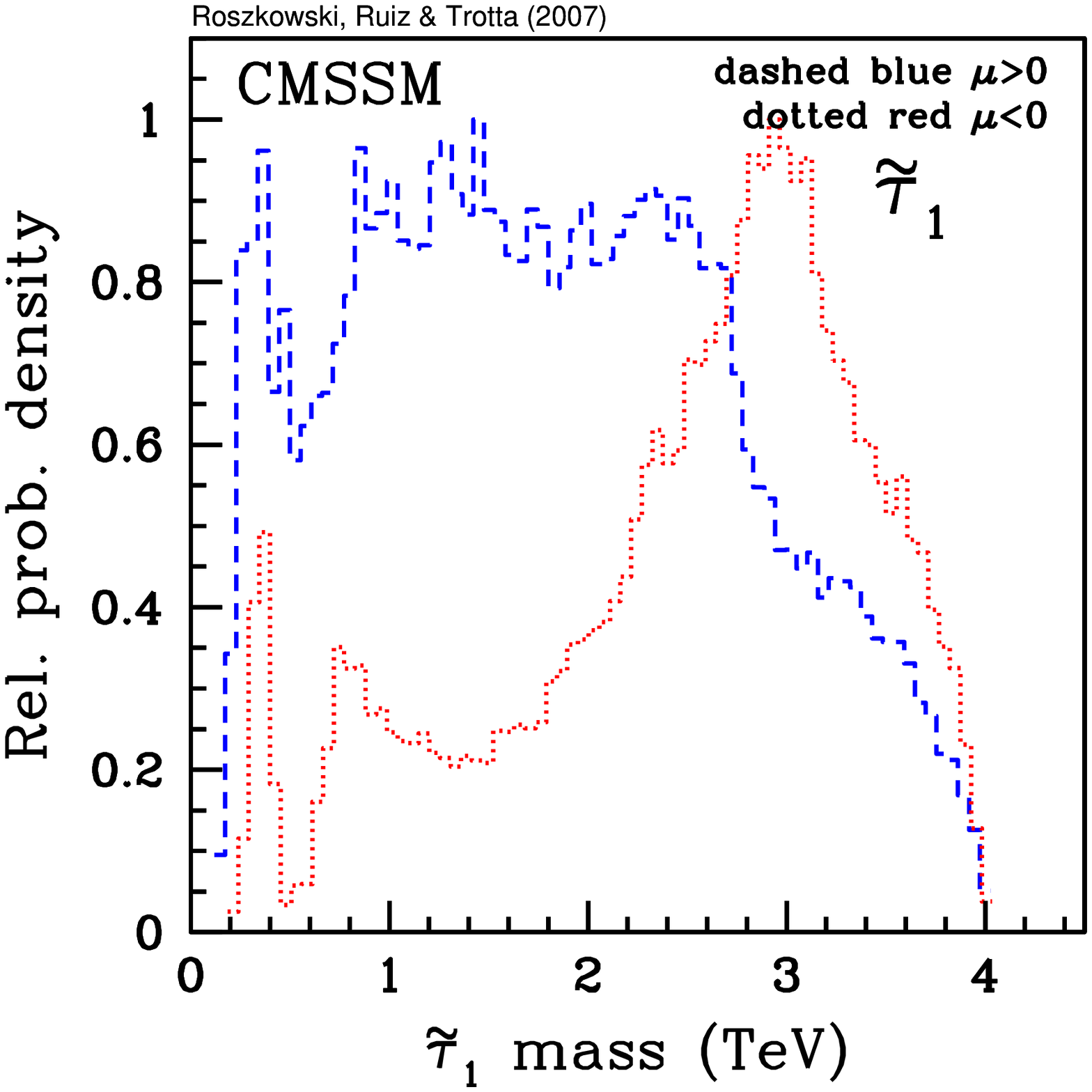}\\
    \includegraphics[width=0.3\textwidth]{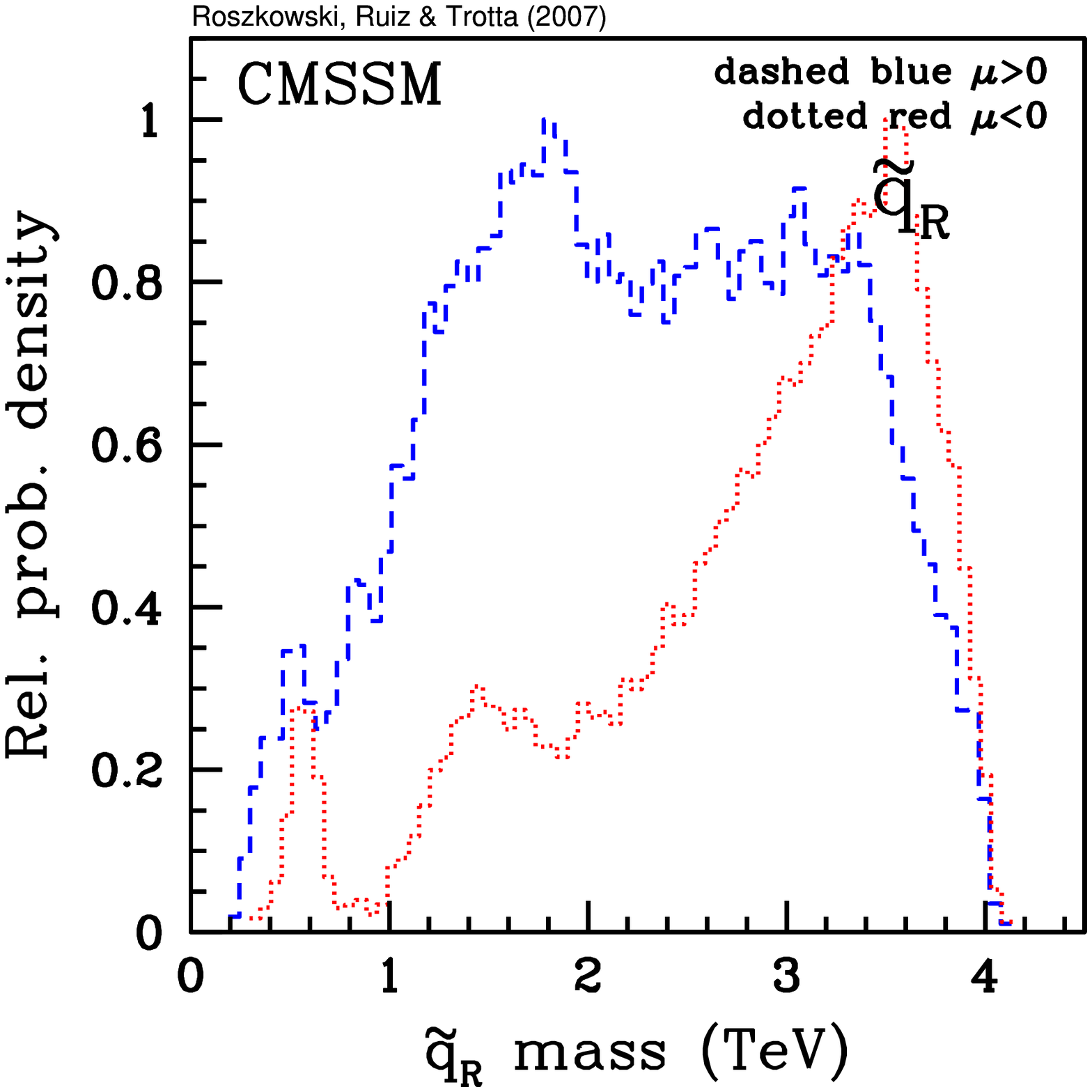}
&   \includegraphics[width=0.3\textwidth]{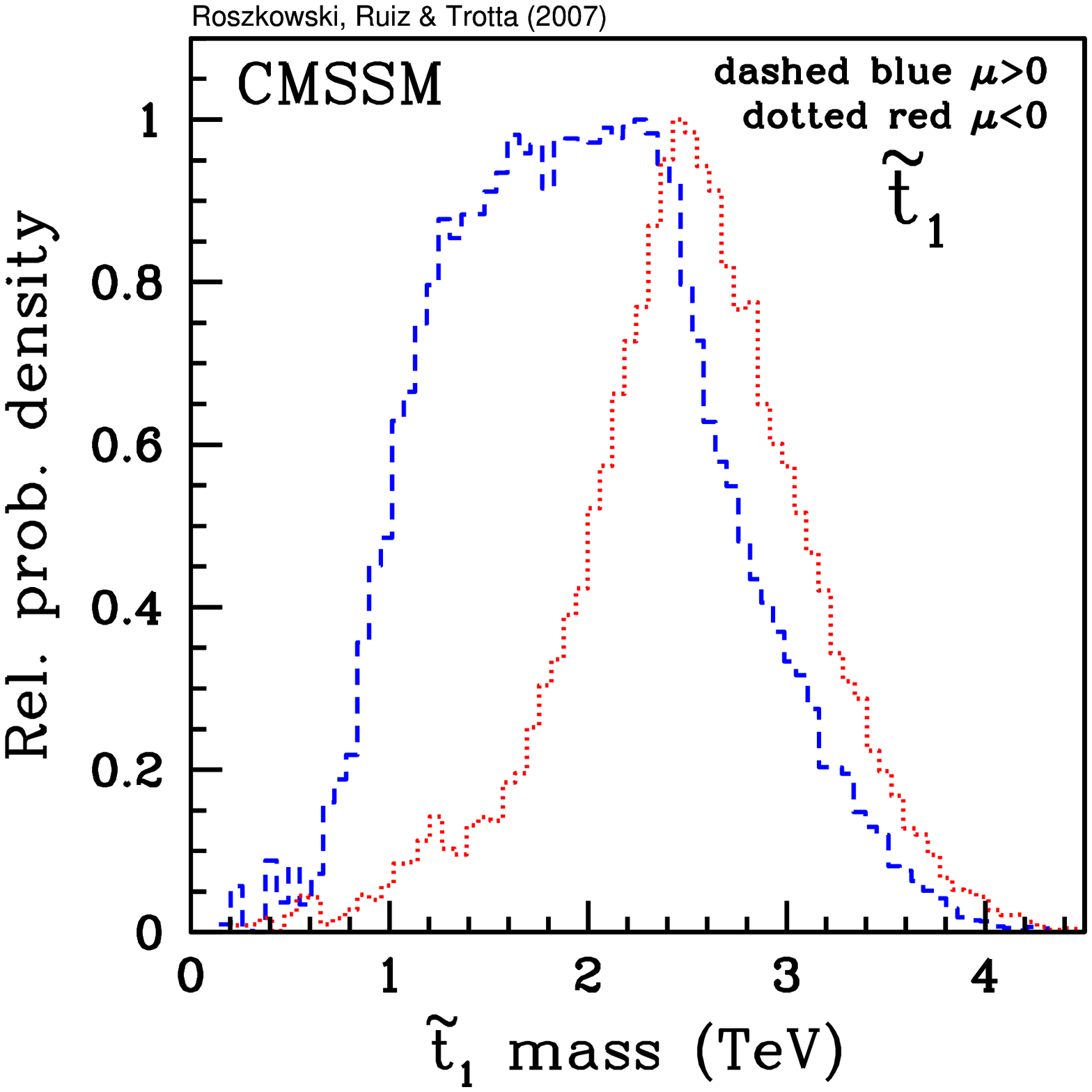}
&   \includegraphics[width=0.3\textwidth]{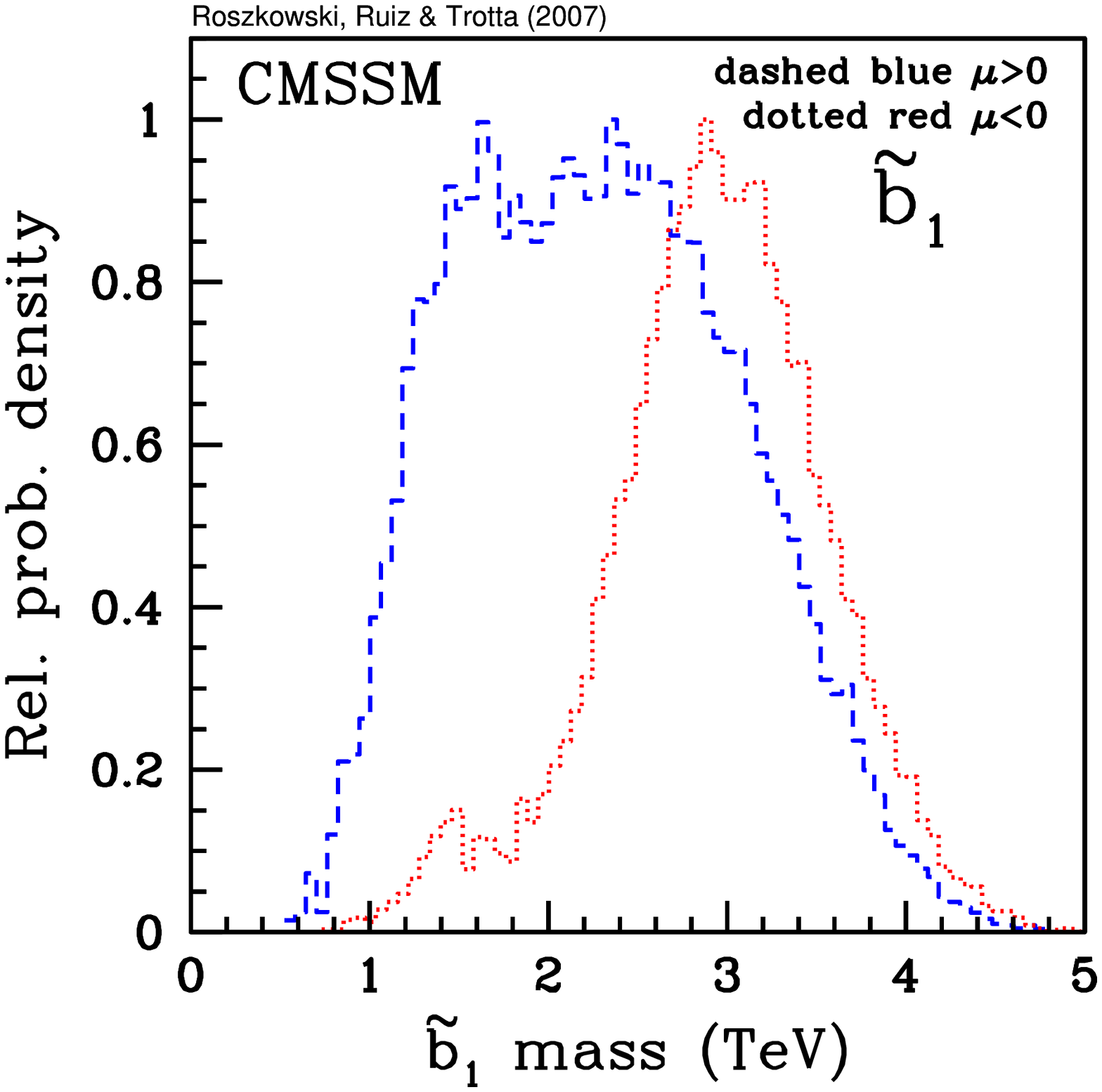}
\end{tabular}
\end{center}
\caption{As in fig.~\protect\ref{fig:cmssm1dpdf}, but for the masses of
several representative superpartners.
This figure should be compared with figure~5 (black solid lines) in
ref.~\protect\cite{rtr1}.
\label{fig:susymasses}
}
\end{figure}

Turning next to superpartners, we show in fig.~\ref{fig:susymasses}
the relative pdf's of the masses of several of them, while in
table~\ref{table:massesTable} we give the corresponding 68\%~\cl\ and
95\%~\cl\ ranges. The blue dashed curves are for $\mu>0$ and the red
dotted ones for $\mu<0$.  Firstly, for $\mu<0$ all the scalar
superpartners are considerably heavier than for the other sign of
$\mu$, as expected based on the discussion of most probable ranges of
$\mhalf$ and especially $\mzero$. In fact, if $\mu<0$ then all the
sleptons and squarks (whose masses, except for the 3rd generation, are
at least as large as $\mzero$) may be beyond the reach of the LHC. For
$\mu>0$ there is a good chance of seeing the gluino (assuming the LHC
reach of some $2.7-3\tev$) and a reasonable chance of seeing some
squarks and sleptons.  Unfortunately, these prospects are considerably
less optimistic than what we found in fig.~5 and table~6 of
ref.~\cite{rtr1}, where the previous SM value of $\brbsgamma$ was
used. A dedicated analysis would be required to derive more detailed
conclusions.

\begin{table}
\centering
\begin{tabular}{|c |c c| c c|}
 \hline
Particle & \multicolumn{2}{|c|}{$\mu<0$} & \multicolumn{2}{|c|}{$\mu>0$}\\
(TeV) & 68\% & 95\% & 68\% & 95\% \\
           \hline
 $\hl^0$ &$(0.1180, 0.1211)$ &$(0.1151,0.1223)$ &$(0.1154,0.1204)$ &$(0.1125,0.1219)$\\
 $\hh^0,\ha^0,\hpm$ & $(1.2,3.1)$  &$(0.91, 3.8)$ &$(0.36,2.5)$   & $(0.21,3.6)$ \\ 
\hline
 $\chi_1^0$     & $(0.23, 0.67)$  &$(0.11, 0.82)$  &$(0.16, 0.49)$ &$(0.06,0.69)$   \\
 $\charone$     & $(0.3, 1.2)$  &$(0.15, 1.4)$   &$(0.25, 0.76)$  &$(0.11, 1.2)$   \\
 $\gluino$      & $(1.4, 3.4)$  &$(0.77, 4.0)$  &$(1.0, 2.6)$  &$(0.41,3.5)$   \\
 $\tilde{e}_R$  & $(1.8, 3.8)$  &$(0.37, 4.0)$  &$(1.5, 3.6)$  &$(0.5, 4.0)$ \\
 $\tilde{\nu}$  & $(1.9, 3.8)$  &$(0.58, 4.0)$  &$(1.6, 3.6)$  &$(0.65,4.0)$   \\
$\tilde{\tau}_1$& $(1.4, 3.3)$  &$(0.34, 3.8)$  &$(0.80, 2.8)$  &$(0.28,3.7)$  \\
 $\tilde{q}_R$  & $(2.9, 4.3)$  &$(1.6, 4.9)$  &$(1.9, 4.0)$  &$(1.3,4.7)$  \\
 $\tilde{t}_1$  & $(1.9, 3.1)$  &$(1.1, 3.6)$   &$(1.3, 2.6)$  &$(0.86,3.3)$   \\
 $\tilde{b}_1$  & $(2.3, 3.5)$  &$(1.4, 4.1)$   &$(1.4, 3.1)$  &$(1.0,3.8)$  \\
\hline
 \end{tabular}
\caption{Higgs boson and selected superpartner mass ranges (in
TeV) containing 68\% and 95\% of posterior probability (with all
other parameters marginalized over) for both signs of $\mu$. Masses
above $1\tev$ have been rounded up to 1 significant digit.
}
\label{table:massesTable}
\end{table}
%

\subsection{Implications for direct detection of Dark
  Matter}\label{sec:dmsearch}


\begin{figure}[!tb]
\begin{center}
\begin{tabular}{c c}
    \includegraphics[width=0.4\textwidth]{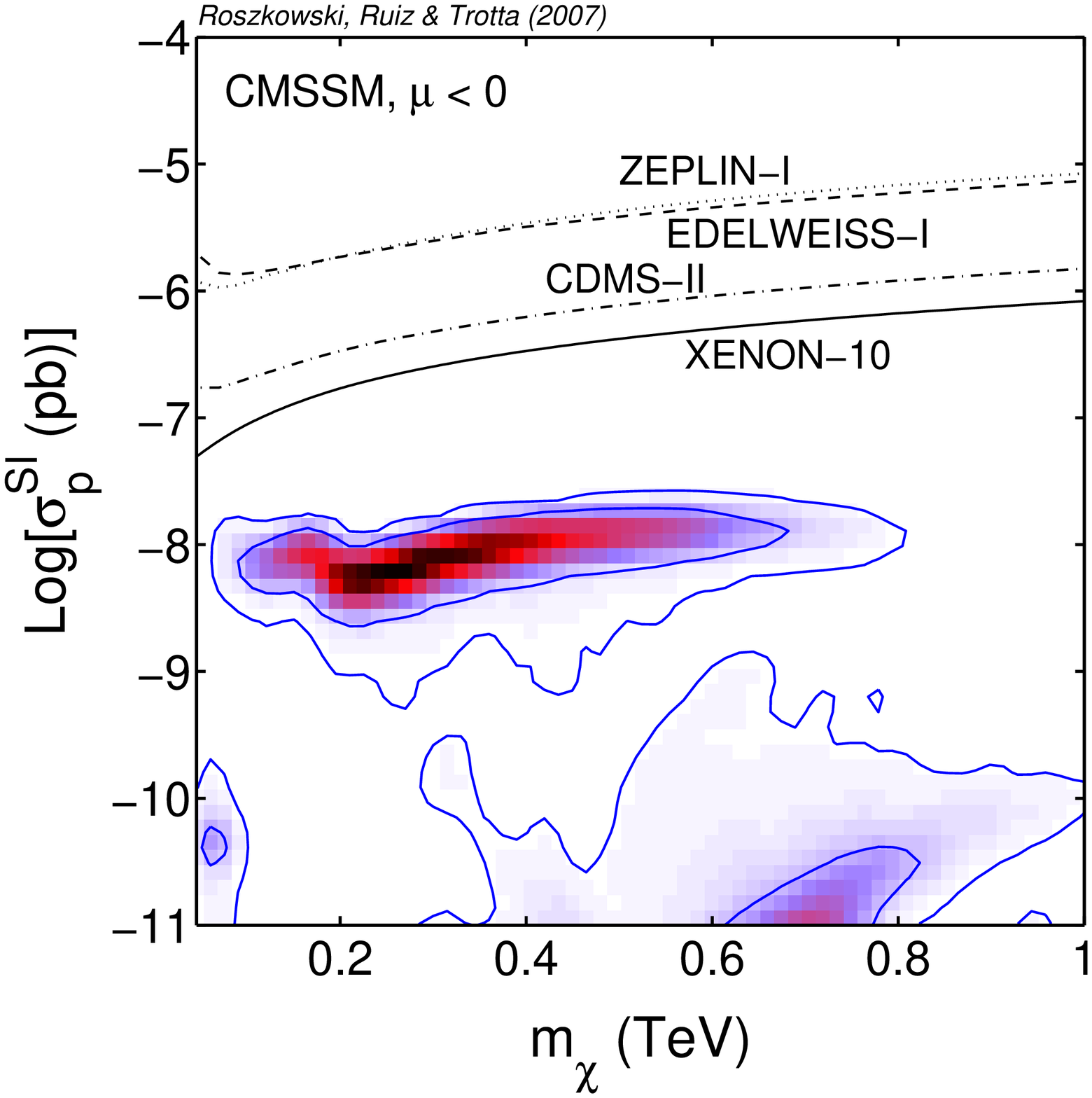}
&   \includegraphics[width=0.4\textwidth]{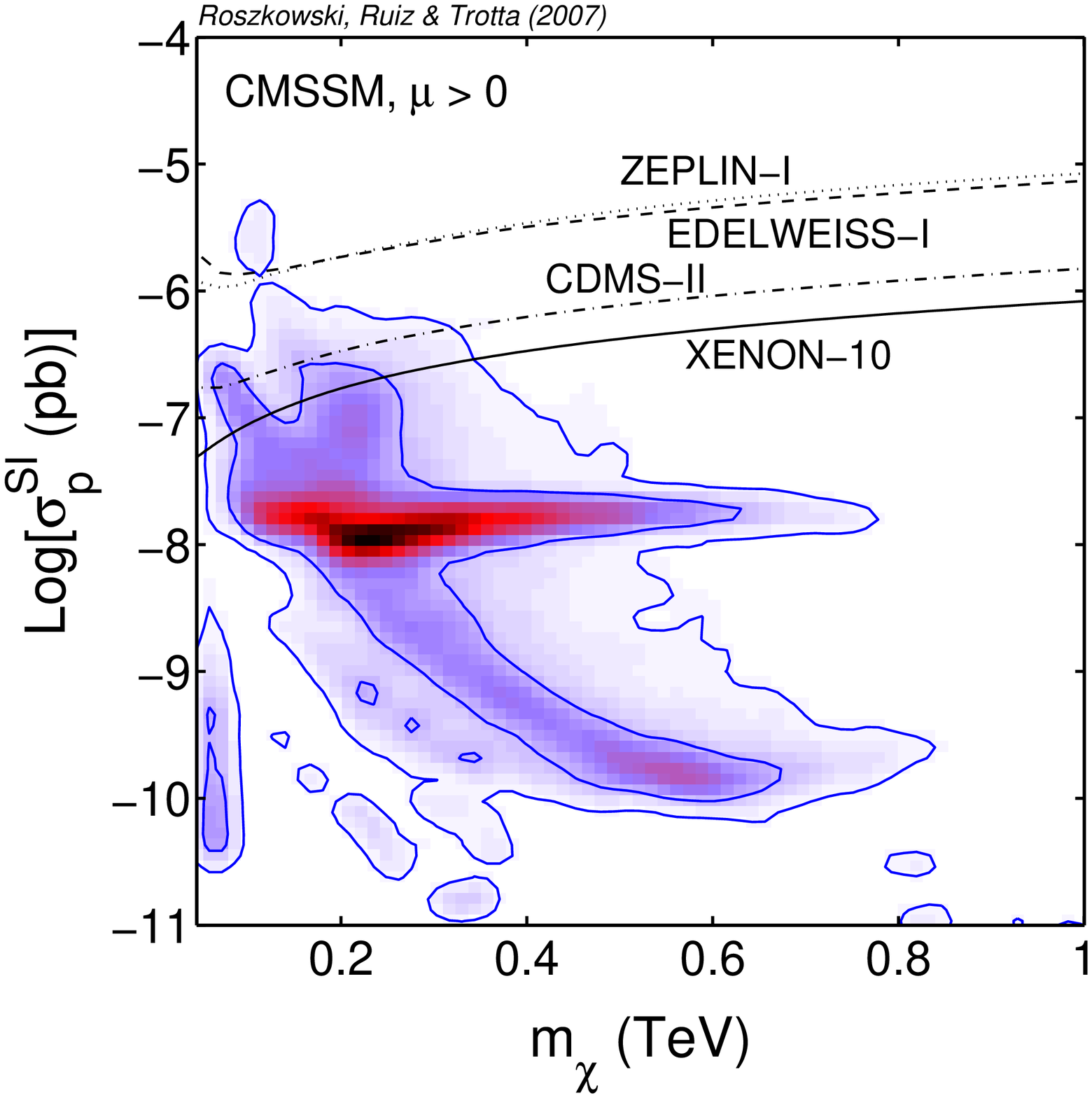}
\end{tabular}
  \includegraphics[width=0.3\textwidth]{rrt3-colorbar.ps}
\end{center}
\caption{The 2-dim relative probability density for $\sigsip$ vs. the
  neutralino mass $\mchi$ for $\mu<0$ (left panel) and $\mu>0$ (right
  panel). The inner (outer) solid contours delimit the regions of 68\%
  and 95\% total probability, respectively. Some current experimental
  upper bounds are also shown.
The right panel should be compared with figure~13 (top) in
ref.~\protect\cite{rtr1}.
\label{fig:sigsipvsmchi-pdf}
}
\end{figure}
\begin{figure}[!tb]
\begin{center}
\begin{tabular}{c c}
    \includegraphics[width=0.4\textwidth]{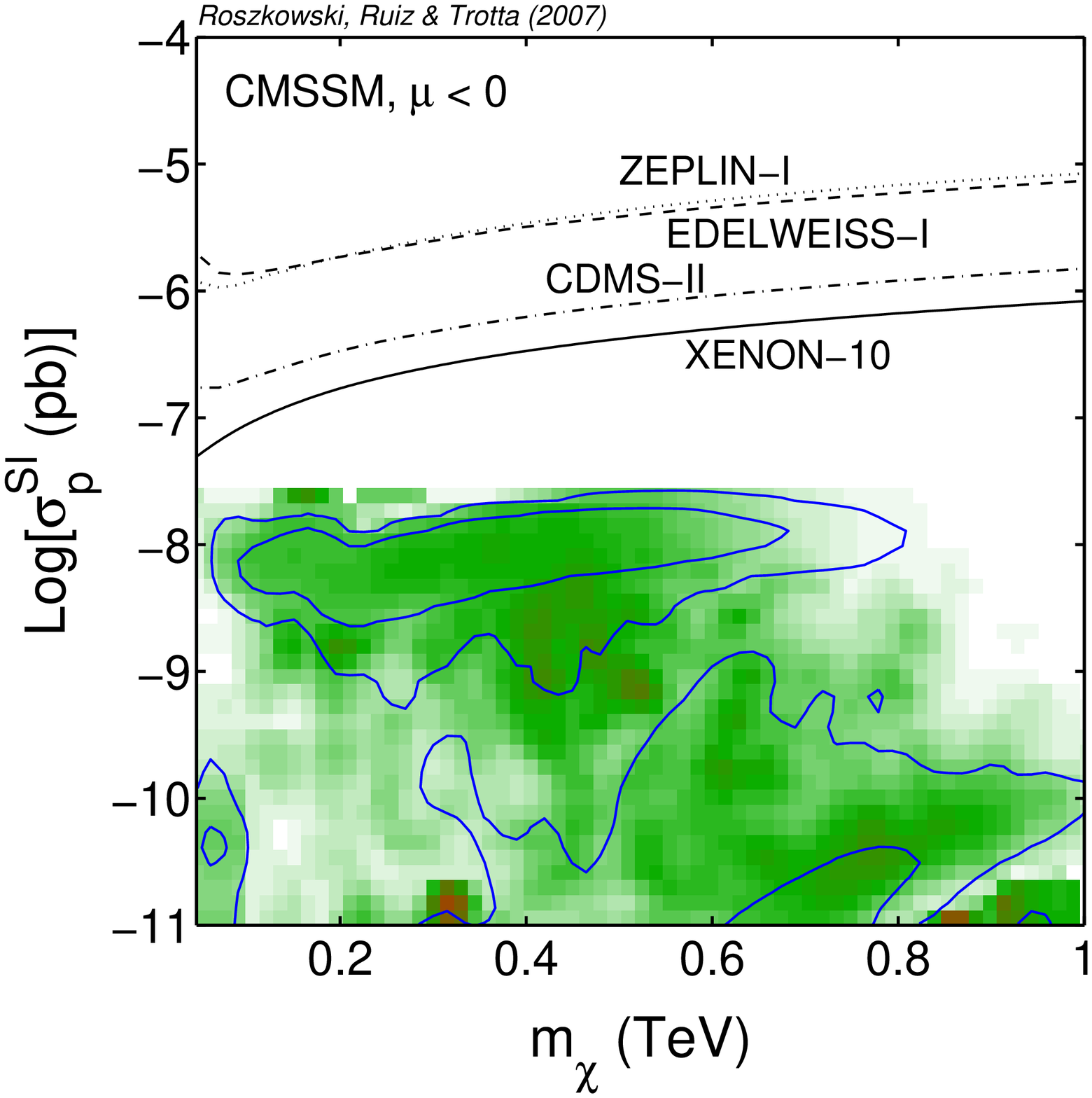}
&   \includegraphics[width=0.4\textwidth]{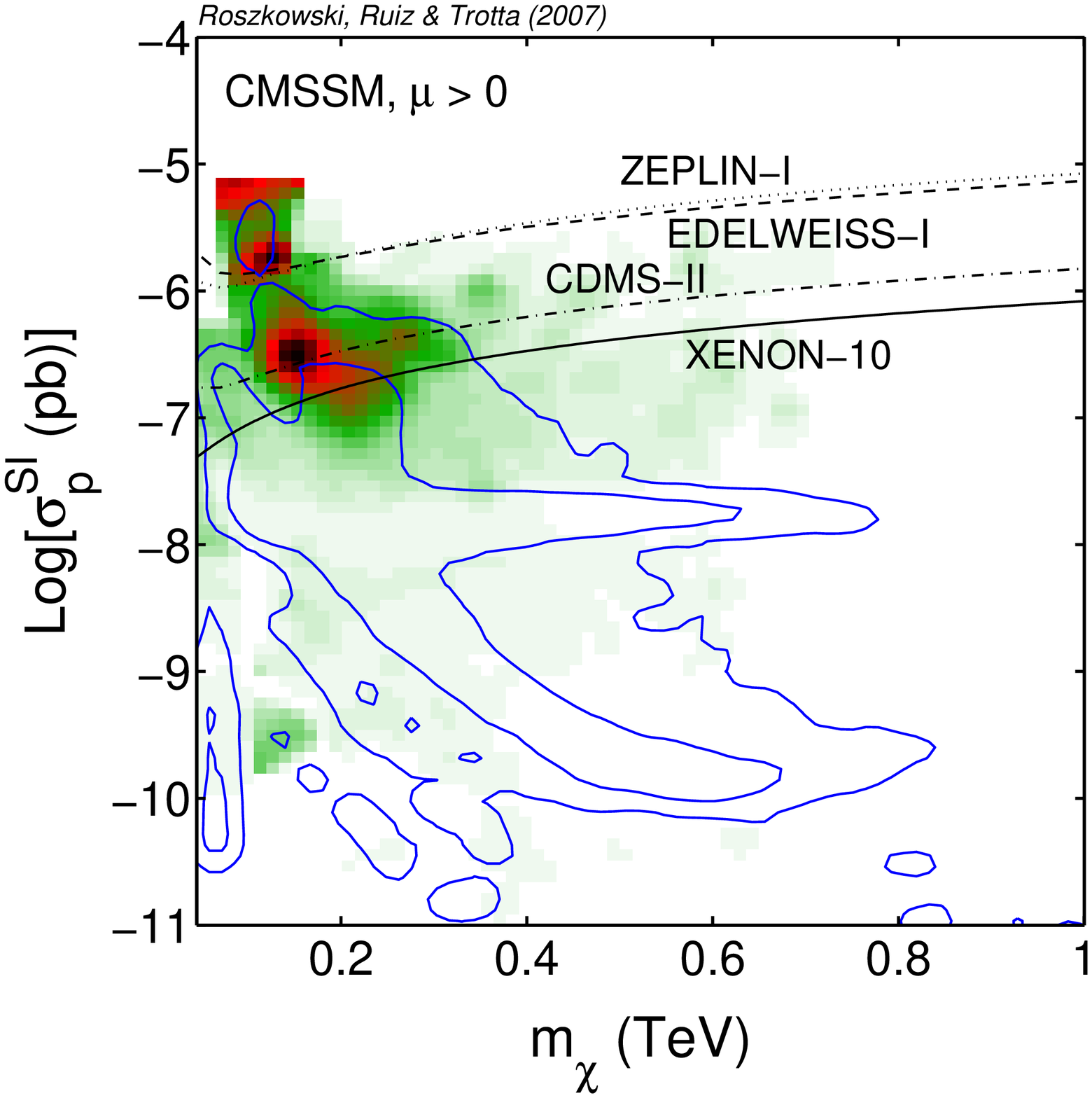}
\end{tabular}
  \includegraphics[width=0.3\textwidth]{rrt3-colorbar-like.ps}
\end{center}
\caption{The mean quality-of-fit for $\sigsip$ vs. the neutralino mass
  $\mchi$ for $\mu<0$ (left panel) and $\mu>0$ (right panel). The
  inner (outer) solid contours delimit the regions of posterior 68\%
  and 95\% total probability, respectively (compare
  fig.~\protect\ref{fig:sigsipvsmchi-pdf}). Some current experimental
  upper bounds are also shown, but have not been included as
  constraints in the likelihood function.
The right panel should be compared with figure~13 (bottom) in
ref.~\protect\cite{rtr1}.
\label{fig:sigsipvsmchi-qof}
}
\end{figure}

We will now examine implications for direct detection of the lightest
neutralino assumed to be the DM in the Universe, via its elastic
scatterings with targets in underground detectors~\cite{susy-dm-reviews}. We
will consider both spin-independent (SI) and spin-dependent (SD)
interactions. The underlying formalism for both types of interactions
can be found in several sources. (See,
\eg,~\cite{dn93scatt:ref,susy-dm-reviews,efo00,knrr1}.) In this analysis we use
the expressions and inputs as presented in ref.~\cite{knrr1}. We only
note here that the SI interactions cross section $\sigsip$ of a WIMP
scattering off a proton in a target nucleus is the same as that of a
neutron and that the total SI interactions cross section of the
nucleus is proportional to $\sigsip$ times the square of the mass
number. In contrast, for the SD interactions, the cross section for a
WIMP scattering off a proton, $\sigsdp$, does not necessarily have to
be the same as the one from a neutron~\cite{lewin+smith96,tggrr00}.

In fig.~\ref{fig:sigsipvsmchi-pdf} we show the Bayesian posterior
relative probability distribution in the usual plane of $\sigsip$
and the DM neutralino mass $\mchi$ for $\mu<0$ (left panel) and
$\mu>0$ (right panel). Starting with $\mu>0$, we can see a big
concentration of probability density at rather
high values of $\sigsip\sim10^{-8}\pb$, characteristic of the FP
region of large $\mzero$~\cite{focuspoint-fmw}, which is favored
by the current theoretical evaluation of $\brbsgamma$, as we have
seen above. In the FP region, the neutralino, while remaining
predominantly bino-like, receives a sizable higgsino component,
which strengthens the dominant Higgs-exchange contribution to
$\sigsip$. In addition, there is a more well-known branch, with
$\sigsip$ decreasing with $\mchi$, which comes from the (now
somewhat disfavored) region of $\mhalf,\mzero\lsim1\tev$ where the
relic abundance $\abundchi$ of the neutralinos is reduced to agree
with WMAP and other determinations by a pseudoscalar Higgs
resonance in their pair annihilation and/or by their
coannihilations with sleptons.  In order to appreciate the change
in the CMSSM predictions for $\sigsip$, the right panel should be
compared with the top panel of fig.~13 in ref.~\cite{rtr1} where a
previous value of the SM prediction for $\brbsgamma$ was used.

The left panel of fig.~\ref{fig:sigsipvsmchi-pdf} (the case of $\mu<0$) also
shows a high preference for $\sigsip\sim10^{-8}\pb$, which corresponds
to the FP region at multi-TeV $\mzero$, as for the other sign of
$\mu$. In addition, we find another rather large 68\%
total probability region at extremely low values of below
$10^{-10}\pb$, which corresponds to the higher probability region of $\mhalf\sim1.5\tev$
in the $(\mhalf,\mzero)$ plane. Such tiny ranges of $\sigsip$ are a
result of cancellation between the Higgs-exchange contribution to up-
and down-type quarks~\cite{dn93scatt:ref,efo00}.

In both panels of fig.~\ref{fig:sigsipvsmchi-pdf} we have marked some
of the current direct experimental upper
limits~\cite{cdms-sep05,edelweiss-one-final,zeplin-one-final},
assuming a default value of $0.3\gev/\cmeter^3$ for the local DM
density. 
It is encouraging that experiments, notably XENON-10 with its very new
limit, are already probing some portions of the CMSSM PS for
$\mu>0$. CDMS-II is currently taking data and is expecting to improve
its limit to a similar level of sensitivity. Clearly, a further
improvement by about an order of magnitude will constitute a 
critical leap as it will hopefully allow one to reach down to the
heart of the SI interactions cross sections favored in the
CMSSM.
Future one-tonne detectors are expected to reach down to
$\sigsip \gsim 10^{-10}\pb$, thus probing most of the favored
parameter space of the CMSSM, at least for the more favored case of
$\mu>0$. 
We note that the probability of $\sigsip>10^{-10}\pb$ in fig.~\ref{fig:sigsipvsmchi-pdf}
is $98.4\%$ for $\mu>0$ and $62.5\%$ for $\mu<0$.

\begin{figure}[!tbh]
\begin{center}
\begin{tabular}{c c}
    \includegraphics[width=0.4\textwidth]{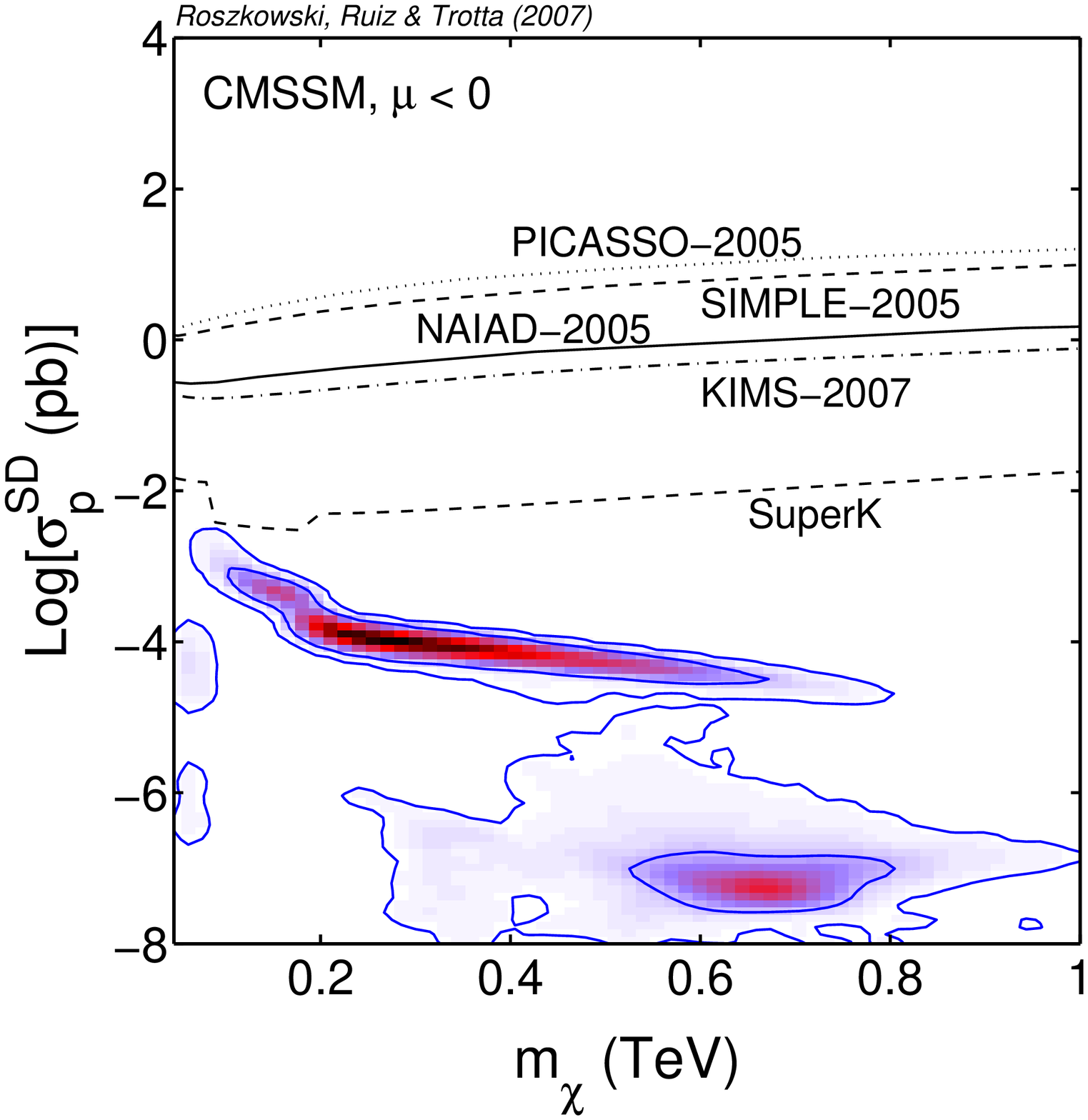}
&   \includegraphics[width=0.4\textwidth]{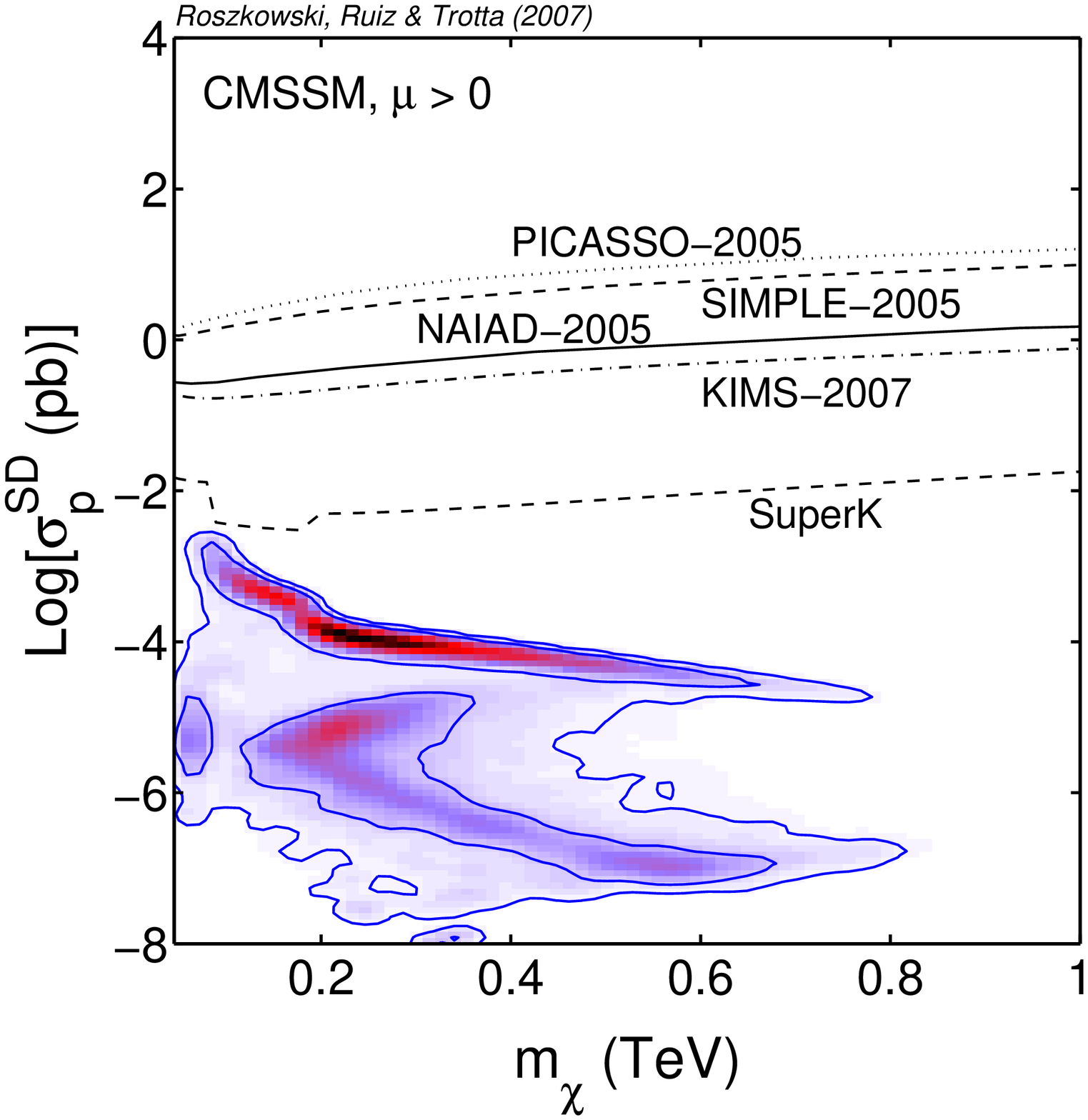}
\end{tabular}
  \includegraphics[width=0.3\textwidth]{rrt3-colorbar.ps}
\end{center}
\caption{The 2-dim relative probability density of $\sigsdp$ vs. the
  neutralino mass $\mchi$ for $\mu<0$ (left panel) and $\mu>0$ (right
  panel). The inner (outer) solid contours delimit the regions of 68\%
  and 95\% total probability, respectively. Some current experimental
  90\%~\cl\ upper bounds are also shown.  Analogous plots for $\sigsdn$
  are basically identical but the experimental limits from direct
  detection are weaker by nearly two orders of magnitude.
\label{fig:sigsdpvsmchi-pdf}
}
\end{figure}

It is worth re-emphasizing that, in addition to light Higgs searches
at the Tevatron and the LHC, direct detection DM searches will
provide an alternative experimental probe of the FP region at multi-TeV
$\mzero$, in which the squarks will be too heavy to be produced at the
LHC, although the gluino mass should remain mostly (partly) accessible
for $\mu>0$ ($\mu<0$).

For comparison, fig.~\ref{fig:sigsipvsmchi-qof} shows the ranges
favored by the alternative measure of the mean quality-of-fit.
Starting with the case of $\mu>0$ (the right panel), we can see a
handful of small regions of a rather large $\sigsip$, above a few
times $10^{-7}\pb$, 
which are already in conflict with the
current limit from the XENON-10. These cases corresponds to the
``islands'' of good fit to the data that we have already seen in
fig.~\ref{fig:cmssm2dcontoursmup-like}, where they are all
concentrated in the region of $\mhalf\lsim 0.5\tev$ and
$\mzero\lsim1.5\tev$. 
With a modest improvement of sensitivity,
DM search experiments will be able to probe the entire region
favored by the mean quality-of-fit in the case of $\mu>0$. (Notice
that neither the posterior pdf nor the mean quality-of-fit give
much preference for very small $\sigsip$.)  In contrast, for
$\mu<0$ (the left panel of fig.~\ref{fig:sigsipvsmchi-qof}) there
is hardly any region giving good quality fits. This is a
reflection of what we have already seen in the $(\mhalf,\mzero)$
plane in fig.~\ref{fig:cmssm2dcontoursmun-like}. 

Turning next to SD interactions, in fig.~\ref{fig:sigsdpvsmchi-pdf} we
present the relative probability density for neutralino-proton
scattering cross section $\sigsdp$ versus $\mchi$ for $\mu<0$ (left
panel) and $\mu>0$ (right panel).  In the FP region the increased
higgsino component of the neutralino in this case leads to a larger
coupling to the dominant $Z$-boson exchange. This is reflected in the
figure where the highest probability regions are, for both signs of
$\mu$, concentrated around $\sigsdp\sim 10^{-4}\pb$. For $\mu>0$ there
is an additional higher probability region which is visible in the
right panel, and which corresponds to the Higgs resonance and
coannihilation region mentioned above.  In the case of $\mu<0$
instead, an additional higher probability region is visible in the
left panel at $\sigsdp\lsim 10^{-7}\pb$ and $\mchi\sim0.7\tev$. It
corresponds to the region of $\mhalf\sim1.5\tev$ in the
$(\mhalf,\mzero)$ plane (fig.~\ref{fig:cmssm2dcontoursmun}).

The current experimental upper
limits~\cite{picasso-sd-05,simple-sd-05,naiad-sdlimit-05,kims-sdlimit-07} 
from direct searches,
assuming a default value of $0.3\gev/\cmeter^3$ for the local DM
density, as well as an indirect limit from neutralino annihilations
into neutrinos in the Sun, the Earth or the Galactic
center~\cite{superk-sdlimit-04}, 
which have also been shown in fig.~\ref{fig:sigsdpvsmchi-pdf}, still
remain a few order of magnitude above the predictions of the CMSSM. On
the other hand, experimental sensitivity has undergone steady progress
also in the case of SD interactions. Eventually, it will be important
to reach down below the level of $\sigsdp\lsim 10^{-4}\pb$, which would allow one
for an independent cross-check of CMSSM predictions for dark
matter. 

In table~\ref{table:DDranges} we have listed the ranges of both
$\sigsip$ and $\sigsdp$ containing 68\% and 95\% of posterior
probability (with all other parameters marginalized over) for both
signs of $\mu$. The ranges cover the whole allowed range of $\mchi$
and provide supplementary information to what one can read out of
figs.~\ref{fig:sigsipvsmchi-pdf} and~\ref{fig:sigsdpvsmchi-pdf}.

\begin{table}
 \centering
\begin{tabular}{|c |c | c|}
 \hline
 & \multicolumn{2}{|c|}{Spin-independent cross section $\sigsip$ ($\pb$)} \\  
 & 68\% & 95\% \\ \hline
 $\mu<0$ & $(2.9\times10^{-12}, 8.1\times10^{-9})$ &
 $(1.6\times10^{-13}, 1.4\times10^{-8})$ \\ \hline
 $\mu>0$ & $(2.8 \times10^{-10}, 3.9 \times10^{-8})$ &
 $(7.2\times10^{-11}, 2.5\times10^{-7})$\\
\hline \hline
 & \multicolumn{2}{|c|}{Spin-dependent cross section $\sigsdp$ ($\pb$)} \\  
 & 68\% & 95\% \\ \hline
 $\mu<0$ & $(5.3\times10^{-8}, 9.6\times10^{-5})$  &
$(1.7\times10^{-8}, 4.4\times10^{-4})$ \\ \hline
 $\mu>0$ & $(2.0\times10^{-7}, 8.9\times10^{-5})$   &
 $(4.2\times10^{-8}, 4.9\times10^{-4})$ \\  
\hline
  \end{tabular}
 \caption{Direct detection of dark matter: ranges of spin-independent
 and spin-dependent cross section corresponding to 68\% and 95\% of
 posterior probability (with all other parameters marginalized over)
 for both signs of $\mu$.  } \label{table:DDranges}
 \end{table}

As mentioned above, in the SD interaction case the cross section of
WIMP scattering from a neutron can in general be very different from
the one from a proton. However, in our scan of the CMSSM, we have
found its relatively probability distribution versus $\mchi$ to be
very close to the case with the proton, and thus do not show it
here. As regards the mean quality-of-fit (not shown for $\sigsdp$), at
$\mu>0$ the best-fit regions in fig.~\ref{fig:cmssm2dcontoursmup-like}
give a rather small value of $\sim 10^{-5}\pb$, while at $\mu<0$ no
range of $\sigsdp$ is favored by the mean quality-of-fit criterion.

\begin{figure}[!tbh]
\begin{center}
\begin{tabular}{c c}
    \includegraphics[width=0.4\textwidth]{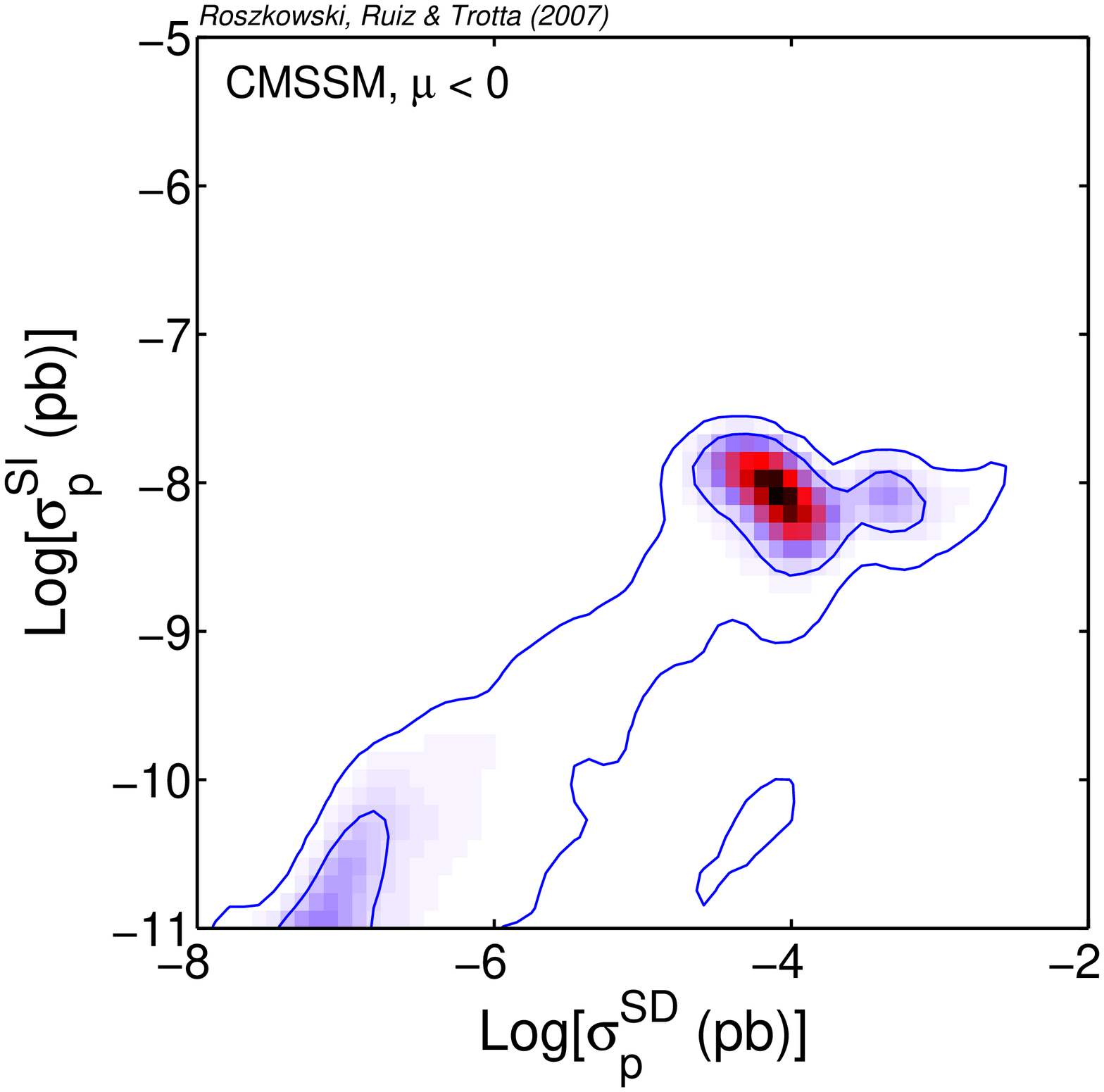}
&   \includegraphics[width=0.4\textwidth]{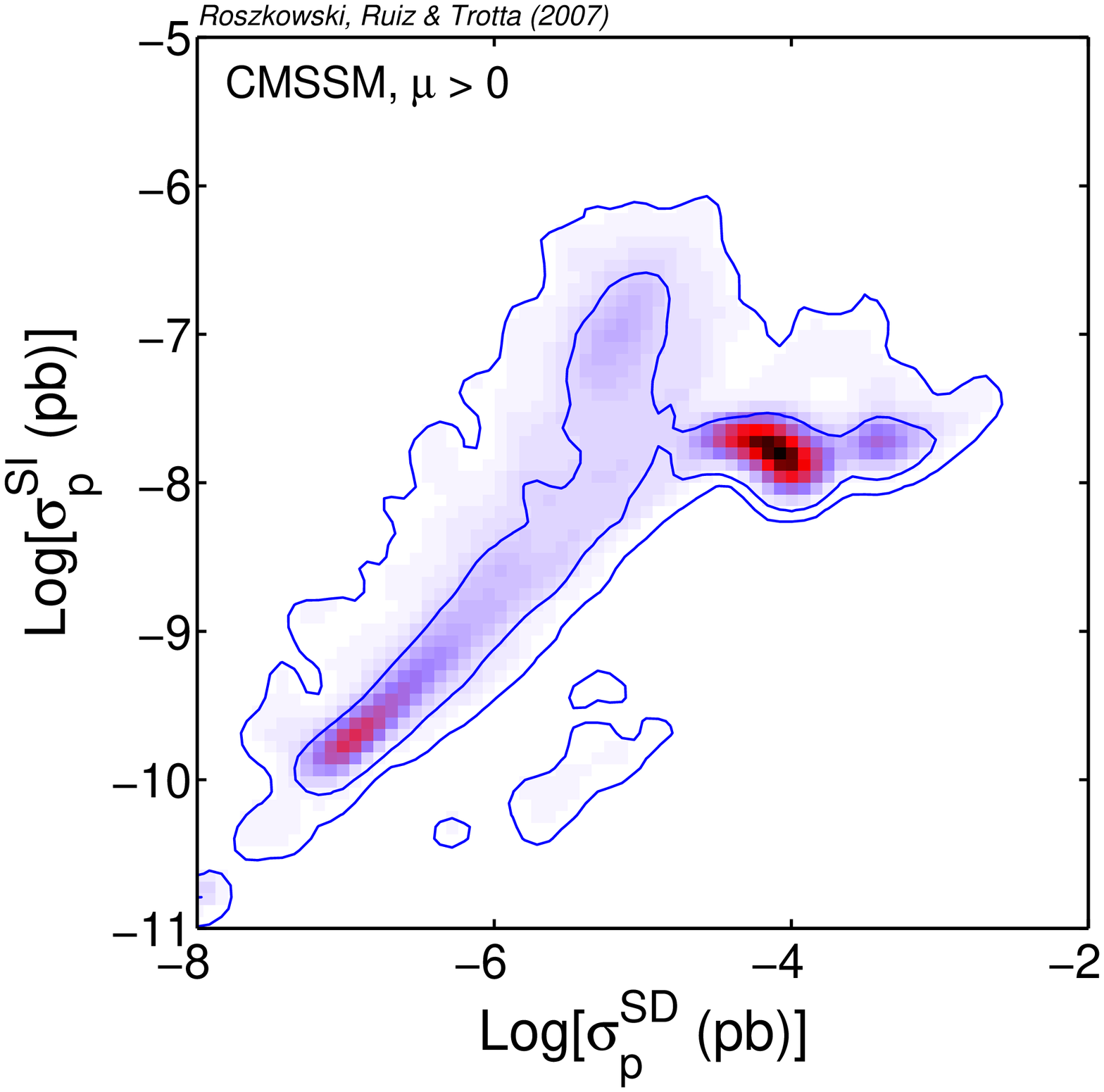}
\end{tabular}
  \includegraphics[width=0.3\textwidth]{rrt3-colorbar.ps}
\end{center}
\caption{The 2-dim relative probability density of $\sigsip$
  vs. $\sigsip$ for $\mu<0$ (left panel) and $\mu>0$ (right
  panel). The inner (outer) solid contours delimit the regions of 68\%
  and 95\% total probability, respectively. All other parameters have
  been marginalized over.
\label{fig:sigsipvssigsdp-pdf}
}
\end{figure}
\begin{figure}[tbh!]
\begin{center}
\begin{tabular}{c c c}
  \includegraphics[width=0.3\textwidth]{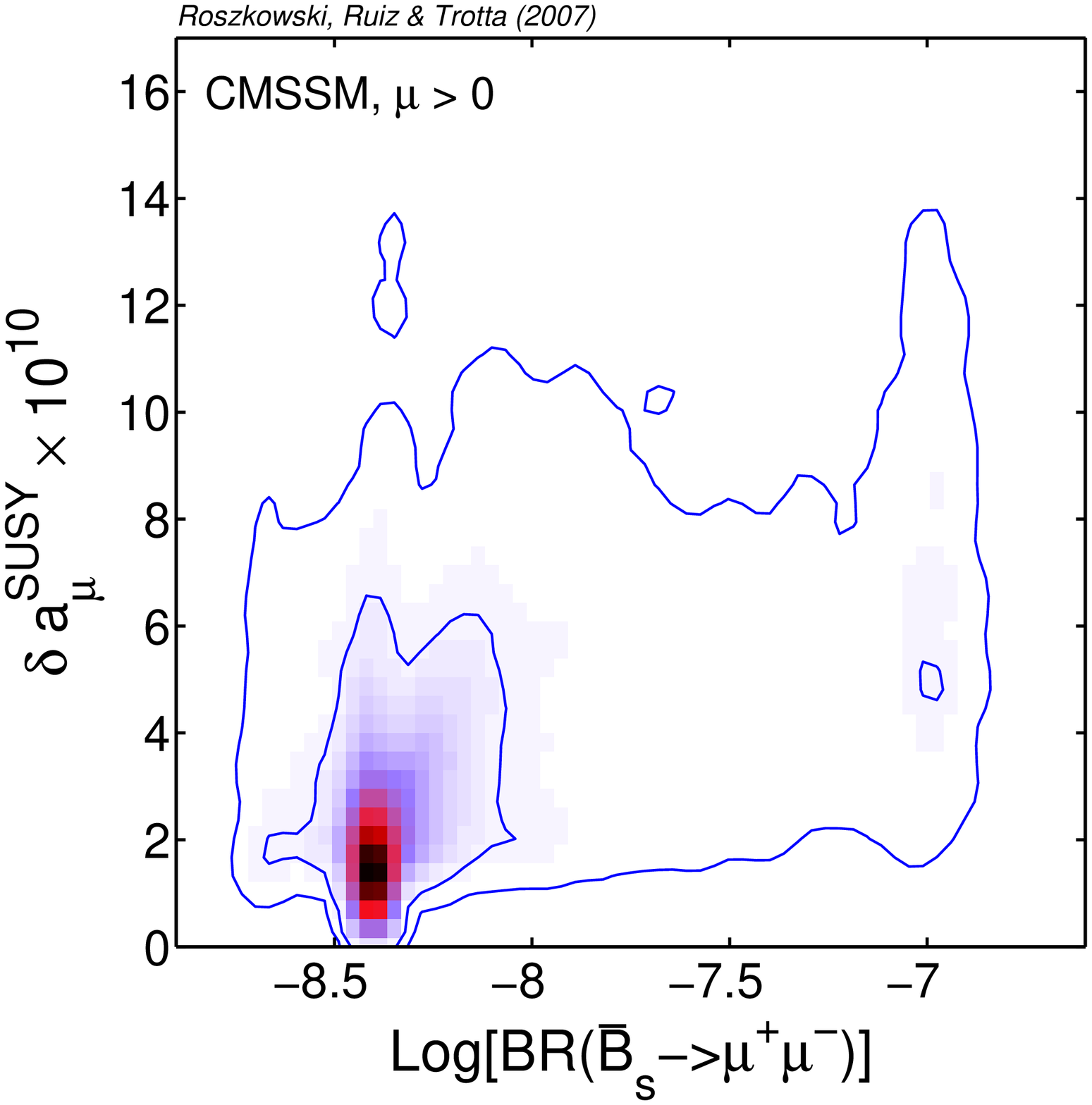}
& \includegraphics[width=0.3\textwidth]{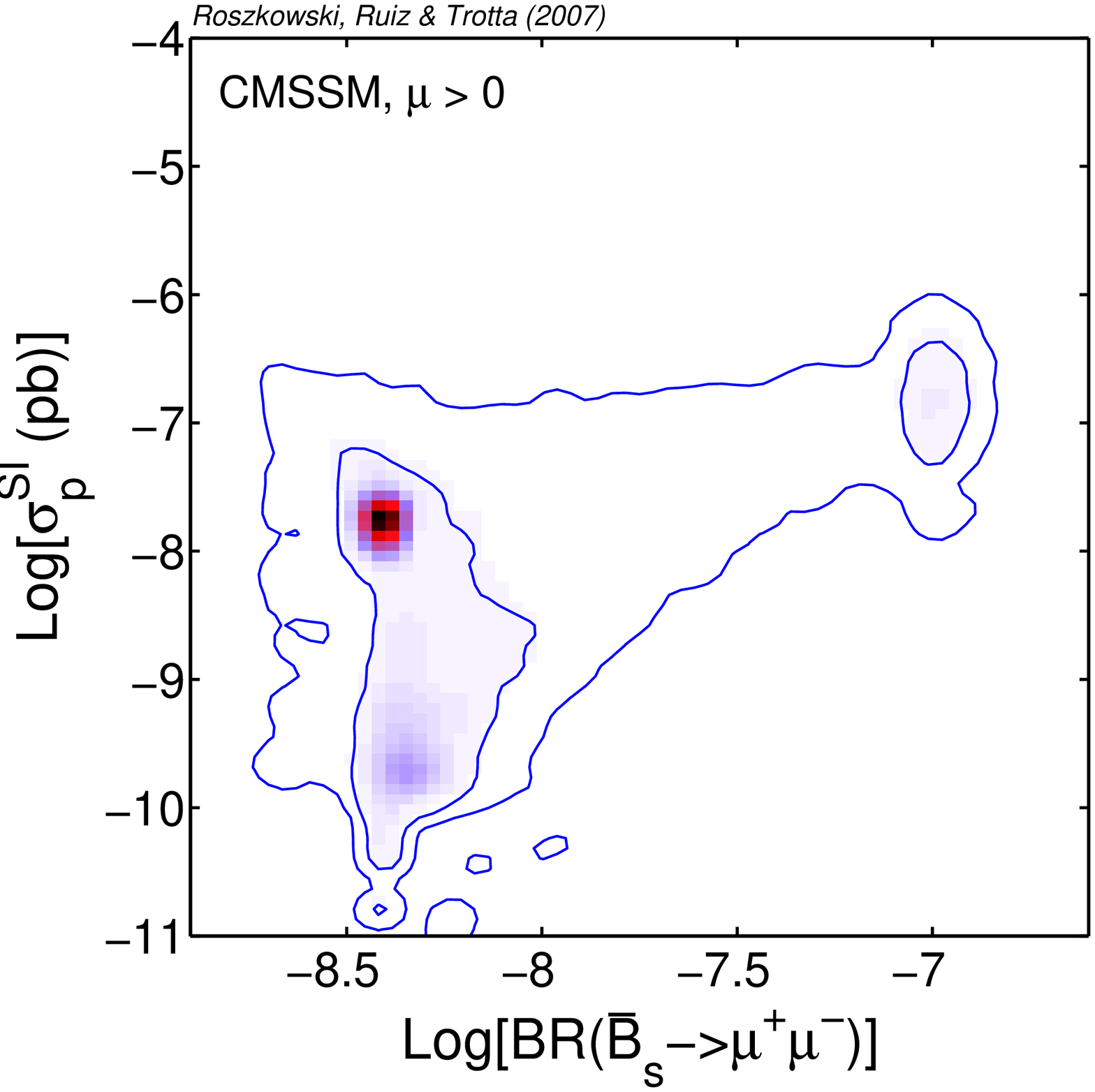}
& \includegraphics[width=0.3\textwidth]{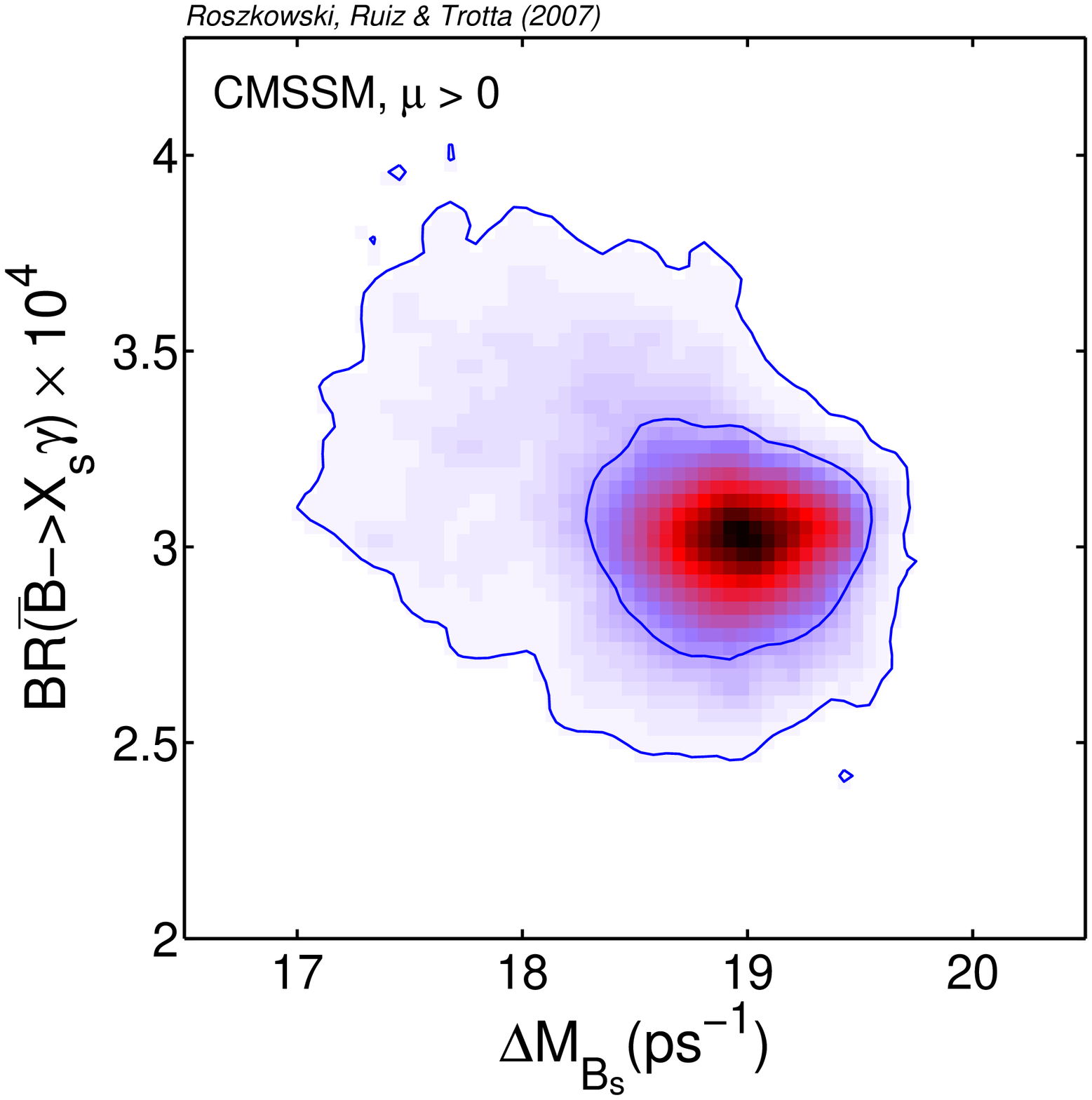}\\
  \includegraphics[width=0.3\textwidth]{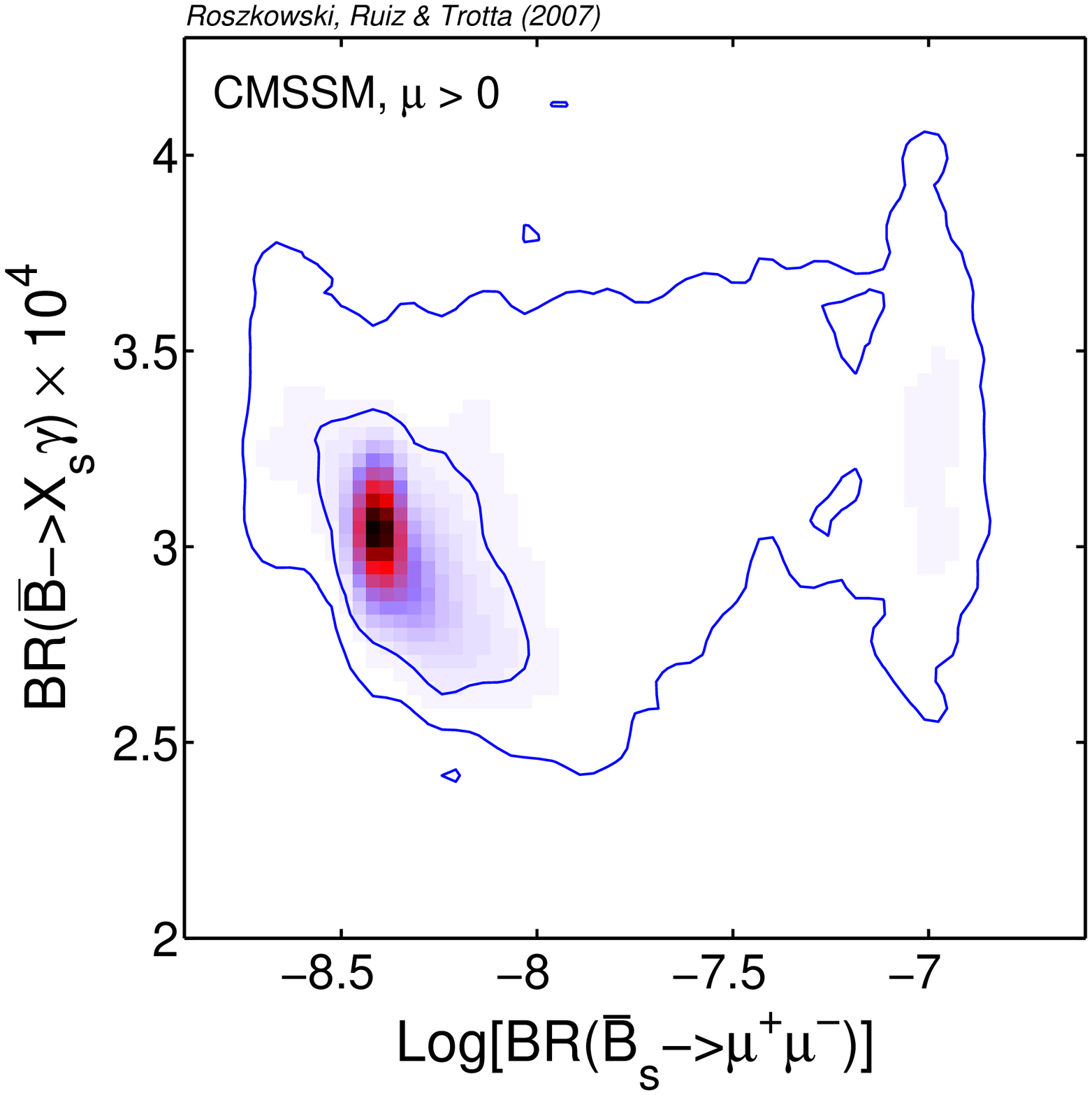}
& \includegraphics[width=0.3\textwidth]{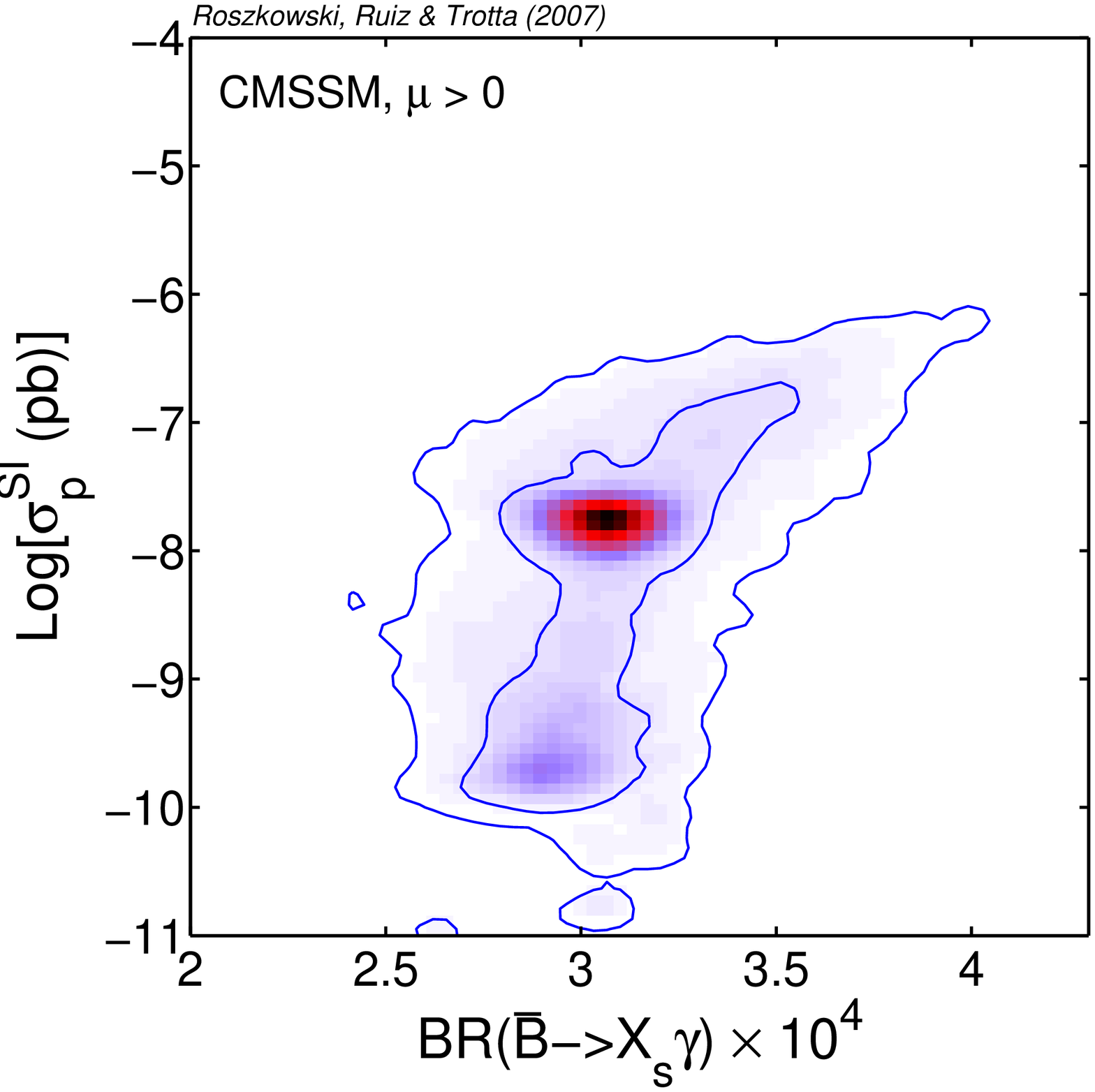}
& \includegraphics[width=0.3\textwidth]{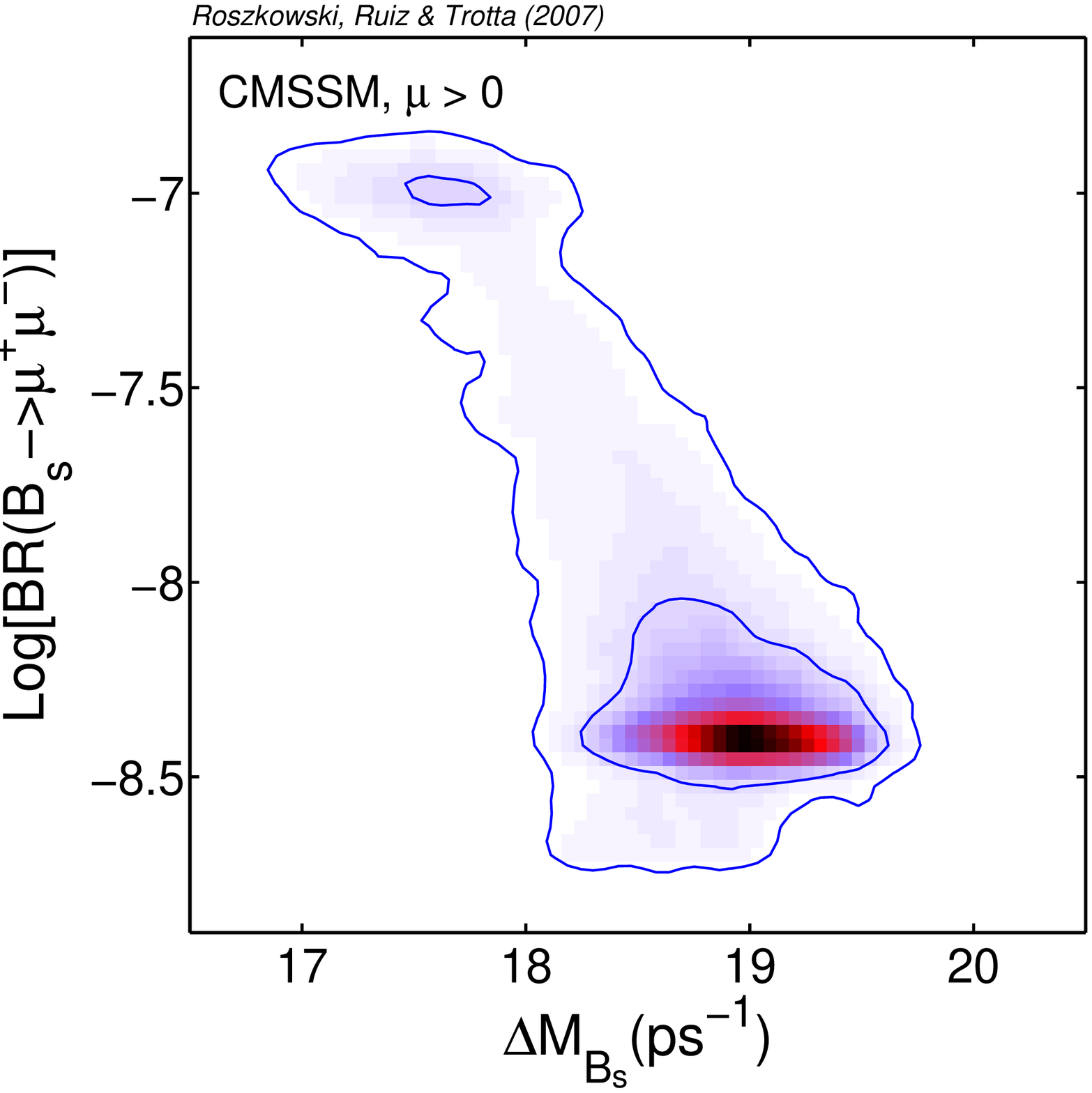}
\end{tabular}
  \includegraphics[width=0.3\textwidth]{rrt3-colorbar.ps}
\end{center}
\caption{The 2-dim relative probability density for pairs of selected
variables for $\mu>0$. The inner (outer) solid contours delimit the
regions of 68\% and 95\% total probability, respectively. All other
parameters have been marginalized over.
This figure should be compared with figure~14 in
ref.~\protect\cite{rtr1}.
\label{fig:correlations-mup-pdf}
}
\end{figure}

\subsection{Correlations among observables}\label{sec:correlations}

The probabilistic approach employed in this analysis makes it very
easy to examine various possible global correlations among different
observables in the CMSSM.  For example, in
fig.~\ref{fig:sigsipvssigsdp-pdf} we present the 2-dim pdf of
$\sigsip$ versus $\sigsdp$ for $\mu<0$ (left panel) and $\mu>0$ (right
panel). For both signs of $\mu$ we can see a big concentration of
probability density on $\sigsip\sim 10^{-8}\pb$ and $\sigsip\sim
10^{-4}\pb$, which is a reflection of their behavior in
fig.~\ref{fig:sigsipvsmchi-pdf}) and fig.~\ref{fig:sigsdpvsmchi-pdf},
respectively -- an effect of the FP region. For $\mu>0$ we can see an
additional feature of a positive correlation which is due to the
contribution from the Higgs resonance and/or coannihilation effects.

More correlations are displayed in fig.~\ref{fig:correlations-mup-pdf}
for the case of $\mu>0$. Again, the effect of the FP region is
overwhelming. In all the observables, other than the  dark matter SI cross
section, SUSY effects likely to be tiny -- the probability density is
clearly peaked at the respective SM values. For $\mu<0$ (not
displayed) the concentration is typically even
stronger. Fig.~\ref{fig:correlations-mup-pdf} should be compared with
fig.~14 of ref.~\cite{rtr1} in order to appreciate the change. Some
correlations which were quite well pronounced in that figure (eg. in
$\sigsip$ versus $\brbsmumu$) are now barely visible.

\section{Summary and conclusions}\label{sec:summary}

We have applied the highly efficient MCMC scanning method to explore
the parameter space of the CMSSM and outlined the regions favored by
the Bayesian posterior probability and, for comparison, by the mean
mean quality-of-fit. We assumed flat priors in the usual CMSSM
parameters, applied and updated all relevant experimental constraints
from colliders and cosmological dark matter abundance, while paying
particular attention to the impact of the recent change in the SM
value of $\brbsgamma$. We examined both signs of $\mu$.  For both
choices, we found strong preference for the focus point region with
large $\mzero$ in the few TeV range (for $\mu<0$ actually saturating
at the assumed prior boundary of $4\tev$), and not as large
$\mhalf\lsim2\tev$. For comparison, the mean quality-of-fit measure
selected a small number of isolated regions giving good fit to the
data for $\mu>0$ but not for $\mu<0$.  It appears that the choice
$\mu<0$ is more at odds with the data, as reflected by its worse
quality-of-fit.

We then examined ensuing implications for Higgs and superpartner
searches and for direct detection of dark matter. Prospects
for the Tevatron of excluding the whole 95\%~\cl\ range of $\mhl$
remain very good, or else there is very good hope to see at least
some evidence of a signal with the expected final integrated
luminosity. Scalar superpartner masses are typically heavier than
$1\tev$ (compare table~\ref{table:massesTable}) but there is a
reasonable chance for at least some squarks (but probably no
sleptons) to be seen at the LHC. On the other hand, the gluino,
while also preferably heavy, should have a much better chance of
discovery at the LHC. Prospects for detection are also promising
in direct detection searches for dark matter which are sensitive
to spin-independent interactions. An improvement in sensitivity
down to $\sigsip\sim10^{-8}\pb$ will allow CDMS-II and other
experiments to reach down to the bulk of the values favored by the
posterior pdf (for both signs of $\mu$), and actually to fully
explore all best-fit regions in the preferred case of $\mu>0$. On
the other hand, an improvement in sensitivity of at least 3 orders
of magnitude will be required before favored ranges of cross
sections for spin-dependent neutralino-proton interactions are
tested by experiment.

We stress that some of our findings do depend
on our choice of the (usual) variables $\mzero$, $\mhalf$,
$\azero$ and, especially, $\tanb$ as CMSSM parameters over which
we scan, and on the choice of taking flat priors on those
variables. Of particular relevance is the upper bound
$\mzero<4\tev$. Other choices are possible and should be examined.
(See refs.~\cite{allanach06,alw07-natpriors} and Note Added
below.) The choice of CMSSM parameters and flat priors that we
have made in the present analysis and the comparison with the
mean-quality-of-fit statistics are meant to
facilitate comparison with fixed-grid studies using similar
assumptions.

\bigskip
{\bf Note added:} Very recently, after our analysis was completed, a
new paper of Allanach \etal,~\cite{alw07-natpriors} has appeared. The
authors argue that, instead of assuming a flat prior in $\tanb$, it is
more natural to use a ``REWSB prior'' where $\mu$ and the bi-linear
soft mass parameter $B$ are taken as inputs with flat
distributions.  Whether this choice (originally advocated by
R.~Ratazzi) is superior to any other is debatable but it is certainly
justifiable to apply it, at least for the sake of examining the
sensitivity of observables to the choice of priors.  We note that with
the REWSB prior the preference for large $\mzero$, well above $1\tev$,
still remains. (A more detailed comparison is difficult because in
ref.~\cite{alw07-natpriors} a previous SM value of $\brbsgamma$ has
been used.) On the other hand, the authors in addition choose to give
strong preference to cases where all the CMSSM mass parameters are of
the same order. In our opinion this assumption does reflect a certain
level of theoretical bias which at the end strongly changes the
conclusions obtained with the REWSB prior only. (In particular it
disfavors the focus point region which, as we have shown, can be seen
as being favored by the current results on $\brbsgamma$.)  We would
prefer to see the lack of the hierarchy of the CMSSM mass parameters
to be an outcome of applying experimental constraints, rather than of
applying theoretical prejudice.

\medskip
{\bf Acknowledgements} \\ We are indebted to M.~Misiak for
providing to us a part of his code computing the SM value of
$\brbsgamma$ and for several clarifying comments about the details
of the calculation. L.R is partially supported by the EC
6th Framework Programmes MRTN-CT-2004-503369 and
MRTN-CT-2006-035505. R.RdA is supported
by the program ``Juan de la Cierva'' of the Ministerio de
Educaci\'{o}n y Ciencia of Spain. RT is supported by the Lockyer Fellowship of
the Royal Astronomical Society and by St Anne's College, Oxford.
The author(s) would like to thank the European Network of
Theoretical Astroparticle Physics ENTApP ILIAS/N6 under contract
number RII3-CT-2004-506222 for financial support. This project
benefited from the CERN-ENTApP joint visitor's programme on dark
matter, 5-9 March 2007.



\end{document}